\documentclass[twocolumn]{aastex701}
\usepackage{multirow}
\usepackage{rotating}
\usepackage{amsmath}
\usepackage{graphicx}

\newcommand{\F}{\color{black} \text{{\scriptsize F}}} % Plum
\newcommand{\HFB}{\color{black} \text{{\scriptsize H}}} % Cyan
\newcommand{\T}{\color{black} \text{{\scriptsize T}}} % ForestGreen
\newcommand{\A}{\color{black} \text{\footnotesize *}} % Red

\begin{document}

\title{Identifying observable MeV lines from the decays of weak and main $r$-process isotopes in mergers}

\correspondingauthor{Nicole Vassh, Xilu Wang}
\email{nvassh@triumf.ca, wangxl@ihep.ac.cn}

\author[0009-0005-1538-7951]{Maude Larivi\`ere}
\affiliation{TRIUMF, 4004 Wesbrook Mall, Vancouver, BC V6T 2A3, Canada}
\affiliation{Department of Physics and Astronomy, University of British Columbia, Vancouver, BC V6T 1Z1, Canada}
\email{maudelar@phas.ubc.ca}

\author[0000-0002-3305-4326]{Nicole Vassh}
\affiliation{TRIUMF, 4004 Wesbrook Mall, Vancouver, BC V6T 2A3, Canada}
\email{nvassh@triumf.ca}

\author{Yanwen Deng}
\affil{State Key Laboratory of Particle Astrophysics, Institute of High Energy Physics, Chinese Academy of Sciences, Beijing 100049, China}
\email{yanwen.deng@ihep.ac.cn}

\author[0000-0002-5901-9879]{Xilu Wang}
\affil{State Key Laboratory of Particle Astrophysics, Institute of High Energy Physics, Chinese Academy of Sciences, Beijing 100049, China}
\email{wangxl@ihep.ac.cn}

\author[0000-0002-4729-8823]{Rebecca Surman}
\affiliation{Department of Physics and Astronomy, University of Notre Dame, Notre Dame, IN 46556, USA}
\email{rsurman@nd.edu}

%% Note that the \and command from previous versions of AASTe$\F$ is now
%% depreciated in this version as it is no longer necessary. AASTe$\F$ 

%% AASTe$\F$ 6.31 has the new \collaboration and \nocollaboration commands to


\begin{abstract}
We consider predictions for the MeV gamma-ray spectrum emitted by the $\beta$ decays of freshly synthesized isotopes from a neutron star merger at timescales of relevance for post-merger (days) and remnant (years) emission. We develop a search algorithm to identify observable spectral peaks and then determine if a specific isotope has a dominant emission line producing the spectral feature. We predict emission spectra using nucleosynthesis calculations which consider nuclear models with distinct masses, $\beta$-decays, and fission properties as well as variations on main ($A>130$) and weak ($A<130$) $r$-process astrophysical conditions. We tabulate all lines from decaying isotopes that our procedure identifies and provide the predicted range in time over which each line could be visible. We find that Rh-106 presents a unique opportunity to distinguish between main and weak $r$-process emission, as our calculated spectrum above $\sim 1$ MeV for an event dominated by the weak $r$ process is identical to the Rh-106 emission spectrum from $\sim$ 0.2 to $\sim$17 years. We further find emission from species such as Hf-181, Ta-182, Ta-184, and Re-188 offers the potential to be able to distinguish between nuclear models. We investigate whether the 2.6 MeV strong gamma-ray line from Tl-208 is predicted to be robustly observable across calculation variations on both timescale of days and years. We find Tl-208 to consistently shine through on the order of years, though it can face competition from Ga-72 and La-140 at early times ($\sim$ days). We additionally highlight numerous isotopes of interest for observation and nuclear experiment.
\end{abstract}

\section{Introduction}\label{sec:intro}

Gamma rays from astrophysical events carry with them the memory of the extreme conditions that forged them. Of particular interest in nuclear astrophysics are the gamma-ray emission lines from the decays of exotic nuclei, which can be used to discern the isotope composition produced by an event. 

Nuclear decays occur on distinct timescales, ranging from seconds to billions of years, and produce experimentally established gamma emission lines ranging from fractions to tens of MeV. Thus the decay timescale along with the emission line features for a given nuclear species provide a potentially unique combination that can be extracted from the the spectrum of an astrophysical event, if the isotope is synthesized and the gammas are observed. Nuclear gamma-ray emission is generally at its strongest in the days after the event. At these early times, many freshly-synthesized, short-lived species decay, and their emission can produce competing signals that may be challenging to untangle. On longer timescales, such as those relevant for remnants, the signals are fainter but fewer species contribute, making extraction of the contributions of specific isotopes more straightforward.

Gamma-ray lines from the decay chains of specific isotopes (e.g.,  $^{44}{\rm Ti}$, $^{56}{\rm Ni}$ and $^{60}{\rm Fe}$) are widely used to probe nucleosynthesis and emissions from supernovae \citep[e.g.,][]{SN1987A, SNIa2014J, Wang2019} and to identify supernovae remnants properties \citep[e.g.,][]{CasA2015,CasA2020, diffusive2007,Liu2025}. The upcoming COSI telescope and prospective next-generation MeV telescopes such as MeVGRO, AMEGO/AMEGO-X, e-ASTROGAM, GRAMS, APT, etc., \citep{COSI, AMEGO, eASTROGAM, GRAMS, APT} provide new opportunities to map the gamma-ray sky, search for remnants, and generally be ready to observe any real-time local merger or supernovae events. Telescope campaigns can only truly address the question of the ultimate origin of elements should they hunt for signatures from the heavier species produced by the rapid neutron capture process ($r$-process). Astrophysical events with $r$-process ejecta present an incredible challenge and opportunity for MeV telescopes. Gamma rays offer the most promising path to direct detection of isotopic composition and yields. However, hundreds of radioactive species are produced in the $r$ process, making the signal potentially difficult to decode. To date gammas explicitly linked to $r$-process production have yet to be observed.

A dynamical understanding of the evolution in composition can help track when gamma rays from given isotopes are produced. Given the rare nature of $r$-process events, providing a possible mapping between the presence of specific elements and visible gamma-ray spectral features is crucial to assist telescope campaigns in their searches. Accordingly, studies of gamma-ray emission from the decays of nuclear species produced by the rapid neutron capture process have begun in recent years  \citep{Hotokezaka2016_MeV,Li2019_MeV,Korobkin2020_MeV,Wang_fgam, Chen2021_MeV, Terada2022_MeV, VasshTl208, Liu2025_CCSNeMeV,Gross2025}. The species at and above the third $r$-process peak are of particular interest, as the GW170817 kilonova indicated production of lanthanides but no direct evidence for heavier elements. Recently two particularly telling gamma signatures of these species were proposed:
(1) the 2.6 MeV emission line from the $\beta$-decay of Tl-208 \citep{VasshTl208} and (2) the dominance of prompt fission gamma emission above $\sim$3.5 MeV \citep{Wang_fgam}. At lower energies, several isotopes which could contribute emission lines to the spectrum have also been noted, with some of the more frequently reported being: Ga-72, Zr-95, Nb-95, Rh-106, Ag-112, Sn-125, Sn-127, Sb-125, Sb-126, Sb-127, Sb-128, I-129, I-131, Xe-135, La-140, Pb-214, Bi-214, Pa-233, Np-239, Am-241, Am-243, Cm-245. Such studies tend to report the nuclei responsible for emission lines seen in the calculated spectrum at a given snapshot in time such as 10 days versus 10 years. However, given the wide range of timescales on which nuclear decays can occur, only a thorough search across time can catch possible emitters of interest. Additionally, the variance in the reported possible emission lines across the literature highlights the relatively sensitive nature of the problem to the exact nuclear data and astrophysical conditions assumed.

In this work we calculate the gamma-ray emission expected from $r$-process production in mergers. Our study goes beyond previous work in this area as we search for important emitters across time from days to thousands of years post merger while explicitly considering the impact of nuclear physics uncertainties on the interpretation of MeV gamma-ray spectra. We start by considering a mass-weighted average of nucleosynthesis calculations representative of a full neutron star merger simulation. We recalculate the nucleosynthesis using three nuclear models with distinct masses, reactions, decays, and fission products, as well as two distinct $\beta$-decay treatments. We generate the gamma-ray spectra as a function of time for each set of simulations and apply a peak finding algorithm to identify interesting features. We then back out which nuclear species dominates the emission in the energy peak of the identified peak at that time and report the duration of its influence on the light curve. We then relax the assumption on the overall abundance pattern produced in a merger event and consider a grid of different astrophysical conditions, repeating the analysis.

We describe our nucleosynthesis calculations, radiation transfer methods, and peak finding procedure in Sec.~2. In Sec.~3 we present results for dominant emitters from the mass weighted sum of ejecta from a neutron star merger event given all nuclear model inputs, both for a real-time event as well as remnant timescale emission. In Sec.~4 we outline the astrophysical variations considered in this work for main and weak $r$-process cases, and report the dominant emitters across conditions, again considering nuclear model variations for post-merger emission as well as remnant timescales. In Sec.~5 we consider in detail the emission in the energy range of the 2.6 MeV line from Tl-208 to explore the overall robustness of Tl signals as well as potential competition with other nuclei. In Sec.~6 we focus on the handful of nuclei undergoing $\beta$-decay found to emit above 3.5 MeV and compare their emission signatures to that of the continuous spectrum expected for gammas from prompt fission. We conclude in Sec.~7.

\section{Methods}\label{sec:findlines}

We aim to point back to the presence of specific nuclei from a predicted total emission spectrum by finding peaks indicative of distinct emission lines. We adopt a similar procedure as that in \cite{Wang_fgam} and \cite{VasshTl208} to estimate the total emission spectrum as follows. We first use the reaction network PRISM \citep{Mumpower2018, Sprouse2020} to calculate the nucleosynthesis for the mass weighted total ejecta from a neutron star merger event using simulation results from \cite{Radice2018}. We use nuclear data inputs which have considered nuclear masses and corresponding fission barriers for each model to calculate fission rates as in \cite{VasshFiss}, here focusing on Finite Range Droplet Model (FRDM2012) \citep{FRDM2012_Moller}, Hartree-Fock-Bogoliubov (HFB27) \citep{HFB27_Goriely}, and Thomas-Fermi (TF) \citep{TF_Myers} cases. The fission yield distributions are taken to be those of GEF16 \citep{GEF_Schmidt} with fission gammas spectra data from GEF16 as well so that our yields and fission emission spectra are consistent. For the reaction rates of lighter elements, we apply the REACLIB database \citep{Cyburt_2010}. Our network also incorporates AME2020 \citep{AME2020_Wang} and NUBASE2020 \citep{NUBASE2020_Kondev} experimental data for the masses and decay rates / branchings of all channels (e.g. $\beta$, $\alpha$, and spontaneous fission), respectively.

\begin{figure}[h!]
    \centering
    \includegraphics[width=0.9\linewidth]{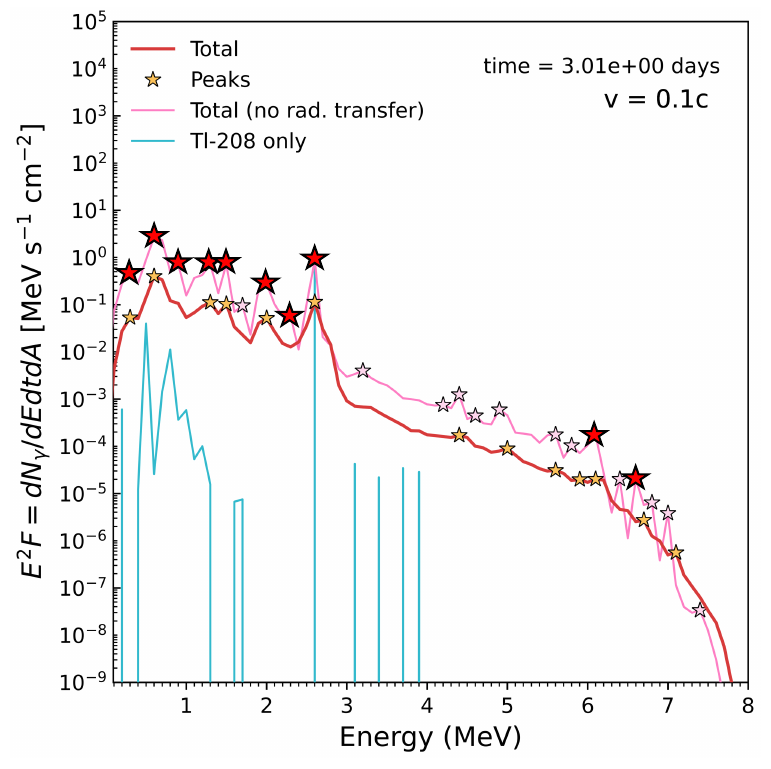}
     \hspace{0.25cm}
    \includegraphics[width=0.9\linewidth]{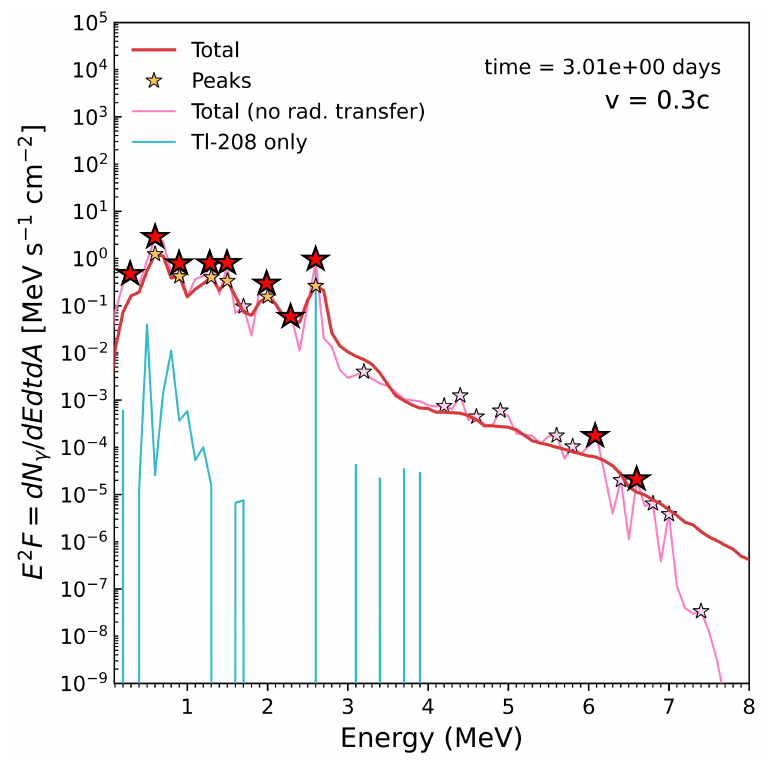}
    \caption{The total emission spectrum from $\beta$-decay at 3 days for a neutron-rich astrophysical trajectory ($Y_e = 0.01$, entropy $s/k=10$, and expansion timescale $\tau = 12$ ms) calculated with FRDM2012 nuclear inputs. We consider predicted spectra both with and without radiation transfer with ejecta expansion velocities of $0.1c$ (top panel) and $0.3c$ (bottom panel). The light-colored smaller stars highlight the peaks determined by the peak finder while the larger red stars indicate cases which satisfy our table criterion that an identified isotope dominates the spectrum in a given bin as well as having a flux above the projected sensitivity limit of near-future telescopes.}
    \label{fig:peakFinder_snap}
\end{figure}

To obtain the predicted total emission spectrum between 0.1 and 20 MeV, we extract the relevant decay spectra from the ENDF/B-VIII.1 database \citep{ENDF8_Brown}. Here we focus on gamma-ray emission from $\beta$-decay, though we also discuss the distinct emission regimes of fission and $\alpha$-decay in Sec.~\ref{sec:fgam} and the Appendix (Sec.~\ref{sec:appendpopmech}), respectively. Our calculations as described above provide predictions for the emission spectrum from a 1.2-1.4 M$_{\odot}$ neutron star merger at 10 kpc across time. In order to search for peaks in the spectrum from early (1 hour) to remnant timescales (taken here to be $>$10 years), we develop an algorithm based on the Python \texttt{scipy.signal} \texttt{find\textunderscore peaks} function. This algorithm identifies local maxima by comparing each value with its neighbors. We apply our peak finder algorithm to the total spectra at each timestep. An example spectrum is given in Fig.~\ref{fig:peakFinder_snap}, where the energy bins containing identified local maxima are highlighted. The blue lines in Fig.~\ref{fig:peakFinder_snap} show the gamma lines of a single emitter and thus illustrate how a detected peak can be caused by the presence of a specific isotope, here Tl-208. 

Our procedure to evaluate whether a predicted spectral peak can be used to identify the presence of an individual isotope is as follows. We first consider each isotope that has a $\beta$ flow at the timestep of interest, where the $\beta$ flow, defined as the $\beta$-decay rate times the abundance for species ${Z,A}$, is a standard PRISM output. We then rank each species with a nonzero $\beta$ flow from highest to lowest in their flux contribution to the total spectrum. This results in an exhaustive list of nuclei contributing to each energy bin. For each bin where a peak has been detected, we then consider the total lightcurve around the energy of interest $\pm 0.1$ MeV (we consider one bin on each side of the peak to account for broadening). We then compare the total lightcurve to the calculation without the contribution from each of our listed isotopes. To report a given case, we require that the contribution from the isotope dominates emission in that it accounts for $\geq 50\%$ of the flux in the energy bin of interest. This enforces that we only report emission lines that can be uniquely tied to a given isotope during a given time interval. We extract a signal duration range by reporting the time interval over which an isotope satisfies both the $\geq 50\%$ of the flux in the energy bin of interest criterion as well as requiring the flux be above a threshold inspired by near-future telescopes. If the flux falls below this threshold, we report the last time the signal is detectable above threshold. For our results on post-merger timescales of days, we enforce a threshold of $1\times 10^{-7} \text{ counts} \text{ s}^{-1} \text{cm}^{-2}$ to the lightcurve. Assuming an observation time of 1 day, this represents an optimistic sensitivity limit for future MeV gamma ray telescopes such as COSI, AMEGO, eAstrogam, MeVGRO, and GRAMS \citep[][also see the table of sensitivity comparisons in \cite{Liu2025}]{COSI, AMEGO, eASTROGAM, MEVGRO, GRAMS}.  Pushing to this limit allows an exploration of the maximal realistic search range of the gamma-ray lines. 

We first run the peak finder on the total emitted spectrum before considering the photon propagation process (radiation transfer) to identify all the possible nuclei that could be dominant emitters at a given timescale. We then perform radiation transfer calculations assuming two different ejecta expansion velocities (maximal velocities at the outmost ejecta radius $v_{\rm max}$ assumed to be $0.1c$ or $0.3c$) to obtain predictions for the spectra after the propagation of photons through the merger ejecta.  We then rerun the peak finder on the results to test which spectral lines will survive the propagation process and are still resolved peaks above threshold.

To model the propagation process, we adopt the shell plus core ejecta model as described in \cite{Wang_fgam} and \cite{VasshTl208} which assumes homologous expansion of the ejecta. Here we consider a total ejecta mass of 0.01 solar mass and a distance of 10 kpc  (to merger or merger remnant). Therefore here our calculations consider galactic events, however we have found signals could potentially be visible for events in a nearby galaxy (i.e. out to the Local Group \citep{VasshTl208}). We build a radiation transfer calculation based on a Monte Carlo radiative transfer code from \cite{Leung2023} with several adjustments applied to be suitable for merger scenario: (1) we expand the 1D code to 3D; (2) we adopt the criteria  of \cite{Milne2004} to determine the photon interaction process at each timestep following photon propagation; (3) the propagating gamma-ray photon interactions in our calculation include Rayleigh scattering, Compton scattering, photoelectric absorption, and pair production. The Doppler effect is also considered during the photon propagation. Instead of using analytical functions to calculate the cross-sections of several fixed isotopes for the ejecta opacity as in \cite{Leung2023}, our opacity calculations for these interactions with the ejecta is based on the ejecta’s composition with a mixture of the opacities of the existing $r$-process isotopes at each timestep. The opacity value of each isotope (from $Z=1$ to $Z=100$; the opacity value for isotopes with $Z>100$ assumed to be the same as $Z=100$) is adopted from the XCOM database\footnote{\url https://www.nist.gov/pml/xcom-photon-cross-sections-database}.
An example of the impact of the gamma-ray photon propagation on the spectral lines can be seen in the two panels of Fig~\ref{fig:peakFinder_snap}. A comparison of the spectra pre- (pink lines) and post- (red lines) radiation transfer demonstrates two main effects. First, on timescales of days before the ejecta becomes optically thin, the gamma-ray flux is attenuated by photon-ejecta interactions. Second, the spectral lines are broadened and smoothed by the Doppler effect. By comparing the upper and lower panels, we can see the influence of the ejecta's expansion velocity. Faster-expanding ejecta ($v_{\rm max} =0.3c$) becomes optically thin sooner as the ejecta density decreases more rapidly; consequently, the observed flux at $\sim3$ days is higher than that of the slower ejecta ($v_{\rm max} =0.1c$). However, this higher velocity also enhances Doppler broadening, resulting in a much smoother spectrum, whereas the 0.1c case retains more distinct spectral lines. Despite these kinematic differences, when a peak is consistently identified by our algorithm, the dominant isotope responsible for the primary gamma-ray emission is also consistent as will be reflected in the results of subsequent sections.

\section{Spectral lines from nuclei produced in mergers considering nuclear physics variations}\label{sec:nsmw3models}

We perform the procedure described in the last section to calculations which consider the mass weighted total ejecta from a neutron star merger event using simulation results from \cite{Radice2018}. The particular simulation set-up from this work that we utilize is for a 1.2-1.4 M$_{\odot}$ progenitor mass case which applied the SFHo equation of state and M0 neutrino treatment. This simulation reports ejecta which is typically low entropy with a range in neutron-richness (i.e. electron fractions between $0.01<Y_e<0.4$) such that full $r$-process pattern from the first peak at $A\sim80$ to the third peak at $A\sim195$ and beyond into the actinides is produced, with high mass ejection of very neutron-rich components ($Y_e<0.1$) corresponding to high actinide production.

\begin{figure}[h]
    \centering
    \includegraphics[width=1.0\linewidth]{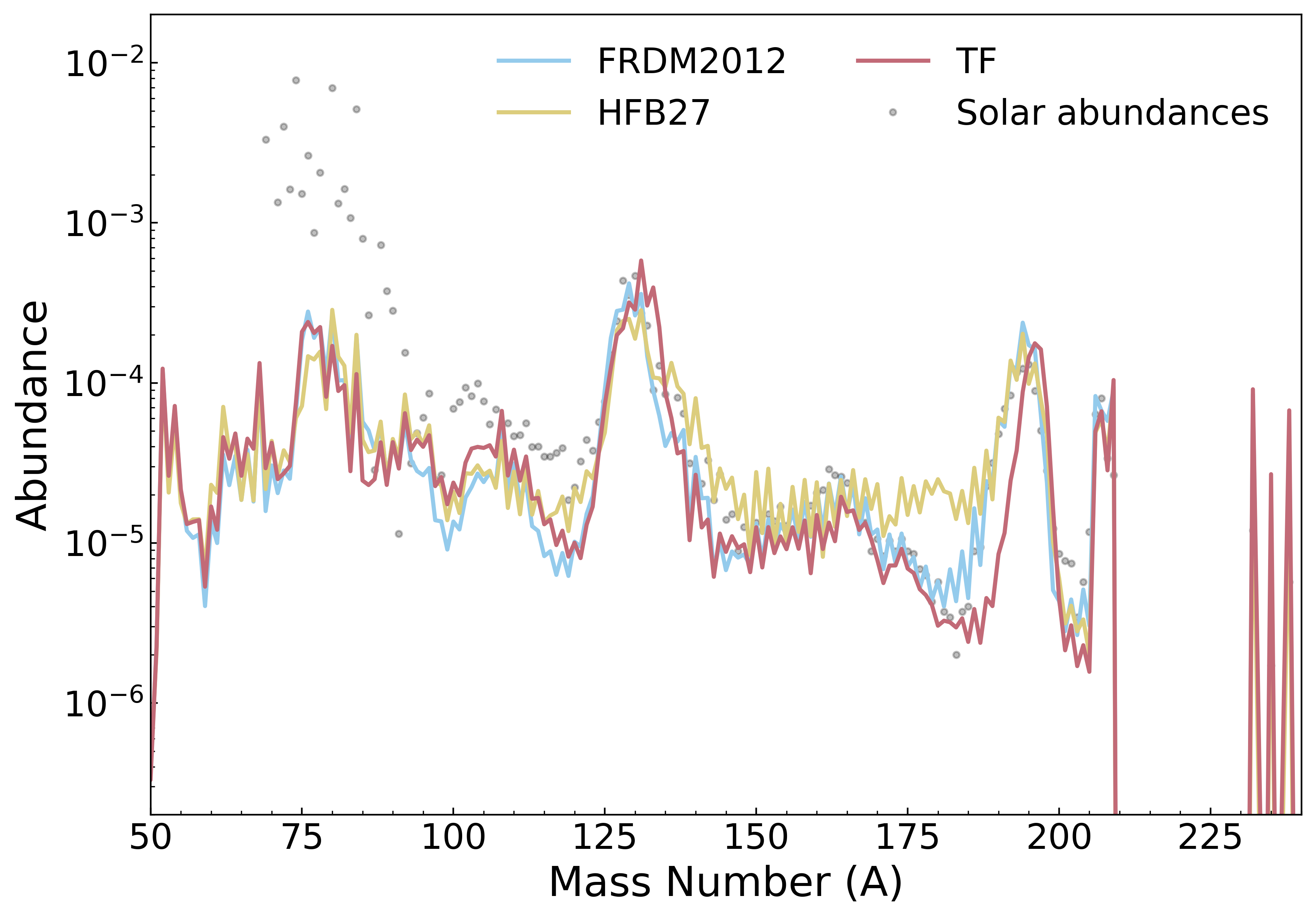}
     \hspace{0.25cm}
    \includegraphics[width=1.0\linewidth]{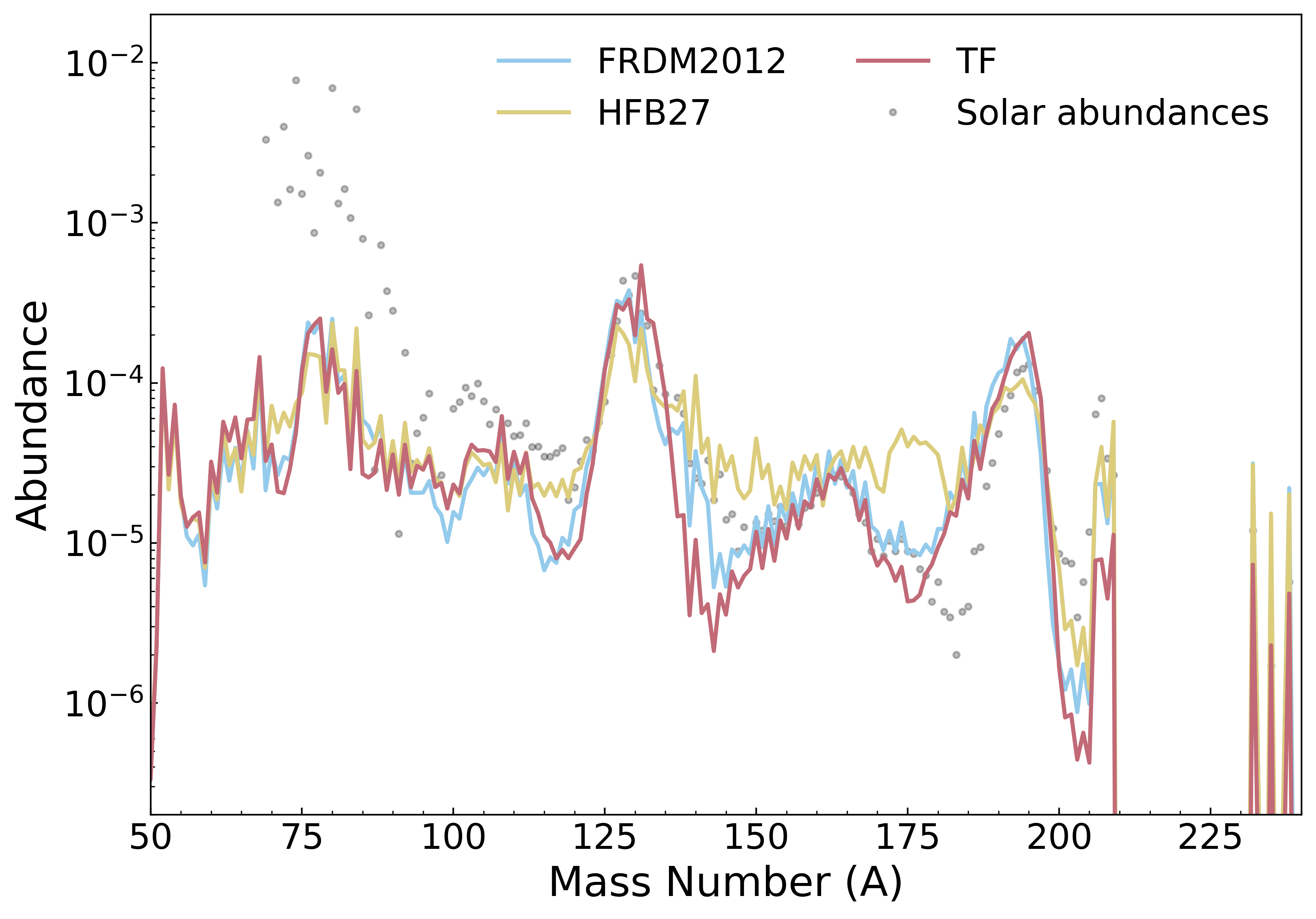}
    \caption{Comparison of predicted final isotopic abundances at 1 Gyr given three different nuclear mass models (FRDM2012, HFB27, TF) and two distinct $\beta$-decay treatments ((top) M\"{o}ller et al. and (bottom) Marketin et al.) for the 1.2 - 1.4 M$_\odot$ neutron star merger results from \cite{Radice2018} (assuming M0 neutrino treatment). The solar abundance data is taken from \cite{Sneden2008}.}
    \label{fig:massWeight}
\end{figure}

We run all astrophysical trajectories from a 1.2-1.4 M$_\odot$ neutron star merger simulation to find the total mass weighted abundances given three nuclear mass model sets (FRDM2012, HFB27, and TF) as well as two distinct $\beta$-decay treatments from M\"{o}ller et al. \citep{MollerSd0} and Marketin et al. \citep{Marketin}. We perform a separate PRISM run for each trajectory up to 20 seconds, when the neutron captures have mostly ceased but before decays have reached the timescales of relevance for observation. We then mass weight the composition of all simulation reported trajectories at 20 s to represent the total ejecta and utilize this as an initial composition for a network calculation running to long timescales. The mass-weighted total abundance patterns for calculations with all three nuclear mass models with two $\beta$-decay treatments are shown in Fig.~\ref{fig:massWeight}. Clear differences in rare-earth element abundances ($A\sim150-180$) as well as third peak species near $A\sim195$ demonstrate the impact of nuclear models becoming increasingly variant for heavier nuclei. The degree of nuclear deformation predicted in the rare-earth region often varies across models having a direct impact on the shape of the rare-earth abundance peak and the population of species close in mass number such as near $A\sim180$. The impact of uncertainties from $\beta$-decay treatments is also evident, particularly for third peak and actinide abundances where a wide third peak and low actinides is commonly seen when using the Marketin et al. model which predicts fast $\beta$-decays past $N=126$, clearing out actinide abundances at early times.

We next run our peak finder algorithm on the predicted spectra from the mass-weighted calculations for all six nuclear variations considered. We present the isotopes which are found to have a contribution $\geq$ 50\% to identified peaks in Table \ref{tab:massweight_days} which includes the predicted time over which the signal is visible, the line energy, and the nuclear models for which these features in the total spectra could be detected. We first focus on signals that are visible shortly after a merger. To account for sky localization and telescope turnover time, we report on signals after 1 hour ($\sim$0.04 days). We denote the start time as the earliest time across all three models at which the isotopes reach a $\geq$ 50\% contribution to the lightcurve in a given energy bin. Similarly, we report the end time as the latest time across all models at which the isotope drops below having a 50\% contribution (or the time at which the total signal falls below detectability, here set at $10^{-7}$ counts s$^{-1}$ cm$^{-2}$). To indicate under which model assumptions a signal is predicted to be observable, we denote each mass model by a letter: F for FRDM2012, H for HFB27 and T for TF. We denote signals produced in all models with a *.  Since our procedure makes use of 0.1 MeV bins, cases exist with multiple lines in a bin so here we bold the strongest. We disregard peaks that are mapped back to theoretical additions in the ENDF spectral data as in this work we aim to report only experimentally measured emission lines in the tables. Note that we do not provide the half-lives of each species in the tables here, since for many cases it is not their half-life which is setting the relevant emission timescale reported in calculations. Rather we report their half-lives in the Appendix Tables~\ref{tab:BemittersBdec} and \ref{tab:BemittersAFdec}, along with providing the relevant decay chain responsible for the emission found to be potentially observable in our calculations.

\begin{longtable*}[]{|c|c|c|c|c|}
    \caption{Signal duration range for post-merger emission lines predicted to dominate the total spectrum which last longer than 1 day and start some time between 1 hr and 10 yrs. Here mass weighted ejecta from a merger simulation which produces both main and weak $r$-process nuclei was considered. Calculation variations included three nuclear mass models FRDM2012 ($\F$), HFB27 ($\HFB$), and Thomas-Fermi ($\T$) along with two distinct $\beta$-decay treatments from M\"{o}ller et al. and Marketin et al. After a peak is identified, to be featured an isotope must dominate the lightcurve ($\geq 50\%$) in the emission bin with flux above a detectability threshold of $10^{-7}$ counts s$^{-1}$ cm$^{-2}$. Entries marked grey will ultimately not be observable, either due to emission being below that of prompt fission gammas (when isotopes and energies are grey) or due to being washed out following radiation transfer (grey denotes peak disappearance assuming 0.1c while italics denotes 0.3c results). All cases labeled in black and non-italic survive to be a visible peak post-radiation transfer. A $*$ denotes when a line was reported given all three mass model variations.}
    \label{tab:massweight_days} \\

    \hline
    \multirow{2}{3em}{\centering Isotope} & \multirow{2}{6em}{\centering Time (days)} & \multirow{2}{4em}{\centering E (keV)} & \multicolumn{2}{|c|}{Nuclear model variations} \\
    \cline{4-5}
     & & & \multicolumn{1}{|c|}{M\"{o}ller $\beta$ decay} & \multicolumn{1}{|c|}{Marketin $\beta$ decay} \\

    \endfirsthead

    \multicolumn{5}{c}
    {{\bfseries \tablename\ \thetable{} -- continued from previous page}} \\
    \hline 
    \multicolumn{1}{|c|}{Isotope} & \multicolumn{1}{|c|}{Time (days)} & \multicolumn{1}{|c}{E (keV)} & \multicolumn{1}{|c|}{M\"{o}ller $\beta$ decay} & \multicolumn{1}{|c|}{Marketin $\beta$ decay} \\
    \hline
    \endhead

    \hline\endlastfoot
    \hline
    Fe-59 & 8.22e+01 -- 1.89e+02 & 1099 & $\T$ & $\T$ \\
     & 6.26e+01 -- 2.48e+02 & 1292 & $\T$ & $\T$ \\
    \hline
    Ga-72 & 2.63 -- 5.26  & 1862 & & $\HFB$ \\
     & 8.53e-01 -- 1.23e+01 & 2202 & $\A$ & $\A$ \\
     & 7.97e-01 -- 1.41e+01 & 2508 & $\HFB\T$ & $\A$ \\
     & 1.56 -- 1.48e+01 & 2844 & \scriptsize{\color{gray}\textit{FT}} & \scriptsize{{\color{gray}\textit{F}}\textit{HT}} \\
     & 1.75 -- 9.88 & \textbf{3325}, 3339 & \scriptsize{\textit{T}} & \scriptsize{\textit{T}} \\
     & 1.33 -- 6.86 & 3679 & & \scriptsize{\textit{T}} \\
    \hline
    Rb-88 & 3.41e-01 -- 1.01 & 3218 & & \scriptsize{\color{gray}\textit{T}} \\
     & 3.87e-01 -- 1.42 & 4036 & $\T$ & $\T$ \\
     & 2.00e-02 -- 1.97 & 4742 & \scriptsize{\textit{FT}} & \scriptsize{\textit{F{\color{gray}\textit{H}}}}$\T$ \\
    \hline
    %Rb-92 & 8.94e-01 -- 1.74e+02 & th & $\F$ & $\F\HFB$ \\
    Nb-95 & 9.85e+01 -- 2.77e+02 & 766 & $\T$ & \\
    \hline
    Ru-103 & 7.67e+01 -- 1.54e+02 & 497 & $\T$ & $\T$ \\
    \hline
    Rh-106 & 2.85e+02 -- 1.06e+03 & 2366 & \scriptsize{\textit{T}} & \scriptsize{\textit{T}} \\
    \hline
    Ag-112 & 8.28e-01 -- 2.98 & 2829 & \scriptsize{\color{gray}\textit{FT}} & \scriptsize{\color{gray}\textit{T}} \\
    \hline
    %Cd-127 & 3.43e+02 -- 3.91e+02  & th & $\HFB$ & \\
    {\color{gray}In-128} & {\color{gray}2.64e-01 -- 1.42} & {\color{gray}4298} & & $\T$ \\
    \hline
    Sn-125 & 1.89e+01 -- 4.31e+01  & \textbf{1067}, 1088, 1089 & & $\F\T$ \\
    \hline
    Sb-125 & 5.03e+02 -- 9.62e+02  & 176 & & $\T$ \\
     & 2.48e+02 -- 3.89e+03 & \textbf{428}, 463 & $\F\T$ & $\A$ \\
     & 3.85e+02 -- 8.44e+03 & \textbf{601}, 636 & $\F\T$ & $\A$ \\
    \hline
    %Sb-126 & 1.98e+04 -- 2.75e+04 & \textbf{666}, 695, 697 & $\F$ & $\F\T$ \\
    %\hline
    Sb-128 & 2.67e-01 -- 1.53 & \textbf{743}, 754 & $\F\HFB$ & $\A$ \\
    \hline
    {\color{gray}Sb-134} & {\color{gray}4.00e-03 -- 9.53} & {\color{gray}6687} & \scriptsize{\textit{H}} & \scriptsize{\textit{H}} \\
     & {\color{gray}5.00e-03 -- 5.99} & {\color{gray}6820} & \scriptsize{\color{gray}\textit{H}} & \scriptsize{\color{gray}\textit{H}} \\
    %Te-137 & 3.69e-01 -- 3.08e+02 & th &  $\A$  &  $\F\HFB$ \\
    \hline
    I-131 & 6.59 -- 5.43e+01 & 364 & $\A$ & $\F\T$ \\
    \hline
    I-132 & 2.56 -- 1.95e+01 & 630, \textbf{668} & $\A$ & $\F\T$ \\
     & 1.96 -- 1.86e+01 & 773 & $\HFB\T$ & \scriptsize{\color{gray}T} \\
     & 3.29 -- 1.27e+01 & 955 & $\T$ & $\T$ \\
     & 2.02 -- 2.15e+01 & 1372, \textbf{1399} & $\A$ & $\F\T$ \\
     & 1.46 -- 1.39e+01 & 2002 & $\F\T$ & $\T$ \\
    \hline
    {\color{gray}I-136} & {\color{gray}1.00e-03 -- 2.12e+02} & {\color{gray}6104} &  $\F\HFB$  &  $\F$\scriptsize{\textit{H}} \\
    \hline
    La-140 & 2.3 -- 1.42e+02 & 1596 &  $\A$  &  $\A$ \\
     & 1.23e+01 -- 1.03e+02 & 2521 & $\A$ & $\A$ \\
     & 1.99e+01 -- 3.84e+01 & 2899.6 & \scriptsize{\textit{T}} & \scriptsize{\textit{T}} \\
     & 1.06e+01 -- 3.84e+01 & 3119 & $\T$ & \scriptsize{\textit{T}} \\
    \hline
    Pr-144 & 2.08e+02 -- 1.65e+03 & 2186 & $\T$ & $\HFB$ \\
    \hline
    Eu-156 & 2.05e+01 -- 5.45e+01 & \textbf{1231}, 1242, 1277 & & $\A$ \\
     & 1.41e+01 -- 1.27e+02 & 1938, \textbf{1966} & \scriptsize{\textit{T}} & $\HFB\T$ \\
     & 1.02e+01 -- 1.35e+02 & 2027, \textbf{2098} & $\A$ & $\A$ \\
     & 1.59e+01 -- 1.35e+02 & 2181, \textbf{2187} & $\HFB$\scriptsize{\textit{T}} & \scriptsize{\textit{HT}} \\
    \hline
    Ir-194 
    %& 3.62e+03 -- 2.48e+04 & 328 & $\F\T$ & $\A$ \\
     & 1.32e+03 -- 1.45e+04 & 939 & & $\T$ \\
     & 5.78e+02 -- 1.57e+04 & \textbf{1151}, 1183 & $\F\T$ & $\A$ \\
     & 3.82e+02 -- 1.10e+04 & 1469 & & $\F\T$ \\
     & 7.14e+02 -- 8.44e+03 & 1622 & & \scriptsize{\textit{T}} \\
     & 7.14e+02 -- 7.23e+03 & 1806 & $\F$\scriptsize{\textit{T}} & $\F$\scriptsize{\textit{T}} \\
    \hline
    Tl-208 & 3.84e-01 -- 1.80e+01 & 2615 & $\A$ & \\
     & 1.39e+02 -- 1.90e+04 & 2615 & $\A$ & $\A$ \\
    \hline
    %Pb-214 & 2.98e+04 -- 2.84e+05 & 352 & $\F$ &  \\
    %\hline
    Ac-228 & 4.45e+02 -- 1.61e+04 & \textbf{911}, 969 & $\F\T$ & $\HFB$ \\
     & 5.04e+02 -- 1.39e+04 & 1588 & $\F\T$ & \\
\end{longtable*}

Across the three mass models and two $\beta$-decay variations, we report a total of 21 isotopes with lines having a contribution $\geq$ 50\% to the total spectrum in a single energy bin at timescales $\sim$ days, many of which have several potentially visible emission lines. We note that the peak finder highlighted two additional isotopes (Rb-92, Te-137) with a single line each arising from theoretical additions to the ENDF spectra, but these are omitted from our table as we focus only on established lines. With the signal duration range considering the earliest start time and latest end time reported across calculations for all three mass models and two $\beta$-decay variations, we note that this range can significantly differ across calculations. Typically the start / stop time during which we find a signal to be detectable varies within a factor of a few when considering different nuclear inputs, but it is not uncommon for these times to differ by an order of magnitude. A handful of cases (e.g. Sb-134) reported up to two orders of magnitude difference in their start or stop times across models. Even when all sets of calculations show production of a particular isotope, the relative abundances of emitting isotopes vary and therefore so does the range of time over which a species produces a signal above threshold. 

The case that reports the largest number of isotopes which could be linked to a viable signal is TF, which includes 19 out of the 21 isotopes in Table~\ref{tab:massweight_days}. Out of these 19, 5 rise above threshold only for TF (Fe-59, Nb-95, Ru-103, Rh-106, and In-128). Across both FRDM2012 and HFB27 calculations, we see 14 isotopes found to have viable emission lines, with 4 isotopes found to be robust across all nuclear data variations: Ga-72, La-140, Eu-156, and Tl-208. Since gamma-rays provide the potential to identify individual isotopes an exciting prospect of observations is to learn the ultimate reach of astrophysical nucleosynthesis. Thus it is interesting to note the highest mass number of a reported isotope for all models considered: Ac-228. The lightest isotope reported in Table~\ref{tab:massweight_days} is Fe-59 in the TF case, with a signal at 1.292 MeV being visible starting as early as 62 days. In the case of FRDM2012 and HFB27 calculations, the lightest isotope with a viable signal, Ga-72, is found to emit early, on the order of few days.

Considering the energies of the lines in Table~\ref{tab:massweight_days}, we note that most are below 3.5 MeV, with only 7 of our reported lines above 3.5 MeV. This is consistent with past work \citep{Wang_fgam} that demonstrated the majority of the contributions to the spectrum above 3.5 MeV will be due to prompt fission gammas, if fissioning nuclei are present in the ejecta. Of the eight lines above 3.5 MeV, roughly half are found between 3.5--4.7 MeV and the other half between 6.1--6.8 MeV, with the highest energy 6.82 MeV line of Sb-134 reported for the HFB27 model only. We note that many of these cases are ultimately due to fission producing neutron-rich species at late times and do not compete over prompt fission gammas, which will be discussed in Sec.~\ref{sec:fgam}.

Having considered a real-time kilonova observation occurring nearby, we next turn our attention to emission lines of interest when searching locally for remnants which may in fact be from neutron star mergers rather than supernova events. We consider remnant timescales ranging from 10 years to 1 Gyr and build a table using the same method outlined in Sec.~\ref{sec:findlines}. Here, however, we implement a lower lightcurve threshold of $10^{-10} \text{ counts s}^{-1} \text{cm}^{-2}$ considering that a neutron star merger remnant may potentially be located at a distance as near as a few $100$pc \citep{Wang2021}, and we aim to present the most complete list of lines that might possibly be visible on the timescale of years. These species are reported in Table \ref{tab:massweight_years}.

\begin{longtable*}[]{|c|c|c|c|c|}
    \caption{Signal duration predictions for emission lines of relevance to remnant searches as they are reported to be visible some time after 10 years post-merger. The same approach and notation as Table~\ref{tab:massweight_days} is also applied here, but with a detectability threshold of $10^{-10}$ counts s$^{-1}$ cm$^{-2}$ given the prospect of identifying nearby remnants.}
    \label{tab:massweight_years} \\
    \hline
    \multirow{2}{3em}{\centering Isotope} & \multirow{2}{6em}{\centering Time (years)} & \multirow{2}{4em}{\centering E (keV)} & \multicolumn{2}{|c|}{Nuclear model variations} \\
    \cline{4-5}
     & & & \multicolumn{1}{|c|}{M\"{o}ller $\beta$ decay} & \multicolumn{1}{|c|}{Marketin $\beta$ decay} \\

    \endfirsthead

    \multicolumn{5}{c}
    {{\bfseries \tablename \thetable{} -- continued from previous page}} \\
    \hline 
    \multicolumn{1}{|c|}{Isotope} & \multicolumn{1}{|c|}{Time (years)} & \multicolumn{1}{|c}{E (keV)} & \multicolumn{1}{|c|}{M\"{o}ller $\beta$ decay} & \multicolumn{1}{|c|}{Marketin $\beta$ decay} \\
    \hline
    \endhead

    \hline\endlastfoot
    \hline
    Co-60 & 4.44e+05 -- 4.82e+06 & 1173 & $\A$ & $\A$ \\
     & 1.56e+05 -- 4.21e+06 & 1332 & $\A$ & $\A$ \\
    \hline
    %Rb-92 & 2.00e-03 -- 1.09e+01 & th &  $\F$  &  $\HFB$ \\
    Rh-106 &  9.17e-01 -- 1.14e+01  & 2366 & & \scriptsize{\textit{{\color{gray}T}}} \\
    \hline
    {\color{gray}In-128} & {\color{gray}3.43e+01 -- 1.55e+02} & {\color{gray}3520} & $\F$ & \\
    \hline
    %Sn-126 & 9.10e+02 -- 1.55e+06 & th &  $\A$  &  $\A$ \\
    Sb-125 &  6.80e-01 -- 1.06e+01  & \textbf{428}, 463 & & $\F\T$ \\
     & 1.05 -- 2.31e+01 & \textbf{601}, 636 & $\F\T$ & $\A$ \\
    \hline
    Sb-126 & 8.74e+01 -- 2.10e+06 & 415 & $\A$ & $\A$ \\
     & 5.42e+01 -- 2.26e+06 & \textbf{666}, 695, 697 & $\A$ & $\A$ \\
     & 1.95e+04 -- 2.10e+05 & 1213 & $\F$ & \\
    \hline
    {\color{gray}Sb-134} & {\color{gray}1.00e-03 -- 8.85e+01} & {\color{gray}6687} & \scriptsize{\textit{H}}  & \\
     & {\color{gray}1.00e-03 -- 3.31e+01} & {\color{gray}6820} & \scriptsize{\textit{\color{gray}H}} & \\
    \hline
    %Te-137 & 5.29 -- 4.40e+02 & th &  $\HFB$  & $\HFB$ \\
    {\color{gray}I-136} & {\color{gray}7.00e-03 -- 3.02e+02} & {\color{gray}6104} & {\color{gray}$\HFB$} & {\color{gray}$\HFB$} \\
    \hline
    Ir-194 & 9.9 -- 7.22e+01 & 328 & $\F\T$ & $\A$ \\
     & 3.62 -- 4.98e+01 & 939 & & $\T$ \\
     & 1.58 -- 7.55e+01 & \textbf{1151}, 1183 & $\F\T$ & $\A$ \\
     & 1.05 -- 7.00e+01 & 1469 & & $\F\T$ \\
     & 1.96 -- 7.65e+01 & 1622 & & \scriptsize{\textit{T}}  \\
     & 1.96 -- 7.99e+01 & 1806 & $\F$\scriptsize{\textit{T}} & \scriptsize{\textit{FT}}  \\
     & 5.02 -- 5.93e+01 & 2044 & & \scriptsize{\textit{FT}}  \\
    \hline
    Tl-208 & 3.81e-01 -- 1.12e+02 & 2615 & $\A$ & $\A$ \\
    \hline
    Tl-209 & 1.01e+02 -- 6.47e+04 & 1567 & $\A$  & $\A$ \\
    \hline
    Pb-214 & 8.17e+01 -- 3.21e+03 & 352 & $\F\T$ & \\
    \hline
    %Bi-213 & 1.77e+06 -- 3.01e+06 & th & $\A$ & $\HFB$ \\
    %Ra-225 &  2.00e+06 -- 3.17e+06  & th & $\T$ & \\
    Ac-228 & 1.22 -- 4.41e+01 & \textbf{911}, 969 & $\F\T$ & $\HFB$ \\
     & 1.38 -- 5.18e+01 & 1588 & $\F\T$ & \\
\end{longtable*}

Even after implementing a lower threshold and considering calculations using three nuclear mass models and two $\beta$-decay variations, here we find only 9 nuclei whose emission lines could cause visible peaks in the emission spectrum from a neutron star merger remnant. Several isotopes (Rh-106, Sb-125, In-128, Sb-134, I-136, Ir-194, Tl-208, and Ac-228) appear in both the post-merger and remnant signal tables since their emission begins before 10 years and continues beyond. Note that the latest time we report the signal to be viable differs between post-merger and remnant tables due to the lower threshold we implement in the remnant tables to capture fainter emission. While many nuclei have emission lines that fall below our threshold on the order of tens to hundreds of years, we also find potentially detectable emission lines well beyond 10,000 years from three isotopes: Co-60, Sb-126, and Tl-209. These lines are especially robust as they are reported across almost all nuclear data variations. We also find Tl-208 emission on the order of years to be consistently predicted across model variations, which will be discussed in detail in Sec.~\ref{sec:tl208}.

\section{Spectral lines from nuclei given both astrophysical and nuclear physics variations}

\subsection{Astrophysical variations for weak and main $r$-process emission}\label{sec:astrovar}

As hydrodynamic simulations evolve, their predictions for the distribution of the ejecta (in terms of neutron-richness and other key outflow properties) also evolve. Additionally, variations in merger progenitor masses, equation of state, and neutrino treatments can significantly adjust the predicted $Y_e$ distribution for a given merger scenario. Thus here we move away from considering a mass weighted sum representing the total ejecta of a merger event towards considering individual trajectories in order to evaluate what nuclei may be reported should predictions change. Furthermore, should some $r$-process scenarios only produce weak $r$-process nuclei below ($A\sim130$) and fail to reach the main $r$-process nuclei beyond, there could be additional nuclear species that emerge in the lighter element regions to have visible emission due to less competition from heavier species emitting in their same energy window.

We thus expand our search for MeV gamma signals by considering an extended set of parametrized trajectories. We consider trajectories utilized to describe merger simulations in \cite{Radice2018} with $Y_e=0.01,0.07,0.13,0.16,0.19,0.25,0.29,0.32$ at three different entropies per baryon $s/k = 10, 32, 75$  (all with a dynamic timescale of $\tau = 12$ ms). This grid generated 8 weak $r$-process conditions and 15 main $r$-process conditions. These astrophysical variations are labeled in Table \ref{tab:astrovar}. With the three mass models considered, we therefore report on a total sample of 24 weak $r$-process and 45 main $r$-process calculations.

\begin{table}[]
\centering
\caption{The electron fraction $Y_e$ (lower when increasingly neutron-rich) and entropy per baryon $s/k$ for the 15 main $r$-process and 8 weak $r$-process conditions in this work. Each of these conditions will be considered alongside three nuclear models (TF, FRDM2012, and HFB27) to explore the impact of nuclear variations.}
 \label{tab:astrovar}
\begin{tabular}{|cclccl|ccl|}
\hline
\multicolumn{6}{|c|}{\textbf{Main}}                                                                                    & \multicolumn{3}{c|}{\textbf{Weak}}                          \\ \hline
\multicolumn{1}{|c|}{id} & \multicolumn{2}{c|}{$Y_e$, $s/k$} & \multicolumn{1}{c|}{id} & \multicolumn{2}{c|}{$Y_e$, $s/k$} & \multicolumn{1}{c|}{id}   & \multicolumn{2}{c|}{$Y_e$, $s/k$} \\ \hline
\multicolumn{1}{|c|}{1}  & \multicolumn{2}{c|}{0.01, 10}   & \multicolumn{1}{c|}{9}  & \multicolumn{2}{c|}{0.13, 75}   & \multicolumn{1}{c|}{1}    & \multicolumn{2}{c|}{0.25, 10}   \\
\multicolumn{1}{|c|}{2}  & \multicolumn{2}{c|}{0.01, 32}   & \multicolumn{1}{c|}{10} & \multicolumn{2}{c|}{0.16, 10}   & \multicolumn{1}{c|}{2}    & \multicolumn{2}{c|}{0.25, 32}   \\
\multicolumn{1}{|c|}{3}  & \multicolumn{2}{c|}{0.01, 75}   & \multicolumn{1}{c|}{11} & \multicolumn{2}{c|}{0.16, 32}   & \multicolumn{1}{c|}{3}    & \multicolumn{2}{c|}{0.29, 10}   \\
\multicolumn{1}{|c|}{4}  & \multicolumn{2}{c|}{0.07, 10}   & \multicolumn{1}{c|}{12} & \multicolumn{2}{c|}{0.16, 75}   & \multicolumn{1}{c|}{4}    & \multicolumn{2}{c|}{0.29, 32}   \\
\multicolumn{1}{|c|}{5}  & \multicolumn{2}{c|}{0.07, 32}   & \multicolumn{1}{c|}{13} & \multicolumn{2}{c|}{0.19, 10}   & \multicolumn{1}{c|}{5}    & \multicolumn{2}{c|}{0.29, 75}   \\
\multicolumn{1}{|c|}{6}  & \multicolumn{2}{c|}{0.07, 75}   & \multicolumn{1}{c|}{14} & \multicolumn{2}{c|}{0.19, 32}   & \multicolumn{1}{c|}{6}    & \multicolumn{2}{c|}{0.32, 10}   \\
\multicolumn{1}{|c|}{7}  & \multicolumn{2}{c|}{0.13, 10}   & \multicolumn{1}{c|}{15} & \multicolumn{2}{c|}{0.19, 75}   & \multicolumn{1}{c|}{7}    & \multicolumn{2}{c|}{0.32, 32}   \\
\multicolumn{1}{|c|}{8}  & \multicolumn{2}{c|}{0.13, 32}   & \multicolumn{1}{c|}{}   & \multicolumn{2}{c|}{}           & \multicolumn{1}{c|}{8}    & \multicolumn{2}{c|}{0.32, 75}   \\ \hline
\end{tabular}
\end{table}

\begin{figure}
    \centering
    \includegraphics[width=0.9\linewidth]{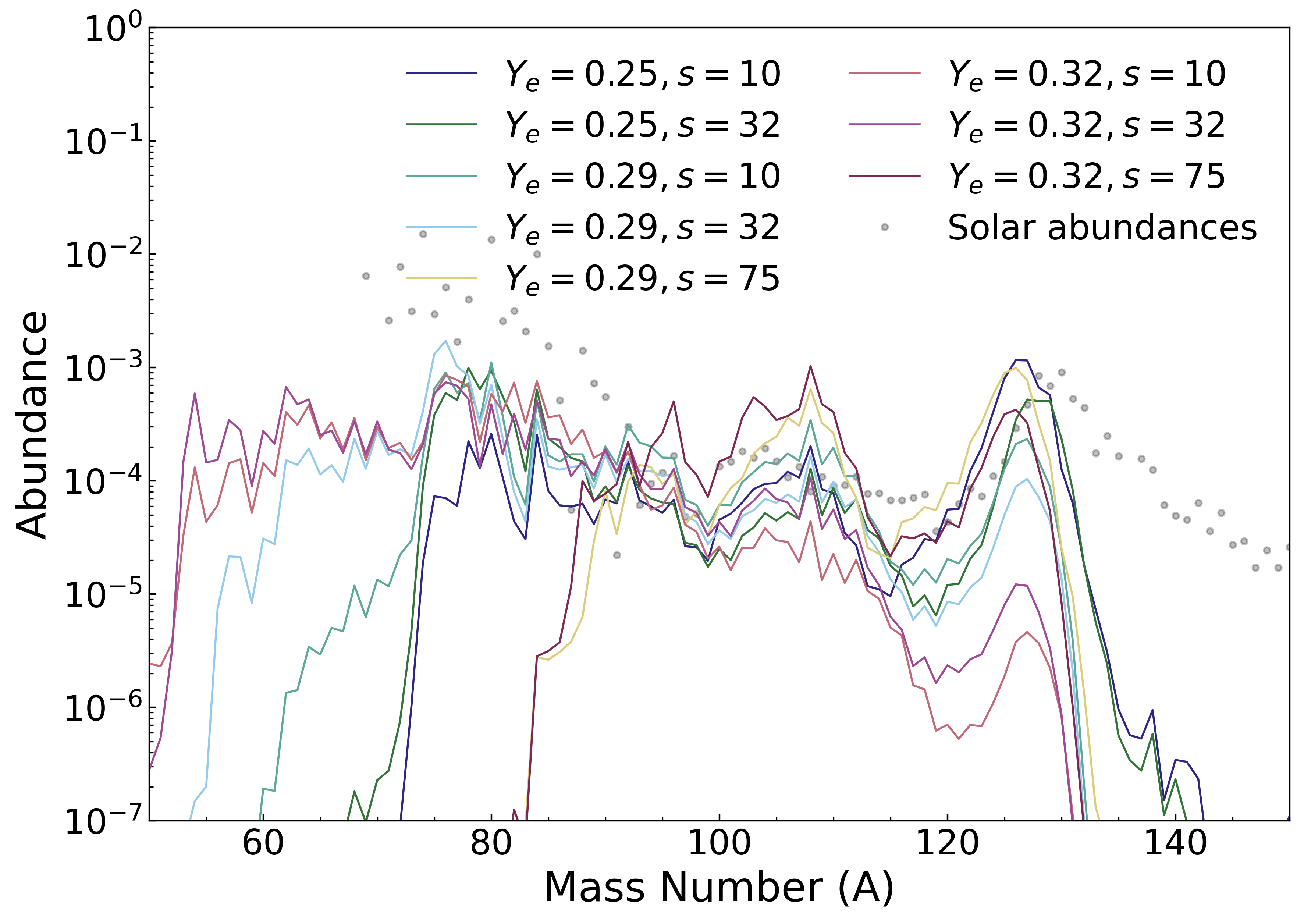}
    \caption{Comparison of predicted final isotopic abundances (assuming the FRDM2012 nuclear model) at 1 Gyr given all the weak $r$-process trajectories considered in Table~\ref{tab:astrovar}.}
    \label{fig:weak_frdm2012_patterns}
\end{figure}

We demonstrate the variation in overall isotopic abundances across astrophysical outflows in Fig.~\ref{fig:weak_frdm2012_patterns} by showing the final abundances (1 Gyr) for all the weak $r$-process trajectories considered. The $Y_e=0.25$ cases produce both first and second peak nuclei for all nuclear models, but do not produce many species lighter than $A\sim$ 75. In contrast, the highest electron fraction considered, $Y_e=0.32$, populates mostly first peak $A\sim80$ species as wells as nuclei at mass numbers around $A \sim 50,60$, with abundances dropping after $A\sim100$. Higher $Y_e\sim0.29$ cases, however, can significantly populate the second peak but only at higher entropy ($s/k=75$). 

\begin{figure}[h!]
    \centering
    \includegraphics[width=0.9\linewidth]{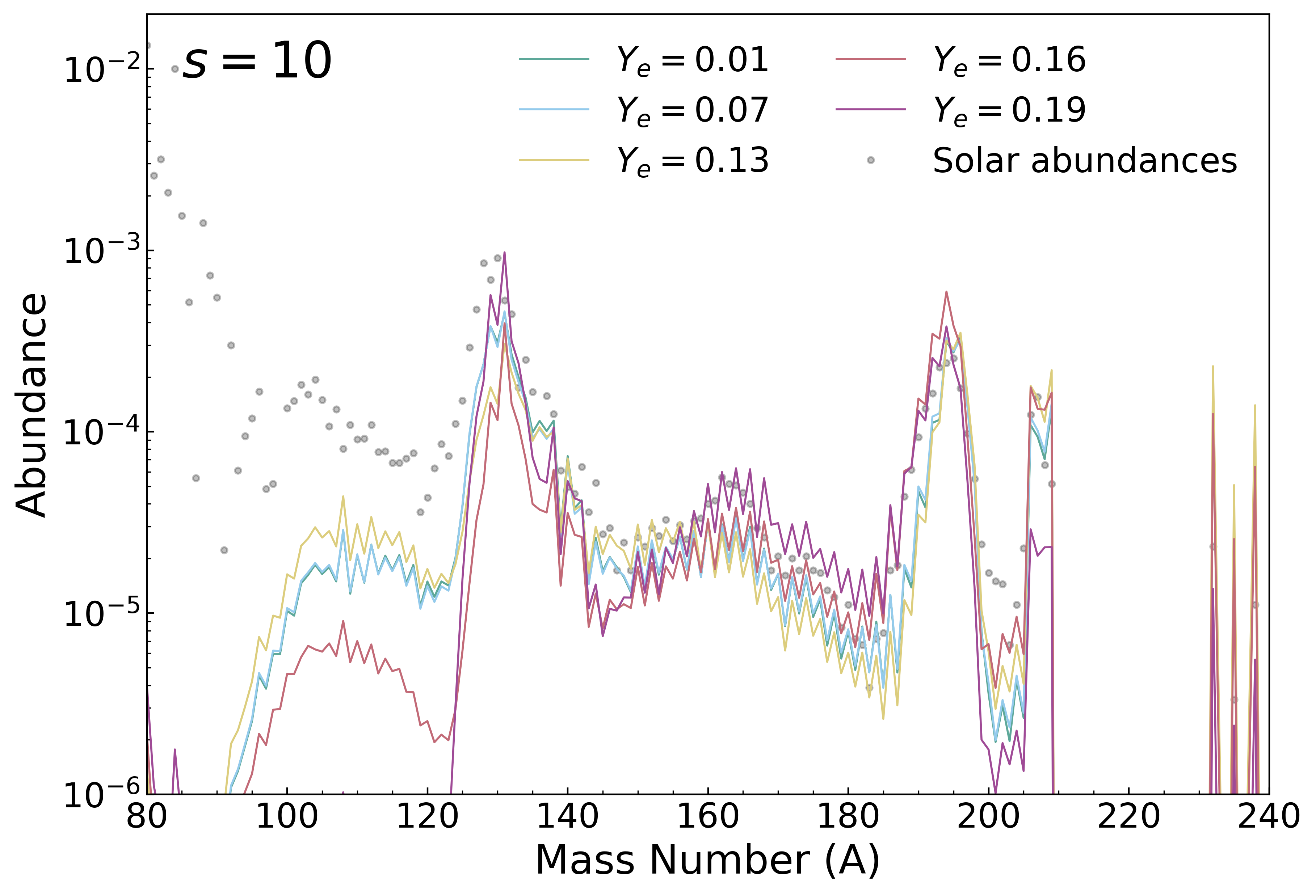}
    \includegraphics[width=0.9\linewidth]{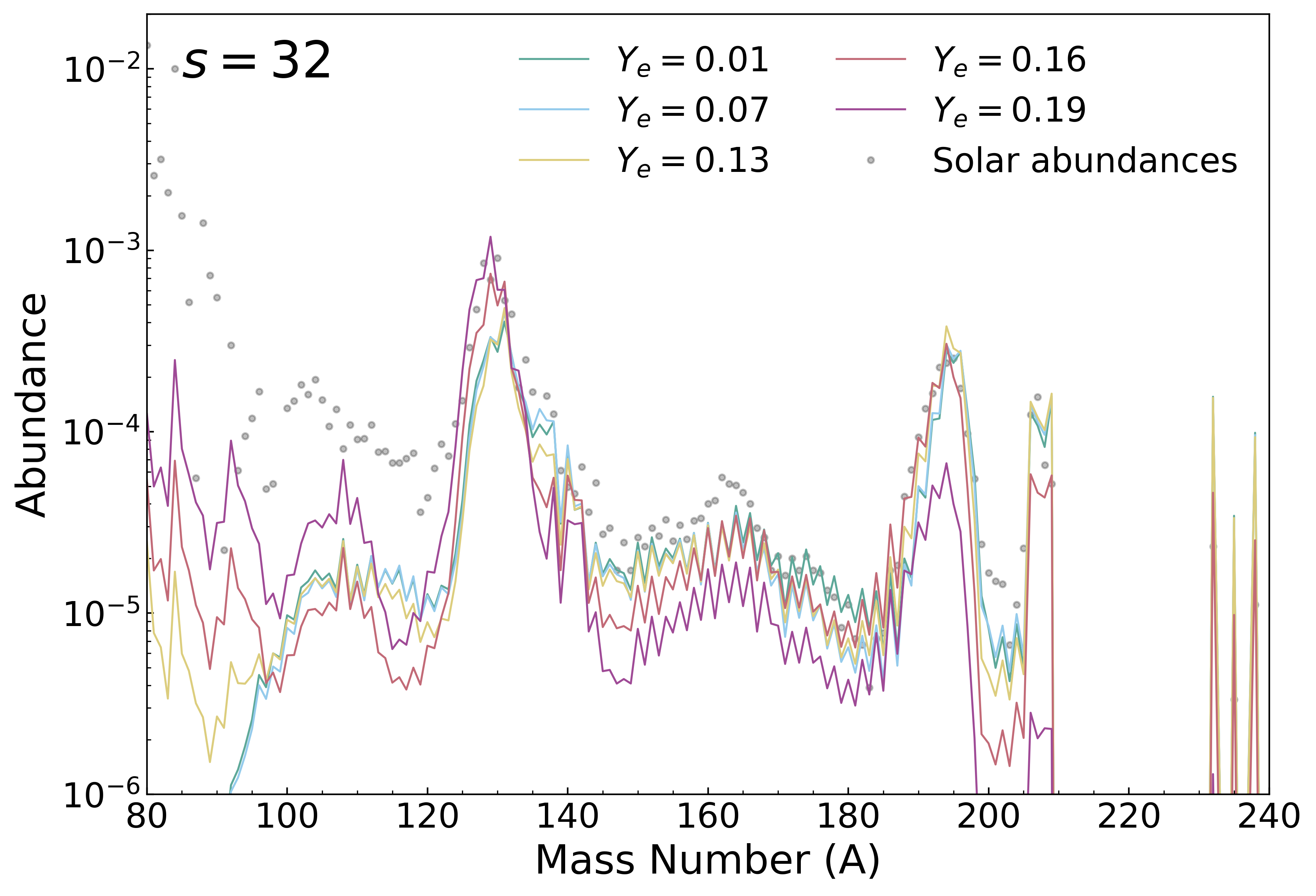}
    \includegraphics[width=0.9\linewidth]{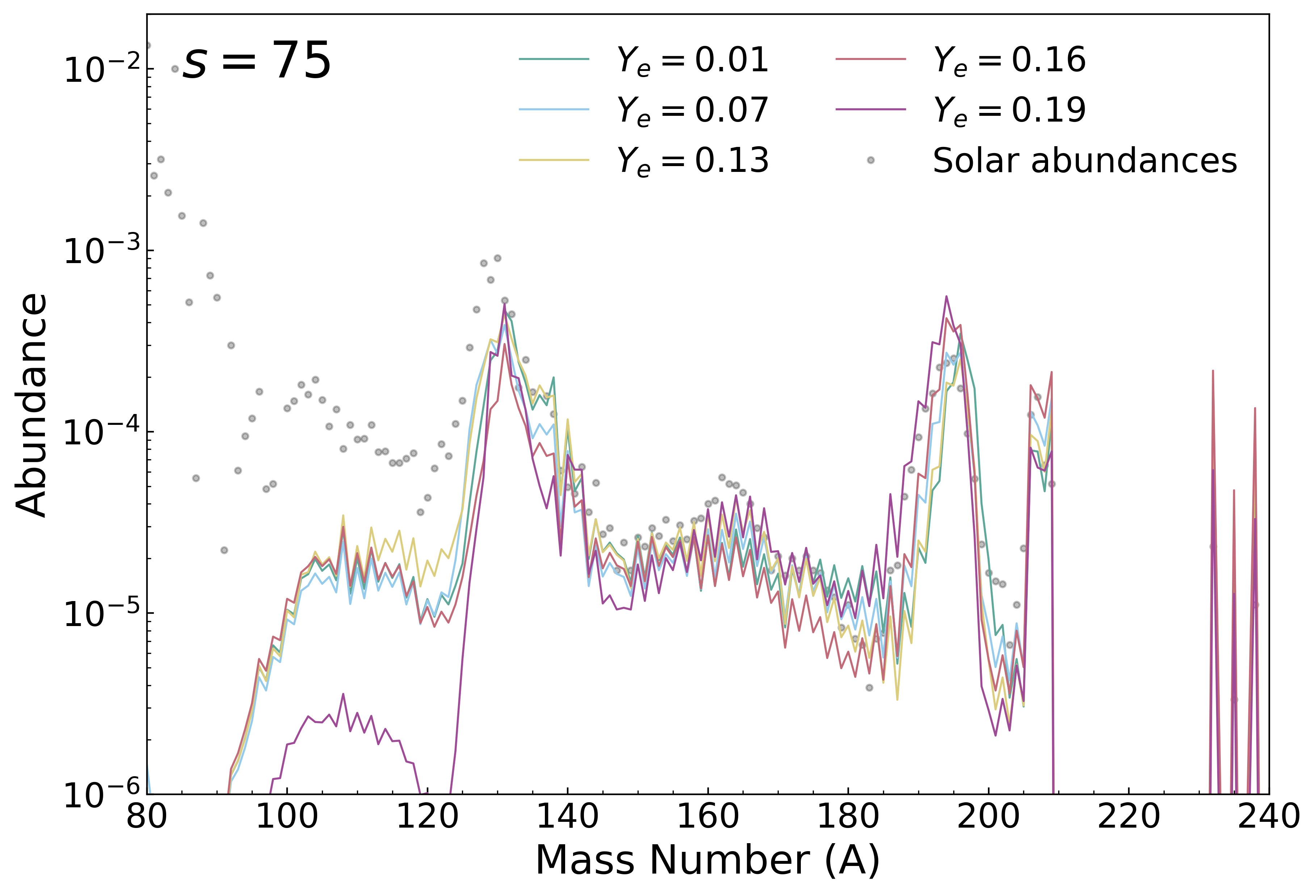}
    \caption{Comparison of predicted final isotopic abundances (assuming the FRDM2012 nuclear model) at 1 Gyr given all the main $r$-process trajectories considered in Table~\ref{tab:astrovar}. Each panel represents $Y_e$ variation results for the three entropies considered: (top) $s/k=10$, (middle) $s/k=32$, and (bottom) $s/k=75$.}
    \label{fig:main_frdm2012_patterns}
\end{figure}

We next show all the 15 main $r$-process abundance patterns found given the conditions on our grid in Fig.~\ref{fig:main_frdm2012_patterns}. For all five trajectories with $s/k=10,75$, the third peak is produced. Most cases push synthesis past the third peak and into the actinides, however significant variance in lead / actinide levels is observed. For $s/k=32$ cases all $Y_e$ values produce actinides, except the $Y_e=0.19$ case which has a very low abundance of third peak and lead isotopes. This behavior of a low entropy outflow producing high actinides, followed by a higher entropy producing less actinides can be seen when the $Y_e$ is at a threshold value such as $0.19$ which separates cases that significantly breach $N=126$ and those that do not (e.g., see Fig.~7 in \cite{Kuske_2025}). This can also be connected to the distinct distributions of neutron-rich seed nuclei for each case, with the $s/k=32$ case having higher abundances held up in the weak $r$-process regime than the $s/k=10,75$ with $Y_e=0.19$.

\begin{figure}[h!]
    \centering
    \includegraphics[width=1.0\linewidth]{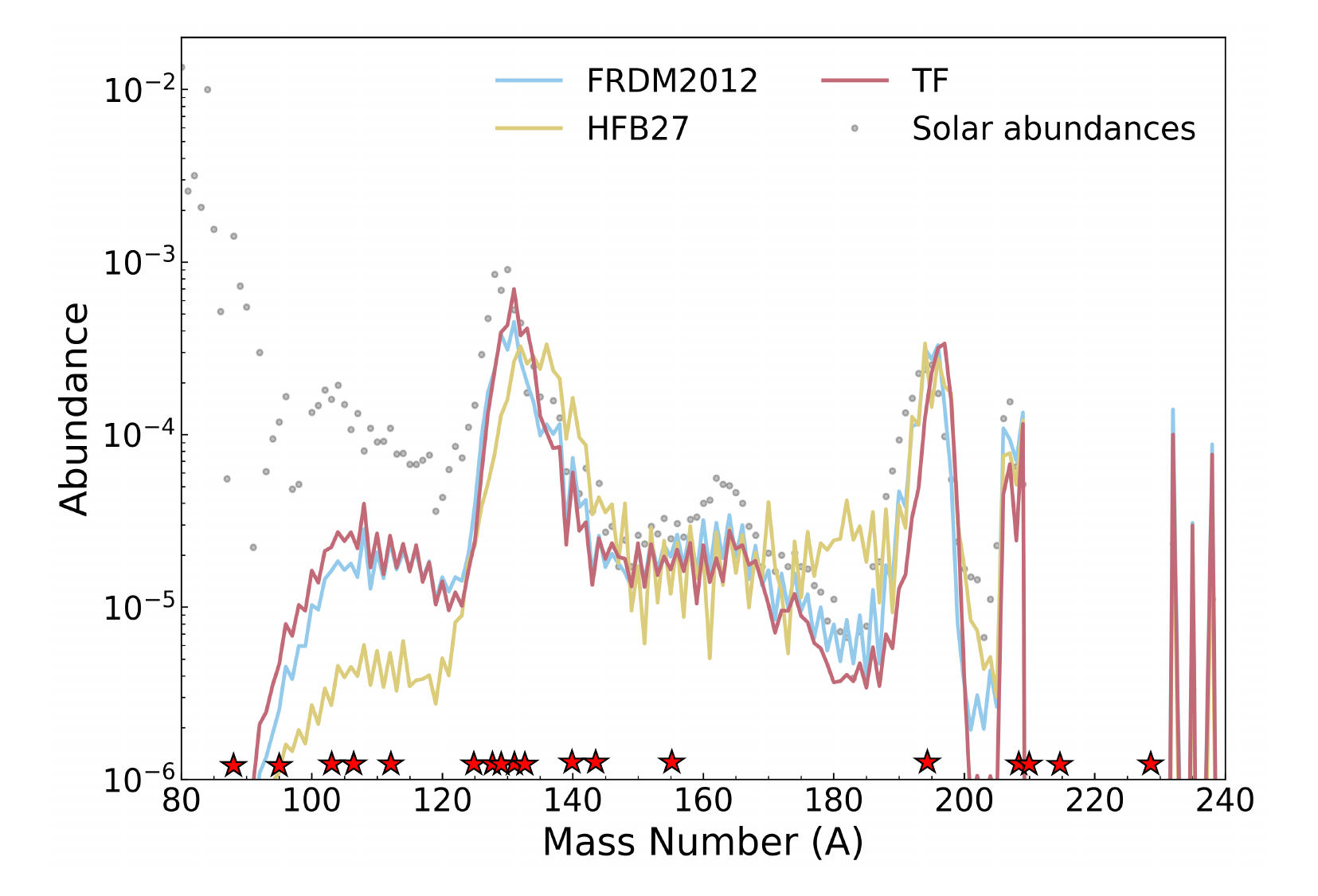}
    \includegraphics[width=1.0\linewidth]{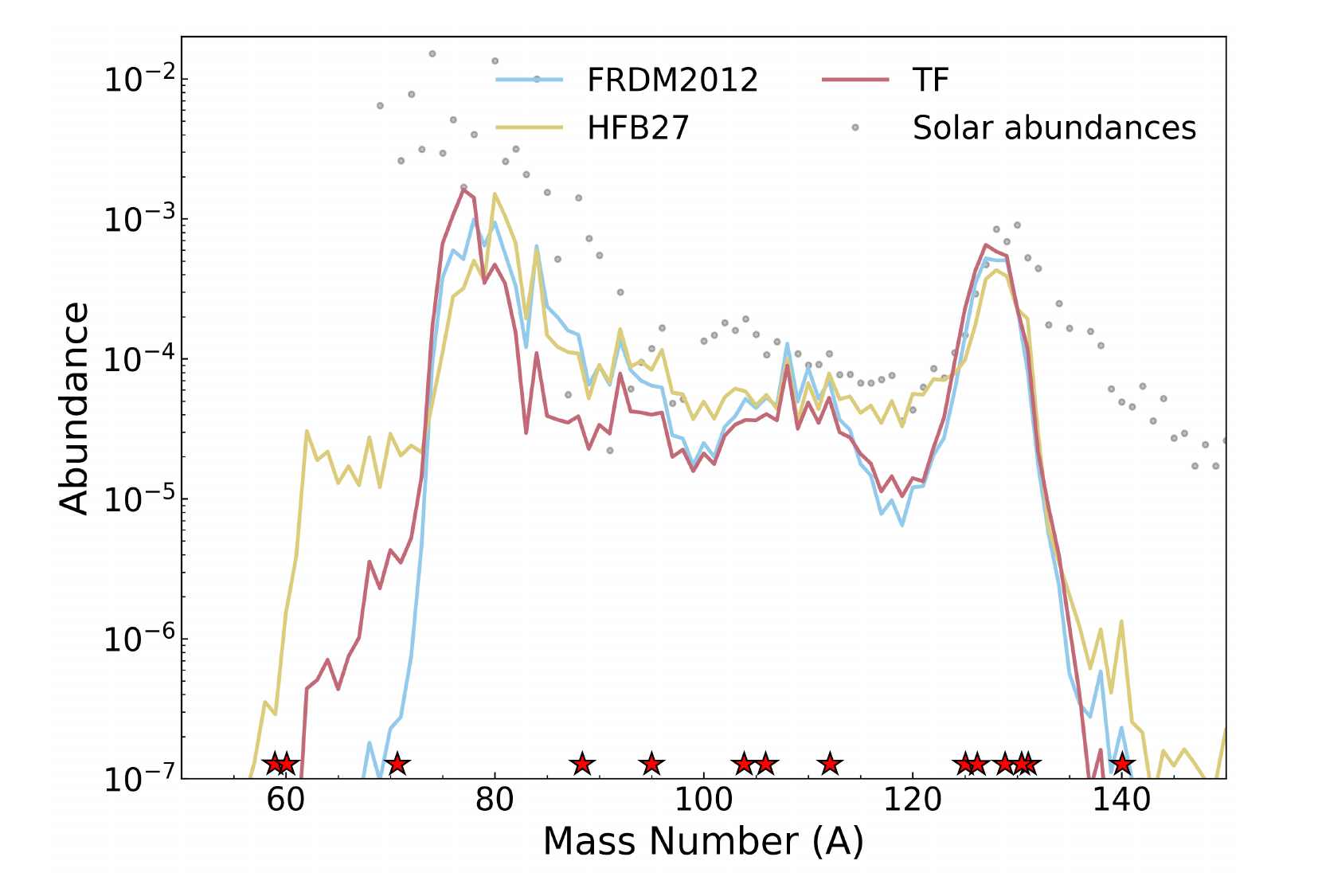}
    \caption{Comparison of predicted final isotopic abundances at 1 Gyr for main (top) and weak (bottom) $r$-process cases given three different nuclear model inputs (TF, FRDM2012, HFB27) (the main $r$-process case shown here is for $Y_e = 0.01$, $s/k = 10$, $\tau = 12$ ms while the weak $r$-process case corresponds to $Y_e = 0.25$, $s/k = 32$, $\tau=12$ ms). Red stars along the x-axis denote the mass numbers of isotopes reported in Tables~\ref{tab:massweight_days} and \ref{tab:massweight_years}.}
    \label{fig:abunpatt_compmodel}
\end{figure}

We next consider nuclear model variations for all trajectories considered here. Note that both the general reach of species produced as well as the relative abundances of key isotopes varies depending on nuclear model inputs, with numerous key abundance features such as the second and third peaks showing variability in their height and width. These abundance variations with the three nuclear models we consider here are highlighted in Figure \ref{fig:abunpatt_compmodel} for the main $r$-process case of $Y_e = 0.01$, $s/k = 10$. For the main $r$ process, TF and FRDM2012 share similar abundance features, however TF has higher abundances than FRDM2012 in the second peak region. The second peak looks significantly different with the HFB27 nuclear model as it has a lower overall abundance, is wider, and is right shifted relative to the second peak produced by TF and FRDM2012. This is in part due to fission deposition from the heavier mass species that can be reached by HFB27 as compared to FRDM2012 and TF. The predicted abundances around the rare-earth peak also show variance, with HFB27 having a more flat abundance profile and instead showing enhanced abundances around $A\sim180$.  

The bottom panel of Figure \ref{fig:abunpatt_compmodel} compares weak $r$-process abundances for the $Y_e = 0.25$, $s/k = 32$ case for the three mass models. In the HFB27 case, similar to the $A\sim180$ behavior seen in main $r$-process abundances, an enhancement in abundances just prior to a major peak is seen in weak $r$-process patterns around mass number 120. HFB27 shows flatter abundances between the first and second peak, while TF and FRDM2012 have a rare-earth peak-like enhancement around $A\sim110$. Here we explicitly see that the distinct nuclear structure of these models imprints on the final relative abundances. Note that nuclear data inputs play an active role in shaping late-time gamma emission via decay and fission rates determining the time evolution of a given radioactive species as well as the relative ratios of species.

We first generate spectra determined with each of the parametrized trajectories outlined in Table~\ref{tab:astrovar} for all three nuclear models. We then run our peak finder algorithm on each calculation set and present the identified isotopes along with their emission lines, the time interval over which the line is above threshold, and which nuclear physics variation reported a given line. Astrophysical conditions are labeled as in Table \ref{tab:astrovar}. Each nuclear model is denoted with its first letter---H for HFB27, T for TF, F for FRDM2012. Signals identified across all nuclear models are denoted with *. We first consider signals appearing between 1 day and 10 years with a threshold of $10^{-7}$ counts s$^{-1}$ cm$^{-2}$. Results are reported for the main $r$-process cases in Table \ref{tab:long_main} in Sec.~\ref{sec:mainemission} with weak $r$-process cases in Table \ref{tab:long_weak} in Sec.~\ref{sec:weakemission}. 
We then consider remnant timescales by repeating the procedure with all combinations of nuclear data and astrophysical conditions but with an updated lightcurve threshold of $10^{-10} \text{ counts s}^{-1} \text{cm}^{-2}$ as described in Sec.~\ref{sec:nsmw3models}. Here we report signals that last beyond 10 years until 1 Gyr. The long-lived signals for the main $r$ process are given in Table \ref{tab:rem_main} in Sec.~\ref{sec:mainemission} with the signals of relevance for weak $r$ process at remnant timescales given in Table \ref{tab:rem_weak} in Sec.~\ref{sec:weakemission}. Before considering radiation transfer, we first apply our peak finding algorithm to the calculated spectrum for all astrophysical conditions considered. We then identify two main $r$-process conditions which collectively report all the isotopes found in all main $r$-process calculations (here condition 1 and 14) and perform radiation transfer calculations for all three nuclear mass model variations. This serves as a means to identify particularly persistent emitting isotopes versus cases which may be washed out when faced with the broadening effects associated with radiation transfer. We do the same for weak $r$-process variations and choose conditions 5 and 6 to perform our radiation transfer calculations.

\subsection{Emission lines from main $r$-process cases}\label{sec:mainemission}

%main post-merger table
\begin{longtable*}[]{|c|c|c|c|c|c|c|c|c|c|c|c|c|c|c|c|c|c|}
    \caption{Signal duration range for post-merger emission lines predicted to dominate the total spectrum which last longer than 1 day and start some time between 1 hr and 10 yrs given main $r$-process variations by considering the astrophysical outflows presented in Table~\ref{tab:astrovar}. The same approach and notation as Table~\ref{tab:massweight_days} is also applied here with the detectability threshold of $10^{-7}$ counts s$^{-1}$ cm$^{-2}$ we require for earlier signals to be reported. Conditions which considered radiation transfer are marked with (r).}
    \label{tab:long_main} \\

    \hline
    \multirow{2}{3em}{\centering Isotope} & \multirow{2}{6em}{\centering Time (days)} & \multirow{2}{4em}{\centering E (keV)} & \multicolumn{15}{|c|}{Conditions} \\
    \cline{4-18}
     & & & \multicolumn{1}{|c|}{1(r)} & \multicolumn{1}{|c|}{2} & \multicolumn{1}{|c|}{3} & \multicolumn{1}{|c|}{4} & \multicolumn{1}{|c|}{5} & \multicolumn{1}{|c|}{6} & \multicolumn{1}{|c|}{7} & \multicolumn{1}{|c|}{8} & \multicolumn{1}{|c|}{9} & \multicolumn{1}{|c|}{10} &
     \multicolumn{1}{|c|}{11} & \multicolumn{1}{|c|}{12} &
     \multicolumn{1}{|c|}{13} & \multicolumn{1}{|c|}{14(r)} & \multicolumn{1}{|c|}{15}\\

    \endfirsthead

    \multicolumn{18}{c}
    {{\bfseries \tablename\ \thetable{} -- continued from previous page}} \\
    \hline 
    \multicolumn{1}{|c|}{Isotope} & \multicolumn{1}{|c|}{Time (days)} & \multicolumn{1}{|c}{E (keV)} & \multicolumn{1}{|c|}{1{(r)}} & \multicolumn{1}{|c|}{2} & \multicolumn{1}{|c|}{3} & \multicolumn{1}{|c|}{4} & \multicolumn{1}{|c|}{5} & \multicolumn{1}{|c|}{6} & \multicolumn{1}{|c|}{7} & \multicolumn{1}{|c|}{8} & \multicolumn{1}{|c|}{9} & \multicolumn{1}{|c|}{10} & \multicolumn{1}{|c|}{11} & \multicolumn{1}{|c|}{12} & \multicolumn{1}{|c|}{13} & \multicolumn{1}{|c|}{14(r)} & \multicolumn{1}{|c|}{15} \\
    \hline
    \endhead

    \hline\endlastfoot

    \hline
    Rb-88 & 6.75e-01 -- 1.06 & 3218 & & & & & & & & & & & & & & \scriptsize{\color{gray}\textit{F}} & \\
     & 3.59e-01 -- 1.54 & 4036 & & & & & & & & & & & & & & $\A$ & \\
     & 2.00e-02 -- 1.99 & 4742 & \scriptsize{\textit{T}} & $\T$ & $\T$ & $\T$ & $\T$ & $\T$ & $\T$ & $\F\T$ & $\T$ & $\F\T$ & $\A$ & $\T$ & $\A$ & \scriptsize{\color{gray}\textit{F}}$\HFB\T$ & $\T$ \\
    \hline
    %Rb-92 & 1.00e-03 -- 2.76e+02 & th & $\F$ & $\F$ & $\F$ & $\F$ & $\F$ & $\F$ & $\F$ & $\F$ & $\F$ & $\F$ & $\F$ & $\F$ & & $\F$ & $\F$ \\
    Nb-95 & 9.59e+01 -- 3.00e+02 & 766 & & & & & & & & & & & & & & $\HFB\T$ & \\
    \hline
    Ru-103 & 9.63e+01 -- 1.53e+02 & 497 & $\T$ & & & $\T$ & & & $\T$ & & & $\T$ & & $\T$ & & $\T$ & \\
    \hline
    Rh-106 & 3.47e+02 -- 9.94e+02 & 622 & & & $\T$ & & & & & $\T$ & & $\T$ & & $\T$ & & & \\
     & 1.26e+02 -- 1.42e+03 & 1050 & & & & & & & & & & & & & & $\A$ & \\
     & 1.69e+02 -- 7.85e+02 & 1562 & & & & & & & & & & & & & & $\T$ & \\
     & 1.67e+02 -- 7.97e+02 & \textbf{1766}, 1797 & & & & & & & & & & & & & & $\HFB\T$ & \\
     & 1.89e+02 -- 1.15e+03 & 1927, \textbf{1988} & & & & & & & & & & & & & & \scriptsize{{\color{gray}\textit{F}}\textit{H}}$\T$ & \\
     & 1.94e+02 -- 7.24e+02 & 2113 & & & & & & & & & & & & & & $\T$ & \\
     & 1.23e+02 -- 1.07e+03 & 2366 & \scriptsize{\textit{T}} & $\T$ & $\T$ & $\T$ & $\T$ & $\T$ & $\T$ & $\T$ & $\T$ & $\T$ & & $\T$ & & \scriptsize{\textit{{\color{gray}F}H}}$\T$ & $\T$ \\
     & 1.72e+02 -- 1.05e+03 & 2406 & & & & & & & & & & & & & & $\HFB$ & \\
     & 7.04e+01 -- 1.54e+02 & 2705, \textbf{2710} & & & & & & & & & & & & & & \scriptsize{\textit{HT}} & \\
    \hline
    Ag-112 & 8.76e-01 -- 3.31 & 2507 & & & & & & & & & & & & & & $\HFB$\scriptsize{\textit{T}} & \\
     & 6.22e-01 -- 7.44 & 2829 & \scriptsize{\color{gray}\textit{T}} & $\T$ & $\T$ & $\T$ & $\F\T$ & $\T$ & $\F\T$ & $\F\T$ & $\T$ & $\T$ & $\A$ & $\T$ & $\T$ & \scriptsize{\textit{{\color{gray}FH}T}} & $\T$ \\
     & 1.01 -- 7.57 & 3393 & \scriptsize{\textit{T}} & & & $\T$ & $\T$ & & $\T$ & $\T$ & $\T$ & $\T$ & $\T$ & $\T$ & & \scriptsize{\textit{F{\color{gray}H}T}} & \\
    %Cd-127 & 1.00e-03 -- 1.21e+03 & th & $\HFB$ & & & $\HFB$ & $\HFB$ & & $\HFB$ & $\HFB$ & & & $\HFB$ & $\HFB$ & & & \\
    %Cd-127 & 1.00e-03 -- 1.63e+03 & th & $\HFB$ & & & $\HFB$ & $\HFB$ & & $\HFB$ & $\HFB$ & & & $\HFB$ & $\HFB$ & & $\HFB$ & \\
    %Cd-127 & 1.00e-03 -- 1.28e+03 & th & & & & $\HFB$ & $\HFB$ & & $\HFB$ & $\HFB$ & & & $\HFB$ & $\HFB$ & $\HFB$ & $\HFB$ & \\
    \hline
    Sn-125 & 1.24e+01 -- 8.47e+01 & \textbf{1067}, 1088, & & & & & & & & & & & & & & $\F\T$ & \\
     & & 1089 & & & & & & & & & & & & & & & \\
     & 1.68e+01 -- 3.99e+01 & 2002 & & & & & & & & & & & & & & $\F$ & \\
    \hline
    Sn-127 & 4.80e-02 -- 1.01 & 2806, \textbf{2846}, & \scriptsize{{\color{gray}\textit{F}}} & $\F$ & & $\F$ & $\F$ & & & $\F$ & & & $\F$ & & $\F$ & \scriptsize{\color{gray}\textit{F}} & \\
    & & 2881 & & & & & & & & & & & & & & & \\
    \hline
    Sb-125 & 1.23e+02 -- 5.11e+03 & 176 & & & & & & & & & & & & & & $\A$ & \\
     & 1.48e+02 -- 8.32e+03 & \textbf{428}, 463 & & & & & & & & & & & $\F\HFB$ & & & $\A$ & \\
     & 3.27e+02 -- 1.12e+04 & \textbf{601}, 636 & & & & & & & & & & & $\F\HFB$ & & $\HFB$ & $\A$ & \\
    \hline
    %Sb-126 & 1.48e+04 -- 3.59e+07 & \textbf{666}, 695, 697 & & & & & & & & & & & $\F$ & & & $\A$ & \\
    %\hline
    Sb-128 & 1.70e-01 -- 1.88 & \textbf{743}, 754 & $\F\T$ & $\F\T$ & $\F$ & $\F\T$ & $\F\T$ & $\F\T$ & $\F\T$ & $\F\HFB$ & $\F\T$ & $\HFB$ & $\F\HFB$ & & $\F\HFB$ & $\A$ & \\
    \hline
    Sb-129 & 1.09e-01 -- 1.46 & 1737 & $\F\T$ & $\F$ & & $\F\T$ & $\F\T$ & $\F$ & & $\F\HFB$ & $\T$ & $\F\HFB$ & $\F\HFB$ & & $\F\HFB$ & $\A$ & $\F\HFB$ \\
    \hline
    {\color{gray}Sb-134} & {\color{gray}1.00e-03 -- 7.31e+01} & {\color{gray}6687} & \scriptsize{\textit{FH}} & $\F\HFB$ & $\F\HFB$ & $\F\HFB$ & $\F\HFB$ & $\F\HFB$ & $\F\HFB$ & $\F\HFB$ & $\F\HFB$ & $\F\HFB$ & $\F\HFB$ & $\F\HFB$ & $\F\HFB$ & \scriptsize{\textit{F}}$\HFB$ & $\F\HFB$ \\
     & {\color{gray}1.00e-03 -- 9.82} & {\color{gray}6820} & \scriptsize{\color{gray}\textit{H}} & $\HFB$ & $\HFB$ & $\HFB$ & $\HFB$ & $\HFB$ & $\HFB$ & $\HFB$ & $\HFB$ & $\HFB$ & $\HFB$ & $\HFB$ & $\HFB$ & $\HFB$ & $\HFB$ \\
    \hline
    %Te-137 & 1.00e-03 -- 4.07e+02 & th & $\A$ & $\A$ & $\A$ & $\A$ & $\A$ & $\A$ & $\A$ & $\A$ & $\A$ & $\A$ & $\A$ & $\A$ & $\A$ & & $\A$ \\
    I-131 & 2.96 -- 8.82e+01 & 364 & $\F\T$ & $\F\T$ & $\F$ & $\F\T$ & $\F\T$ & $\F\T$ & $\F\T$ & $\F\T$ & $\F\T$ & $\F\T$ & $\A$ & $\F\T$ & $\A$ & $\A$ & $\A$ \\
     & 3.08e+01 -- 6.20e+01 & 637 & & & & & & & & & & & $\T$ & & $\F\T$ & & \\
    \hline
    I-132 & 1.07 -- 2.78e+01 & 630, \textbf{668} & $\A$ & $\A$ & $\A$ & $\A$ & $\A$ & $\A$ & $\A$ & $\A$ & $\A$ & $\A$ & $\F\T$ & $\A$ & $\A$ & $\T$ & $\A$ \\
     & 4.43e-01 -- 2.71e+01 & 773 & $\A$ & $\A$ & $\A$ & $\A$ & $\A$ & $\A$ & $\A$ & $\A$ & $\A$ & $\F\T$ & $\F\T$ & $\A$ & $\A$ & $\T$ & $\F\T$ \\
     & 1.24 -- 2.56e+01 & 955 & $\A$ & $\A$ & $\A$ & $\A$ & $\A$ & $\A$ & $\A$ & $\F\T$ & $\A$ & $\F\T$ & $\F\T$ & $\F\T$ & $\A$ & $\A$ & $\F\T$ \\
     & 1.95 -- 2.20e+01 & \textbf{1136}, 1143, & \scriptsize{FH\color{gray}\textit{T}} & $\A$ & $\A$ & $\A$ & $\A$ & $\HFB\T$ & $\T$ & $\F\T$ & $\A$ & $\T$ & $\F\T$ & & $\A$ & $\T$ & $\T$ \\
          & & 1173 & & & & & & & & & & & & & & & \\    
     & 1.43 -- 2.35e+01 & 1291, \textbf{1295} & & $\F\HFB$ & $\A$ & & $\F$ & & & $\F$ & $\HFB$ & & & & & & \\
     & 1.36 -- 3.32e+01 & 1372, \textbf{1399} & $\A$ & $\A$ & $\A$ & $\A$ & $\A$ & $\A$ & $\A$ & $\A$ & $\A$ & $\F\T$ & $\A$ & $\F\T$ & $\A$ & $\A$ & $\A$ \\
     & 2.35 -- 2.45e+01 & 1757 & \scriptsize{\color{gray}\textit{T}} & $\T$ & $\F\T$ & $\T$ & $\T$ & $\T$ & $\T$ & $\T$ & $\HFB\T$ & $\T$ & $\F\T$ & $\T$ & $\F\T$ & \scriptsize{\color{gray}\textit{F}}$\T$ & $\T$ \\
     & 5.22e-01 -- 2.01e+01 & 2002 & $\A$ & $\A$ & $\A$ & $\A$ & $\A$ & $\A$ & $\A$ & $\F\T$ & $\A$ & $\F\T$ & $\F\T$ & $\F\T$ & $\A$ & $\A$ & $\F\T$ \\
     & 9.63e-01 -- 1.29e+01 & 2223 & & & & & & & & & $\HFB$ & & & & $\T$ & & \\
     & 1.23 -- 2.01e+01 & 2390 & \scriptsize{\textit{FT}} & $\F\T$ & $\F\T$ & $\F\T$ & $\F\T$ & $\F\T$ & $\T$ & $\F\T$ & $\F\T$ & $\F\T$ & $\F\T$ & $\T$ & $\A$ & \scriptsize{\textit{F{\color{gray}T}}} & $\F\T$ \\
     & 4.40 -- 1.31e+01 & 2525 & & & & & & & & & & & & & & \scriptsize{\textit{T}} & \\
    \hline
    I-133 & 2.32e-01 -- 3.02 & 530 & & & & & & & & $\T$ & & $\T$ & $\T$ & $\T$ & $\T$ & & $\T$ \\
    \hline
    I-135 & 2.17e-01 -- 1.35 & 1132 & $\HFB$ & $\HFB$ & $\HFB$ & $\HFB$ & $\HFB$ & $\HFB$ & $\HFB$ & & $\HFB$ & & & & & & \\
     & 1.64e-01 -- 1.17 & 1260 & $\HFB$ & $\HFB$ & $\HFB$ & $\HFB$ & $\HFB$ & $\HFB$ & $\HFB$ & & $\HFB$ & & & & & & \\
     & 2.22e-01 -- 2.13 & 1706, \textbf{1791} & $\HFB$ & $\HFB\T$ & $\A$ & $\HFB$ & $\HFB$ & $\HFB\T$ & $\A$ & $\T$ & $\F\HFB$ & $\T$ & $\T$ & $\F\T$ & $\T$ & & $\T$ \\
     & 2.78e-01 -- 1.27 & 2409 & & & & & & $\T$ & $\T$ & $\T$ & & $\T$ & $\T$ & $\T$ & $\T$ & & $\T$ \\
    \hline
    {\color{gray}I-136}  & {\color{gray}1.00e-03 -- 3.38} & {\color{gray}5800} & $\HFB$ & $\HFB$ & $\HFB$ & $\HFB$ & $\HFB$ & $\HFB$ & $\HFB$ & $\HFB$ & $\HFB$ & $\HFB$ & $\HFB$ & $\HFB$ & $\HFB$ & $\HFB$ & $\HFB$ \\
     & {\color{gray}1.00e-03 -- 8.15e+02} & {\color{gray}6104} & $\F\HFB$ & $\F\HFB$ & $\F\HFB$ & $\F\HFB$ & $\F\HFB$ & $\F\HFB$ & $\F\HFB$ & $\F\HFB$ & $\F\HFB$ & $\F\HFB$ & $\F\HFB$ & $\F\HFB$ & $\F\HFB$ & $\F\HFB$ & $\F\HFB$ \\
    \hline
    Xe-133 & 8.04 -- 3.25e+01 & 161 & & & & & & & & & & & & & $\T$ & & $\T$ \\
    \hline
    La-140 & 1.68e+01 -- 1.06e+02 & 816 & $\HFB\T$ & $\HFB\T$ & $\A$ & $\T$ & $\HFB\T$ & $\HFB\T$ & $\T$ & & $\A$ & $\HFB$ & & $\T$ & $\F\HFB$ & & $\F\HFB$ \\
     & 1.14 -- 1.85e+02 & 1596 & $\A$ & $\A$ & $\A$ & $\A$ & $\A$ & $\A$ & $\A$ & $\A$ & $\A$ & $\A$ & $\A$ & $\A$ & $\A$ & $\A$ & $\A$ \\
     & 8.79 -- 8.30e+01 & 2348 & \scriptsize{\textit{H}} & $\T$ & $\HFB\T$ & & $\T$ & $\T$ & & & $\HFB\T$ & & & & & \scriptsize{\textit{H}} & \\
     & 1.76 -- 1.98e+02 & 2521 & $\A$ & $\A$ & $\A$ & $\A$ & $\A$ & $\A$ & $\T$ & $\F\T$ & $\A$ & $\HFB\T$ & $\A$ & $\T$ & $\A$ & $\A$ & $\A$ \\
     & 2.18 -- 1.28e+02 & 2899.6 & \scriptsize{\textit{T}} & $\T$ & $\T$ & $\T$ & $\T$ & $\T$ & $\T$ & $\T$ & $\T$ & $\HFB\T$ & $\HFB\T$ & $\T$ & $\F\HFB$ & \scriptsize{\textit{{\color{gray}F}HT}} & $\HFB\T$ \\
     & 1.01 -- 1.08e+02 & 3119 & $\T$ & $\T$ & $\T$ & $\T$ & $\T$ & $\T$ & $\T$ & $\T$ & $\T$ & $\T$ & $\HFB\T$ & $\T$ & $\A$ & \scriptsize{\textit{FHT}} & $\HFB\T$ \\
     & 1.19 -- 7.18e+01 & 3320 & & & $\T$ & & & & & & & & & & $\HFB$ & \scriptsize{\textit{{\color{gray}F}HT}} & $\HFB$ \\
    \hline
    La-142 & 1.70e-02 -- 1.06 & 3314 & \scriptsize{\color{gray}\textit{FH}} & $\F\HFB$ & $\F\HFB$ & $\F\HFB$ & $\F\HFB$ & $\F\HFB$ & $\F$ & $\F\HFB$ & $\F\HFB$ & $\F\HFB$ & $\F\HFB$ & $\F$ & $\F\HFB$ & \scriptsize{\color{gray}\textit{F}} & $\F\HFB$ \\
     & 9.00e-03 -- 1.39 & 3612, \textbf{3633} & \scriptsize{\textit{F{\color{gray}HT}}} & $\A$ & $\A$ & $\A$ & $\A$ & $\A$ & $\A$ & $\A$ & $\A$ & $\A$ & $\A$ & $\F\T$ & $\A$ & \scriptsize{\textit{F{\color{gray}H}T}} & $\A$ \\
    \hline
    Pr-144 & 2.46e+02 -- 4.66e+02 & 1489 & & & & & & & & & & & & & $\HFB$ & & \\
     & 1.39e+02 -- 2.31e+03 & 2186 & $\F\T$ & $\T$ & $\F\T$ & $\F\T$ & $\F\T$ & $\T$ & $\T$ & $\T$ & $\F\T$ & $\T$ & $\A$ & $\T$ & $\A$ & $\F\HFB$ & $\A$ \\
    \hline
    Eu-155 & 9.44e+02 -- 1.41e+04 & 105 & & & $\T$ & & & & & & & & & & $\HFB$ & $\HFB$ & \\
    \hline
    Eu-156 & 8.10 -- 1.63e+02 & \textbf{1154}, \textbf{1154} & $\T$ & & & & & & & $\T$ & & $\HFB\T$ & $\HFB\T$ & $\T$ & $\A$ & & $\A$ \\
     & 1.23e+01 -- 1.54e+02 & \textbf{1231}, 1242, & $\T$ & $\T$ & $\T$ & $\T$ & $\T$ & $\T$ & $\T$ & $\T$ & $\T$ & $\A$ & $\A$ & $\T$ & $\A$ & $\A$ & $\A$ \\
     &  & 1277 & &  &  &  &  &  &  &  &  &  &  &  &  &  &  \\
     
     & 1.41e+01 -- 1.84e+02 & 1938, \textbf{1966} & \scriptsize{\textit{T}} & $\T$ & $\T$ & $\T$ & $\T$ & $\T$ & $\T$ & $\T$ & $\T$ & $\T$ & & $\T$ & & \scriptsize{\textit{FHT}} & \\
     & 1.77 -- 1.60e+02 & 2027, \textbf{2098} & $\A$ & $\A$ & $\F\T$ & $\A$ & $\A$ & $\A$ & $\F\T$ & $\A$ & $\A$ & $\A$ & $\A$ & $\F\T$ & $\A$ & $\A$ & $\A$ \\
     & 1.36 -- 1.64e+02 & 2181, \textbf{2187} & $\HFB\T$ & $\T$ & $\T$ & $\T$ & $\HFB\T$ & $\HFB\T$ & $\T$ & $\HFB\T$ & $\HFB\T$ & $\HFB\T$ & $\HFB\T$ & $\T$ & $\A$ & $\A$ & $\A$ \\
    \hline
    Hf-181 & 5.86e+01 -- 2.63e+02 & 482 & & & & & & & & & & $\HFB$ & $\HFB$ & & $\HFB$ & & $\HFB$ \\
    \hline
    Ta-184 & 5.28e-01 -- 1.56 & 903, \textbf{921} & & & & & & & & & & $\HFB$ & & & & & $\HFB$ \\
    \hline
    Re-188 & 1.68e+02 -- 2.78e+02 & 931 & & & & & & & & & & & & & $\HFB$ & & $\HFB$ \\
     & 1.98e+02 -- 2.68e+02 & 1802 & & & & & & & & & & & & & $\HFB$ & & \\
    \hline
    Ir-194 & 1.90e+02 -- 2.73e+04 & 328 & $\F\T$ & $\F\T$ & $\F$ & $\F\T$ & $\F\T$ & $\F\T$ & $\F\T$ & $\F\T$ & $\F\T$ & $\A$ & $\A$ & $\F\T$ & $\A$ & $\A$ & $\A$ \\
     & 5.20e+02 -- 1.97e+04 & 645 & & & & & & & & & & & & & $\F\HFB$ & & $\HFB$ \\
     & 3.49e+02 -- 1.43e+04 & 939 & & & & & & & & & & & & & $\HFB$ & $\A$ & $\HFB$ \\
     & 1.65e+02 -- 1.91e+04 & \textbf{1151}, 1183 & $\F\T$ & $\F\T$ & $\F\T$ & $\F\T$ & $\F\T$ & $\F\T$ & $\F\T$ & $\F\T$ & $\F\T$ & $\A$ & $\A$ & $\F\T$ & $\A$ & $\A$ & $\A$ \\
     & 2.80e+02 -- 1.46e+04 & 1469 & & & & & & & & & & $\HFB$ & $\A$ & & $\A$ & $\A$ & $\A$ \\
     & 2.90e+02 -- 1.15e+04 & 1622 & & & & & & & & & & & & & $\F\HFB$ & \scriptsize{\textit{{\color{gray}F}HT}} & $\HFB$ \\
     & 3.14e+02 -- 1.23e+04 & 1797 & & & & & & & & & & & $\HFB$ & & $\HFB$ & & $\F\HFB$ \\
     & 2.97e+02 -- 1.00e+04 & 1806 & $\F$\scriptsize{\textit{T}} & $\F\T$ & $\F$ & $\F\T$ & $\F\T$ & $\F\T$ & $\F\T$ & $\F\T$ & $\F\T$ & $\A$ & $\A$ & $\F\T$ & $\A$ & \scriptsize{\textit{FH}} & $\A$ \\
     & 1.64e+03 -- 4.18e+03 & 2044 & & & & & & & & & & & & & $\F$ & & $\A$ \\
    \hline
    Tl-208 & 1.96e+03 -- 1.71e+04 & 511, \textbf{583} & $\F\T$ & $\F\T$ & $\F\T$ & $\F\T$ & $\F\T$ & $\F\T$ & $\F\T$ & $\F\T$ & $\F\T$ & $\F\T$ & & $\F\T$ & $\T$ & & $\T$ \\
     & 8.50e-02 -- 2.52e+01 & 2615 & $\A$ & $\A$ & $\A$ & $\A$ & $\A$ & $\A$ & $\A$ & $\A$ & $\A$ & $\A$ & $\A$ & $\A$ & $\A$ & & $\A$ \\
     & 1.15e+02 -- 2.16e+04 & 2615 & $\A$ & $\A$ & $\A$ & $\A$ & $\A$ & $\A$ & $\A$ & $\A$ & $\A$ & $\A$ & $\A$ & $\A$ & $\A$ & $\F$ & $\A$ \\
     & 1.18 -- 1.68 & 3708 & & & & & & & & & & & $\T$ & & $\T$ & & $\T$ \\
     & 9.11e-01 -- 1.79 & 3961 & & & & & & & & & & & $\T$ & & $\T$ & & $\T$ \\
    \hline
    %Pb-211 & 2.48e+04 -- 2.90e+04 & 832 & & & & & & & & & & $\F$ & & & & & \\
    %\hline
    %Pb-214 & 2.79e+04 -- 9.48e+05 & 352 & $\F$ & $\F$ & $\F$ & $\F$ & $\F$ & $\F$ & $\F$ & $\F$ & $\F$ & $\F$ & $\F$ & $\F$ & & & $\F\T$ \\
    %\hline
    Ac-228 & 1.81e+02 -- 1.78e+04 & \textbf{911}, 969 & $\F\T$ & $\F\T$ & $\F\T$ & $\F\T$ & $\F\T$ & $\F\T$ & $\F\T$ & $\F\T$ & $\F\T$ & $\F\T$ & $\F\T$ & $\F\T$ & $\T$ & & $\F\T$ \\
     & 3.83e+02 -- 1.67e+04 & 1588 & $\F\T$ & $\F\T$ & $\F\T$ & $\F\T$ & $\F\T$ & $\F\T$ & $\F\T$ & $\F\T$ & $\F\T$ & $\F\T$ & & $\F\T$ & & & $\T$ \\
    %\hline
    %Am-246 & 3.18e+04 -- 1.18e+05 & 205 & & & & $\T$ & & & $\T$ & & & & & & & & \\

\end{longtable*}

%main remnants table
\begin{longtable*}[]{|c|c|c|c|c|c|c|c|c|c|c|c|c|c|c|c|c|c|}
    \caption{Signal duration predictions for emission lines of relevance to remnant searches as they are reported to be visible some time after 10 years post-merger given main $r$-process variations by considering the astrophysical outflows presented in Table~\ref{tab:astrovar}. The same approach and notation as Table~\ref{tab:long_main} is also applied here, but with a detectability threshold of $10^{-10}$ counts s$^{-1}$ cm$^{-2}$ given the prospect of identifying nearby remnants.} 
    \label{tab:rem_main} \\

    \hline
    \multirow{2}{3em}{Isotope} & \multirow{2}{6em}{Time (years)} & \multirow{2}{4em}{E (keV)} & \multicolumn{15}{|c|}{Conditions} \\
    \cline{4-18}
     & & & \multicolumn{1}{|c|}{1(r)} & \multicolumn{1}{|c|}{2} & \multicolumn{1}{|c|}{3} & \multicolumn{1}{|c|}{4} & \multicolumn{1}{|c|}{5} & \multicolumn{1}{|c|}{6} & \multicolumn{1}{|c|}{7} & \multicolumn{1}{|c|}{8} & \multicolumn{1}{|c|}{9} & \multicolumn{1}{|c|}{10} &
     \multicolumn{1}{|c|}{11} & \multicolumn{1}{|c|}{12} &
     \multicolumn{1}{|c|}{13} & \multicolumn{1}{|c|}{14(r)} & \multicolumn{1}{|c|}{15}\\

    \endfirsthead

    \multicolumn{18}{c}
    {{\bfseries \tablename\ \thetable {} -- continued from previous page}} \\
    \hline 
    \multicolumn{1}{|c|}{Isotope} & \multicolumn{1}{|c|}{Time (years)} & \multicolumn{1}{|c}{E (keV)} & \multicolumn{1}{|c|}{1(r)} & \multicolumn{1}{|c|}{2} & \multicolumn{1}{|c|}{3} & \multicolumn{1}{|c|}{4} & \multicolumn{1}{|c|}{5} & \multicolumn{1}{|c|}{6} & \multicolumn{1}{|c|}{7} & \multicolumn{1}{|c|}{8} & \multicolumn{1}{|c|}{9} & \multicolumn{1}{|c|}{10} & \multicolumn{1}{|c|}{11} & \multicolumn{1}{|c|}{12} & \multicolumn{1}{|c|}{13} & \multicolumn{1}{|c|}{14(r)} & \multicolumn{1}{|c|}{15} \\
    \hline
    \endhead

    \hline\endlastfoot

    \hline
    %Rb-92 & 1.00e-03-1.38e+01 & th & $\F$ & $\F$ & $\F$ & $\F$ & $\F$ & $\F$ & $\F$ & $\F$ & $\F$ & $\F$ & & $\F$ & & & \\
    Rh-106 & 3.37e-01 -- 1.40e+01 & 2366 & & & & & & & & & & & & & & \scriptsize{\color{gray}\textit{FHT}} & \\
     & 4.70e-01 -- 1.39e+01 & 2406 & & & & & & & & & & & & & & \scriptsize{\color{gray}\textit{H}} & \\
    \hline
    {\color{gray} In-128} & {\color{gray} 3.44e+01 -- 1.38e+02} & {\color{gray} 3520} & \scriptsize{\textit{F}} & $\F$ & $\F$ & $\F$ & $\F$ & $\F$ & $\F$ & $\F$ & $\F$ & & $\F$ & $\F$ & & & $\F$ \\
     & {\color{gray} 8.22e+01 -- 8.64e+01} & {\color{gray} 4298} & $\F$ & & & & & & & & & & & & & & \\
    \hline
    %Sn-126 & 6.53e+01-1.88e+06 & th & $\F\T$ & $\F\T$ & $\T$ & $\F\T$ & $\F\T$ & $\F\T$ & $\T$ & $\F\HFB$ & $\A$ & & $\A$ & & $\F\HFB$ & $\A$ & \\
    Sb-125 & 3.37e-01 -- 1.40e+01 & 176 & & & & & & & & & & & & & & $\F\T$ & \\
     & 4.06e-01 -- 2.28e+01 & \textbf{428}, 463 & & & & & & & & & & & & & & $\A$ & \\
     & 8.95e-01 -- 3.07e+01 & \textbf{601}, 636 & & & & & & & & & & & $\F\HFB$ & & & $\A$ & \\
    \hline
    Sb-126 & 6.32e+01 -- 2.18e+06 & 415 & $\A$ & $\A$ & $\A$ & $\A$ & $\A$ & $\A$ & $\F\T$ & $\A$ & $\A$ & $\HFB\T$ & $\A$ & $\T$ & $\F\HFB$ & $\A$ & $\F\HFB$ \\
     & 4.06e+01 -- 2.56e+06 & \textbf{666}, 695, & $\A$ & $\A$ & $\A$ & $\A$ & $\A$ & $\A$ & $\A$ & $\A$ & $\A$ & $\A$ & $\A$ & $\A$ & $\A$ & $\A$ & $\A$ \\
     & & 697 & & & & &  & &  &  &  &  & &  &  &  &  \\
     & 4.66e+01 -- 1.72e+06 & 954, \textbf{990} & & & & & & & & & & & & & & \scriptsize{\textit{FHT}} & \\
     & 7.25e+01 -- 8.86e+05 & 1213 & $\F\T$ & $\F\T$ & & $\F\T$ & $\F\T$ & $\F\T$ & $\A$ & $\A$ & $\A$ & & $\F\HFB$ & $\T$ & $\F\HFB$ & $\A$ & \\
     & 8.41e+01 -- 6.87e+04 & 1477 & & & & & & & & & & & & & & $\F\HFB$ & \\
    \hline
    {\color{gray} Sb-134} & {\color{gray} 1.00e-03 -- 1.98e+02} & {\color{gray} 6687} & \scriptsize{\textit{H}} & $\HFB$ & $\HFB$ & $\HFB$ & $\HFB$ & $\HFB$ & $\HFB$ & $\HFB$ & $\HFB$ & & & $\HFB$ & & & \\
     & {\color{gray} 1.00e-03 -- 1.16e+02} & {\color{gray} 6820} & \scriptsize{\color{gray}\textit{H}} & $\HFB$ & $\HFB$ & $\HFB$ & $\HFB$ & $\HFB$ & $\HFB$ & $\HFB$ & $\HFB$ & & & $\HFB$ & & & \\
    \hline
    %Te-137 & 1.06e+01-1.36e+02 & th & $\HFB$ & $\HFB$ & $\HFB$ & $\HFB$ & $\HFB$ & $\HFB$ & $\HFB$ & $\HFB$ & $\HFB$ & $\HFB$ & $\HFB$ & $\HFB$ & & & $\HFB$ \\
    {\color{gray} I-136} & {\color{gray} 1.00e--03 -- 5.99e+02} & {\color{gray} 6104} & $\F$\scriptsize{\textit{H}} & $\F\HFB$ & $\F\HFB$ & $\F\HFB$ & $\F\HFB$ & $\F\HFB$ & $\F\HFB$ & $\F\HFB$ & $\F\HFB$ & $\HFB$ & & $\F\HFB$ & & & \\
    \hline
    Eu-155 & 2.59 -- 3.87e+01 & 105 & & & $\T$ & & & & & & & & & & $\HFB$ & $\HFB$ & \\
    \hline
    Ta-182 & 1.55e+03 -- 3.74e+06 & \textbf{1121}, 1189 & $\HFB$ & $\HFB$ & $\HFB$ & $\HFB$ & $\HFB$ & $\HFB$ & & $\HFB$ & & $\HFB$ & $\HFB$ & & $\HFB$ & & $\HFB$ \\
     & 1.48e+03 -- 3.47e+06 & \textbf{1221}, 1231 & $\HFB$ & $\HFB$ & $\HFB$ & $\HFB$ & $\HFB$ & $\HFB$ & & $\HFB$ & & $\HFB$ & $\HFB$ & & $\HFB$ & & $\HFB$ \\
    \hline
    Ir-194 & 5.20e-01 -- 8.27e+01 & 328 & $\F\T$ & $\F\T$ & $\F$ & $\F\T$ & $\F\T$ & $\F\T$ & $\F\T$ & $\F\T$ & $\F\T$ & $\A$ & $\A$ & $\F\T$ & $\A$ & $\A$ & $\A$ \\
     & 1.42 -- 5.78e+01 & 645 & & & & & & & & & & & & & $\F\HFB$ & & $\HFB$ \\
     & 9.55e-01 -- 6.47e+01 & 939 & & & & & & & & & & & & & $\HFB$ & $\A$ & $\HFB$ \\
     & 4.51e-01 -- 8.99e+01 & \textbf{1151}, 1183 & $\F\T$ & $\F\T$ & $\F\T$ & $\F\T$ & $\F\T$ & $\F\T$ & $\F\T$ & $\F\T$ & $\F\T$ & $\A$ & $\A$ & $\F\T$ & $\A$ & $\A$ & $\A$ \\
     & 7.66e-01 -- 8.27e+01 & 1469 & & & & & & & & & & $\HFB$ & $\A$ & & $\A$ & $\A$ & $\A$ \\
     & 7.95e-01 -- 8.11e+01 & 1622 & & & & & & & & & & & & & $\HFB$ & \scriptsize{\textit{FHT}} & $\HFB$ \\
     & 8.59e-01 -- 9.21e+01 & 1797 & & & & & & & & & & & $\HFB$ & & $\F\HFB$ & & $\F\HFB$ \\
     & 8.15e-01 -- 8.56e+01 & 1806 & $\F$\scriptsize{\textit{T}} & $\F\T$ & $\F$ & $\F\T$ & $\F\T$ & $\F\T$ & $\F\T$ & $\F\T$ & $\F\T$ & $\A$ & $\A$ & $\F\T$ & $\A$ & \scriptsize{\textit{FHT}} & $\A$ \\
     & 4.49 -- 6.77e+01 & 2044 & & & & & & & & & & & $\F\T$ & & $\A$ & \scriptsize{\textit{FHT}} & $\A$ \\
    \hline
    Tl-208 & 5.37 -- 4.69e+01 & 511, \textbf{583} & $\F\T$ & $\F\T$ & $\F\T$ & $\F\T$ & $\F\T$ & $\F\T$ & $\F\T$ & $\F\T$ & $\F\T$ & $\F\T$ & & $\F\T$ & $\T$ & & $\T$ \\
     & 3.15e-01 -- 1.15e+02 & 2615 & $\A$ & $\A$ & $\A$ & $\A$ & $\A$ & $\A$ & $\A$ & $\A$ & $\A$ & $\A$ & $\A$ & $\A$ & $\A$ & $\A$ & $\A$ \\
     & 9.36 -- 2.23e+01 & 3708 & $\T$ & & & & $\T$ & & & & & & $\T$ & & $\T$ & & $\T$ \\
     & 1.02e+01 -- 1.43e+01 & 3961 & & & & & & & & & & & & & $\T$ & & $\T$ \\
    \hline
    Tl-209 & 9.64e+01 -- 5.40e+05 & 1567 & $\A$ & $\A$ & $\A$ & $\A$ & $\A$ & $\A$ & $\A$ & $\A$ & $\A$ & $\A$ & $\A$ & $\A$ & $\F\T$ & & $\A$ \\
    \hline
    Pb-211 & 6.29e+01 -- 1.14e+02 & 832 & & $\T$ & & & $\T$ & $\T$ & & $\T$ & $\T$ & $\F\T$ & $\T$ & $\T$ & $\T$ & & $\F\T$ \\
    \hline
    Pb-214 & 2.07e+02 -- 8.08e+02 & 295 & & & $\F$ & & & & & & & & & & & & \\
     & 7.64e+01 -- 6.95e+05 & 352 & $\A$ & $\A$ & $\A$ & $\A$ & $\A$ & $\A$ & $\A$ & $\A$ & $\F\T$ & $\A$ & $\F\T$ & $\A$ & $\F\T$ & & $\A$ \\
    \hline
    %Bi-213 & 8.25e+05-6.37e+06 & th & $\A$ & $\A$ & $\A$ & $\A$ & $\A$ & $\A$ & $\A$ & $\A$ & $\A$ & $\F\T$ & $\T$ & $\A$ & $\T$ & & $\F\T$ \\
    %Ra-225 & 7.52e+05-4.70e+06 & th & $\T$ & $\T$ & & $\T$ & $\T$ & & $\HFB\T$ & $\T$ & $\T$ & $\T$ & & $\HFB\T$ & $\T$ & & $\T$ \\
    Ac-228 & 4.96e-01 -- 4.86e+01 & \textbf{911}, 969 & $\F\T$ & $\F\T$ & $\F\T$ & $\F\T$ & $\F\T$ & $\F\T$ & $\F\T$ & $\F\T$ & $\F\T$ & $\F\T$ & $\F\T$ & $\F\T$ & $\T$ & & $\F\T$ \\
     & 1.05 -- 5.55e+01 & 1588 & $\F\T$ & $\F\T$ & $\F\T$ & $\F\T$ & $\F\T$ & $\F\T$ & $\F\T$ & $\F\T$ & $\F\T$ & $\F\T$ & & $\F\T$ & & & $\T$ \\
     & 1.11e+01 -- 1.81e+01 & 1823, 1835, & & & $\T$ & & & & & & & & & & & & \\
     & & 1842, 1871, & & & & & & & & & & & & & & & \\
     & & \textbf{1887} & & & & & & & & & & & & & & & \\
    \hline
    Pu-243 & 5.93e+06 -- 2.51e+07 & 109 & & $\HFB$ & $\HFB$ & & & $\HFB$ & $\F\HFB$ & & $\HFB$ & & & $\F\HFB$ & & & \\
    \hline
    Am-246 & 8.70e+01 -- 3.24e+02 & 205 & & & & $\T$ & & & $\T$ & & & & & & & & \\
     & 1.22e+02 -- 3.68e+02 & \textbf{834}, 839 & & $\T$ & & $\T$ & $\T$ & $\T$ & $\T$ & $\T$ & & $\T$ & & $\T$ & & & \\

\end{longtable*}

Often potential gamma emitters near the second $r$-process peak ($A\sim$130) are noted in the literature, and we also report many such species. Interestingly we also find a potential signature from a rare-earth isotope, Eu-156, with its 2.027 / 2.098 MeV lines robustly reported across most nuclear models and astrophysical variations. HFB27 tends to produce high abundances for species around $A = 140$, which can be seen in Figure \ref{fig:abunpatt_compmodel} with the second peak shifted to the right. This results in an extended and stronger signal for species like La-140 ($Z=57$) (further discussed in Section \ref{sec:tl208}). We also report some isotopes with mass numbers numbers between $A = 180-190$, e.g., Hf-181, Ta-182, Ta-184, and Re-188, but they are only produced in the HFB27 case. An important isotope to indicate third $r$-process peak ($A=195$) production is Ir-194, especially the 0.328 MeV, 1.151 / 1.183 MeV and 1.806 MeV lines which are reported to be produced in nearly all astrophysical conditions by at least two models. 

Other than the two Tl isotopes (Tl-208 and Tl-209), five nuclei past the third peak at $A\sim195$ are flagged as having potentially detectable $\beta$-gamma signals at either post-merger or remnant timescales (Pb-211, Pb-214, Ac-228, Pu-243, and Am-246). During early post-merger timescales, the heaviest isotope to produce a signal in the HFB27 case is Tl-208. Every nuclear model predicts the 2.615 MeV line of Tl-208 to be above threshold at some point on both post-merger and remnant timescales for all astrophysical conditions except 14, the details of which are further discussed in Section \ref{sec:tl208}. Out of the two lead isotopes of Pb-211 and Pb-214 reported in the remnant emission timescale table, Pb-214 is the most consistently predicted to be present expect for in condition 14 (which has a lower abundance past the third peak as discussed in Sec.~\ref{sec:astrovar}). The 0.911 / 0.969 MeV emission lines from Ac-228 ($Z=89$) are found to be robust across FRDM2012 and TF, with only the conditions with fewer actinides not producing this signal. The heaviest isotope reported to have a potentially observable line is Am-246 ($Z=95$), predicted only in the TF case, and is not reported in Table~\ref{tab:massweight_years} for the mass weighted ejecta scenarios. We note that this is due to the mass weighted case having similar actinide abundance levels to condition 1, which does not report an Am-246 signal that satisfies our table criterion. Furthermore, we find Am-246 to appearing on the order of hundreds of years ultimately due to theoretical $\alpha$ and $\beta$ decays, explaining why this species is not robustly predicted to be an important emitter across nuclear models.

Nuclei emitting signals we report to be above threshold on remnant timescales are found to either have their signal fade away after tens to hundreds of years or emission can be found as late as a million years later. Looking at the longer lived signals ($\geq 10^{5},10^{6}$ years) for the main $r$ process, the only case that is found to be above threshold across all nuclear models and astrophysical variations is the 0.666 / 0.695 / 0.697 MeV emission from Sb-126, which is also reported in the mass weighted scenarios. Other isotopes with signals lasting beyond  $10^{5},10^{6}$ years are Ta-182, Tl-209, and Pb-214 and Pu-243. Tl-209 is often reported, found across all nuclear models for nearly all astrophysical variations. Pu-243 is the heaviest species reported to emit on the order of millions of years, found in calculations with both HFB27 and FRDM2012. Pu-243 is a special case in that its signal can last as long as 25 million years but does not arise until 6 million years post event. All nuclei other than Pu-243 and Ta-182 have signals that rise above the detectable limit on the order of decades or earlier, and several of these cases can continue to maintain emission above our implemented threshold out to thousands or millions of years. Species of interest for searches of younger remnants (i.e. signals out to tens or hundreds of years) were found to be Tl-208, Ir-194, and Ac-228, which have at least one line reported to be detectable across most nuclear models and astrophysical conditions. Note that with fewer species remaining as strong emitters, the prospects for individual isotope identification are more clean to analyze in the remnant phase. Therefore measurements of the gamma spectra of known remnants could not only shed light on the composition / origin of the event that produced the remnant but could also be used to assist in age estimates of remnants since the emission from some species would not be visible past a given time. 

\subsection{Emission lines from weak $r$-process cases}\label{sec:weakemission}

We next consider combinations of conditions that do not significantly produce main $r$-process nuclei and instead emission is dominated by weak $r$-process nuclei. We select parameterized cases that were reported to be representative of merger ejecta in \cite{Radice2018}. While investigations of prospective $r$-process sites that can take place earlier in cosmic history than mergers have been a focus of recent $r$-process literature (e.g. collapsars, magneto-rotationally driven supernovae, magnetars), most recent simulations find ejecta for such events to be dominated by weak $r$-process nuclei. Therefore deciphering weak $r$-process signals from those of the main $r$-process is especially timely. Note that while here we keep the radiative transfer model set-up described in Sec.~\ref{sec:findlines}, to properly account for specific weak $r$-process environments we would need to tailor our model assumptions accordingly. Nevertheless, the nuclei reported by our peak finding procedure given the variety of weak $r$-process conditions explored here can highlight signals that may be of common interest towards studying numerous types of $r$-process events. We provide the nuclei reported by applying our procedure to all the weak $r$-process calculations presented in Sec.~\ref{sec:astrovar} for all three nuclear models, first reporting on post-merger signals in Table~\ref{tab:long_weak}. We then investigate remnant timescales and report in Table~\ref{tab:rem_weak} which species we find to be of potential use in hunting for remnants that may originate from an $r$-process event. Here we aim to bring forward any species that may have been washed out in a mass weighted calculation by main $r$-process emitters in order to look for distinct emitters capable of indicating that weak $r$-process ejecta dominated the event.

%weak post-merger table
\begin{longtable*}[]{|c|c|c|c|c|c|c|c|c|c|c|}
    \caption{Signal duration range for post-merger emission lines predicted to dominate the total spectrum which last longer than 1 day and start some time between 1 hr and 10 yrs given weak $r$-process variations by considering the astrophysical outflows presented in Table~\ref{tab:astrovar}. The same approach and notation as Table~\ref{tab:long_main} is also applied here with the detectability threshold of $10^{-7}$ counts s$^{-1}$ cm$^{-2}$ we require for earlier signals to be reported.}
    \label{tab:long_weak} \\

    \hline
    \multirow{2}{4em}{\centering Isotope} & \multirow{2}{8em}{\centering Time (days)} & \multirow{2}{4em}{\centering E (keV)} & \multicolumn{8}{|c|}{Conditions} \\
    \cline{4-11}
     & & & \multicolumn{1}{|c|}{1} & \multicolumn{1}{|c|}{2} & \multicolumn{1}{|c|}{3} & \multicolumn{1}{|c|}{4} & \multicolumn{1}{|c|}{5(r)} & \multicolumn{1}{|c|}{6(r)} & \multicolumn{1}{|c|}{7}  & \multicolumn{1}{|c|}{8} \\

    \endfirsthead

    \multicolumn{11}{c}
    {{\bfseries \tablename\ \thetable{} -- continued from previous page}} \\
    \hline 
    Isotope & Time (days) & E (keV) & \multicolumn{1}{|c|}{1} & \multicolumn{1}{|c|}{2} & \multicolumn{1}{|c|}{3} & \multicolumn{1}{|c|}{4} & \multicolumn{1}{|c|}{5(r)} & \multicolumn{1}{|c|}{6(r)} & \multicolumn{1}{|c|}{7}  & \multicolumn{1}{|c|}{8} \\
    \hline
    \endhead

    \hline\endlastfoot
    \hline
    Na-24 & 3.83e-01 -- 1.91 & 3866 & & & & & \scriptsize{\textit{FH}} & & & $\F\HFB$ \\
    \hline
    Fe-59 & 2.90e+01 -- 4.20e+02 & 143, \textbf{192} & & & & & & $\A$ & $\A$ & \\
     & 9.34 -- 4.97e+02 & 1099 & & & & $\A$ & & $\A$ & $\A$ & \\
     & 5.08 -- 5.82e+02 & 1292 & & & $\HFB$ & $\A$ & & $\A$ & $\A$ & \\
     & 3.03e+01 -- 1.15e+02 & 1482 & & & & & & \scriptsize{{\color{gray}\textit{FHT}}} & $\A$ & \\
    \hline
    Cu-66 & 3.37 -- 8.54 & 1039 & & & & & & $\T$ & $\T$ & \\
    \hline
    Cu-67 & 2.85 -- 2.92e+01 & 185 & & & & $\T$ & & $\A$ & $\A$ & \\
    \hline
    Zn-72 & 2.40 -- 1.48e+01 & 145 & & & $\HFB$ & $\F\HFB$ & & $\F\HFB$ & $\HFB$ & \\
    \hline
    Ga-72 & 2.63 -- 1.07e+01 & 630 & & & & $\HFB$ & & $\F\HFB$ & $\F\HFB$ & \\
     & 1.04 -- 1.44e+01 & \textbf{834}, 894 & & & $\HFB\T$ & $\A$ & & $\A$ & $\A$ & \\
     & 2.98 -- 9.22 & 1051 & & & & $\HFB$ & & & & \\
     & 2.62 -- 1.29e+01 & 1231, 1260, \textbf{1277} & & & $\HFB$ & $\A$ & & $\F\HFB$ & $\F\HFB$ & \\
     & 1.56 -- 3.07e+01 & 1464 & & & $\A$ & $\A$ & & \scriptsize{{\color{gray}\textit{F}}\textit{H}{\color{gray}\textit{T}}} & $\A$ & \\
     & 1.14 -- 3.12e+01 & 1597 & & & $\A$ & $\A$ & & $\A$ & $\A$ & \\
     & 1.17 -- 3.61e+01 & 1862 & & $\HFB$ & $\A$ & $\A$ & & $\A$ & $\A$ & \\
     & 6.44e-01 -- 4.16e+01 & 2202 & & $\HFB\T$ & $\A$ & $\A$ & & $\A$ & $\A$ & \\
     & 4.84e-01 -- 4.49e+01 & 2508 & $\HFB\T$ & $\A$ & $\A$ & $\A$ & & $\A$ & $\A$ & \\
     & 6.84e-01 -- 3.54e+01 & 2844 & & $\HFB\T$ & $\A$ & $\A$ & & \scriptsize{\textit{FHT}} & $\A$ & \\
     & 1.12 -- 2.21e+01 & \textbf{3325}, 3339 & $\T$ & $\HFB\T$ & $\A$ & $\A$ & & \scriptsize{\textit{FHT}} & $\A$ & \\
     & 7.87e-01 -- 1.45e+01 & 3679 & & $\HFB\T$ & $\A$ & $\A$ & & $\A$ & $\A$ & \\
    \hline
    Ge-77 & 2.44e-01 -- 3.16 & 211, 216, \textbf{264} & & $\F\T$ & $\A$ & $\A$ & & $\A$ & $\A$ & \\
     & 1.64e-01 -- 3.85 & 416 & & & $\A$ & $\A$ & & \scriptsize{\textit{F{\color{gray}HT}}} & $\A$ & \\
     & 5.29e-01 -- 1.79 & \textbf{632}, 634 & & & & & & $\F$\scriptsize{\color{gray}\textit{H}} & $\F$ & \\
     & 5.65e-01 -- 2.79 & 1085 & & $\T$ & & $\F\T$ & & $\F$ & $\F$ & \\
     & 6.29e-01 -- 3.42 & 1368 & & $\F\T$ & $\F\T$ & $\A$ & & $\A$ & $\F\T$ & \\
     & 6.58e-01 -- 3.35 & 1539, \textbf{1574} & & $\T$ & & & & & & \\
     & 1.33 -- 2.96 & 1710, \textbf{1720}, 1727 & & $\T$ & & & & & & \\
     & 6.14e-01 -- 2.84 & \textbf{2000}, 2077, 2090 & & $\T$ & & $\F\T$ & & \scriptsize{{\color{gray}\textit{FT}}} & $\F\T$ & \\
     & 6.29e-01 -- 2.33 & 2342 & & $\F\T$ & & & & & & \\
    \hline
    Kr-85 & 3.09e+03 -- 1.96e+04 & 514 & & $\HFB$ & $\HFB$ & $\F\HFB$ & & $\A$ & $\A$ & \\
    \hline
    Kr-88 & 3.50e-02 -- 1.17 & 2392 & $\F$ & $\F$ & $\F$ & $\F$ & & \scriptsize{\color{gray}\textit{F}} & $\F$ & $\F$ \\
    \hline
    Rb-88 & 1.35e-01 -- 1.11 & 3010 & $\F$ & $\F$ & $\F$ & & {\color{gray} \scriptsize{\textit{F}}} & {\color{gray} \scriptsize{\textit{F}}} & & $\F$ \\
     & 1.45e-01 -- 1.32 & 3218 & $\A$ & $\A$ & $\A$ & $\A$ & {\color{gray} \scriptsize{\textit{F}}} & \scriptsize{{\color{gray}\textit{FHT}}} & $\A$ & $\F\HFB$ \\
     & 1.37e-01 -- 1.23 & 3486 & $\A$ & $\A$ & $\A$ & $\A$ & & \scriptsize{{\color{gray}\textit{FHT}}} & $\HFB\T$ & $\F\HFB$ \\
     & 1.79e-01 -- 1.89 & 4036 & $\A$ & $\A$ & $\A$ & $\A$ & \scriptsize{\textit{FHT}} & \scriptsize{\textit{FHT}} & $\A$ & $\A$ \\
     & 1.90e-02 -- 2.36 & 4742 & $\A$ & $\A$ & $\A$ & $\A$ & $\A$ & $\A$ & $\A$ & $\A$ \\
    \hline
    Y-91 & 7.69e+01 -- 1.37e+02 & 1205 & & $\F$ & & & & & & \\
    \hline
    Zr-95 & 1.17e+01 -- 6.42e+01 & 724, \textbf{757} & & $\HFB$ & $\A$ & $\A$ & & $\A$ & $\A$ & $\F\HFB$ \\
    \hline
    Nb-95 & 6.93e+01 -- 7.01e+02 & 766 & $\HFB$ & $\F\HFB$ & $\A$ & $\A$ & & $\A$ & $\A$ & $\A$ \\
    \hline
    Ru-103 & 6.13e+01 -- 1.06e+02 & 295 & & & $\HFB$ & & & & & \\
     & 6.09 -- 2.71e+02 & 497 & $\A$ & $\A$ & $\A$ & $\A$ & $\A$ & $\A$ & $\A$ & $\A$ \\
    \hline
    Rh-106 & 2.33e+02 -- 2.73e+03 & 512 & & & & & & $\A$ & $\A$ & \\
     & 4.69e+02 -- 2.95e+03 & 622 & & & $\HFB$ & $\HFB$ & & & $\A$ & $\HFB$ \\
     & 7.78e+02 -- 3.59e+03 & 873 & $\A$ & $\A$ & $\A$ & $\A$ & \scriptsize{\textit{{\color{gray}F}H{\color{gray}T}}} & $\A$ & $\A$ & $\A$ \\
     & 7.57e+01 -- 4.39e+03 & 1050 & $\A$ & $\A$ & $\A$ & $\A$ & $\A$ & $\A$ & $\A$ & $\A$ \\
     & 4.38e+01 -- 3.02e+03 & 1562 & $\A$ & $\A$ & $\A$ & $\A$ & $\A$ & $\A$ & $\A$ & $\A$ \\
     & 3.52e+01 -- 2.38e+03 & \textbf{1766}, 1797 & $\A$ & $\A$ & $\A$ & $\A$ & $\A$ & $\A$ & $\A$ & $\A$ \\
     & 5.88e+01 -- 2.15e+03 & 1927, \textbf{1988} & $\A$ & $\A$ & $\A$ & $\A$ & $\A$ & $\A$ & $\A$ & $\A$ \\
     & 5.88e+01 -- 2.10e+03 & 2113 & $\A$ & $\A$ & $\A$ & $\A$ & $\A$ & $\A$ & $\A$ & $\A$ \\
     & 3.97e+01 -- 2.29e+03 & 2366 & $\A$ & $\A$ & $\A$ & $\A$ & $\A$ & $\A$ & $\A$ & $\A$ \\
     & 1.02e+01 -- 2.25e+03 & 2406 & & & & & $\F$ & & & $\HFB$ \\
     & 1.25e+01 -- 1.26e+03 & 2705, \textbf{2710} & $\A$ & $\A$ & $\A$ & $\A$ & $\A$ & $\T$ & $\A$ & $\A$ \\
     & 8.80 -- 5.22e+02 & 3037 & $\F\T$ & & $\A$ & $\HFB\T$ & \scriptsize{\textit{F}}$\HFB\T$ & & & $\A$ \\
    \hline
    %Pd-112 & 7.77e-01 -- 3.18 & th & & & & & & & & $\T$ \\
    Ag-112 & 7.77e-01 -- 6.35 & 1312, \textbf{1388} & & & & & $\HFB\T$ & & & $\A$ \\
     & 5.30e-01 -- 1.07e+01 & 1614 & $\HFB\T$ & $\F$ & & & $\A$ & & & $\A$ \\
     & 9.31e-01 -- 8.42 & 1798 & & & & & & & & $\HFB$ \\
     & 4.16e-01 -- 5.07 & 2106 & $\HFB$ & & & & $\HFB\T$ & & & $\A$ \\
     & 4.42e-01 -- 1.31e+01 & 2507 & $\A$ & $\F$ & & & $\A$ & & & $\A$ \\
     & 4.29e-01 -- 1.35e+01 & 2829 & $\A$ & $\A$ & $\A$ & $\T$ & $\A$ & & & $\A$ \\
     & 2.64e-01 -- 9.85 & 3393 & $\A$ & $\A$ & $\F$ & & \scriptsize{\textit{FHT}} & & & $\A$ \\
    \hline
    Sn-125 & 1.44 -- 1.06e+02 & \textbf{1067}, 1088, 1089 & $\A$ & $\A$ & $\A$ & $\F\T$ & $\A$ & & & $\A$ \\
     & 9.27 -- 8.06e+01 & 1221 & $\A$ & $\F\T$ & $\F$ & & \scriptsize{{\color{gray}\textit{FT}}} & & & $\F$ \\
     & 4.50 -- 9.41e+01 & 1420 & $\F\T$ & $\F\T$ & $\A$ & $\A$ & $\A$ & & & $\A$ \\
     & 3.48 -- 9.24e+01 & 1807 & $\A$ & $\A$ & $\A$ & $\A$ & \scriptsize{\textit{FHT}} & & & $\A$ \\
     & 1.35 -- 1.31e+02 & 2002 & $\A$ & $\A$ & $\A$ & $\A$ & $\A$ & $\A$ & $\A$ & $\A$ \\
     & 2.98 -- 9.50e+01 & 2276 & $\A$ & $\A$ & $\A$ & $\F\T$ & \scriptsize{\textit{FH{\color{gray}T}}} & & & $\A$ \\
    \hline
    %Sn-126 & 1.51e+04 -- 1.66e+04 & th & $\F\T$ & & & & $\F$ & & & \\
    Sb-125 & 3.63e+01 -- 1.50e+04 & 176 & $\A$ & $\A$ & $\A$ & $\A$ & $\A$ & $\A$ & $\A$ & $\A$ \\
     & 1.16e+02 -- 1.65e+04 & \textbf{428}, 463 & $\A$ & $\A$ & $\A$ & $\A$ & $\A$ & & $\T$ & $\A$ \\
     & 2.29e+02 -- 1.43e+04 & \textbf{601}, 636 & $\A$ & $\A$ & $\A$ & $\A$ & $\A$ & & $\F\T$ & $\A$ \\
    \hline
    %Sb-126 & 1.60e+04 -- 1.05e+07 & 415 & $\A$ & & & & $\A$ & & & \\
    % & 1.37e+04 -- 1.50e+08 & \textbf{666}, 695, 697 & $\A$ & $\F\T$ & $\F$ & & $\A$ & & & $\A$ \\
    %\hline
    Sb-127 & 2.12 -- 1.97e+01 & 473 & $\A$ & $\F\T$ & $\F$ & $\F$ & $\A$ & & & \\
     & 2.15 -- 2.83e+01 & 686 & $\A$ & $\A$ & $\F$ & & $\A$ & & & $\A$ \\
     & 2.86 -- 2.26e+01 & 784 & & $\F$ & $\F$ & & & & & \\
     & 4.29 -- 1.56e+01 & 1290 & $\F\T$ & $\F$ & $\F$ & & \scriptsize{{\color{gray}\textit{FH}}} & & & $\HFB$ \\
    \hline
    Sb-128 & 3.52e-01 -- 1.17 & 314 & $\F\HFB$ & & & & \scriptsize{{\color{gray}\textit{H}}} & & & \\
     & 1.35e-01 -- 2.08 & \textbf{743}, 754 & $\A$ & $\A$ & & & $\A$ & & & $\HFB$ \\
     & 7.68e-01 -- 1.80 & 1113, 1158, \textbf{1182} & $\HFB$ & & & & \scriptsize{\textit{H}} & & & \\
    \hline
    Sb-129 & 2.00e-01 -- 1.26 & 1737 & $\F$ & $\F$ & & & $\F$ & & & \\
    \hline
    I-131 & 1.30e+01 -- 3.67e+01 & 364 & & $\HFB$ & & & & & & \\
    \hline
    I-132 & 8.38 -- 1.12e+01 & 1372, \textbf{1399} & & $\HFB$ & & & & & & \\
    \hline
    La-140 & 1.27e+01 -- 1.15e+02 & 1596 & $\HFB$ & $\HFB$ & & & & & & \\
     & 1.29e+01 -- 9.13e+01 & 2521 & $\F\HFB$ & $\F\HFB$ & & & & & & \\
     & 2.20e+01 -- 2.36e+01 & 2899.6 & $\HFB$ & & & & & & & \\
     & 6.41 -- 2.82e+01 & 3119 & $\HFB$ & $\HFB$ & & & & & & \\

 \end{longtable*}

%weak remnants table
\begin{longtable*}[]{|c|c|c|c|c|c|c|c|c|c|c|}
    \caption{Signal duration predictions for emission lines of relevance to remnant searches as they are reported to be visible some time after 10 years post-merger given weak $r$-process variations by considering the astrophysical outflows presented in Table~\ref{tab:astrovar}. The same approach and notation as Table~\ref{tab:long_weak} is also applied here, but with a detectability threshold of $10^{-10}$ counts s$^{-1}$ cm$^{-2}$ given the prospect of identifying nearby remnants.}
    \label{tab:rem_weak} \\

    \hline
    \multirow{2}{3em}{\centering Isotope} & \multirow{2}{6em}{\centering Time (years)} & \multirow{2}{4em}{\centering E (keV)} & \multicolumn{8}{|c|}{Conditions} \\
    \cline{4-11}
     & & & \multicolumn{1}{|c|}{1} & \multicolumn{1}{|c|}{2} & \multicolumn{1}{|c|}{3} & \multicolumn{1}{|c|}{4} & \multicolumn{1}{|c|}{5(r)} & \multicolumn{1}{|c|}{6(r)} & \multicolumn{1}{|c|}{7}  & \multicolumn{1}{|c|}{8} \\

    \endfirsthead

    \multicolumn{8}{c}
    {{\bfseries \tablename\ \thetable{} -- continued from previous page}} \\
    \hline 
    Isotope & Time (years) & E (keV) & \multicolumn{1}{|c|}{1} & \multicolumn{1}{|c|}{2} & \multicolumn{1}{|c|}{3} & \multicolumn{1}{|c|}{4} & \multicolumn{1}{|c|}{5(r)} & \multicolumn{1}{|c|}{6(r)} & \multicolumn{1}{|c|}{7}  & \multicolumn{1}{|c|}{8} \\
    \hline
    \endhead

    \hline\endlastfoot
    \hline
    K-42 & 9.76 -- 2.89e+02 & 1525 & & & & & $\F\HFB$ & & & $\A$ \\
     & 1.75e+01 -- 1.82e+01 & 1921 & & & & & & & & $\F$ \\
    \hline
    Co-60 & 1.43e+01 -- 1.40e+07 & 1173 & & & $\HFB$ & $\A$ & & $\A$ & $\A$ & \\
     & 8.10 -- 1.40e+07 & 1332 & & & $\HFB$ & $\A$ & & $\A$ & $\A$ & \\
    \hline
    Kr-85 & 8.47 -- 1.51e+02 & 514 & & $\HFB$ & $\HFB$ & $\F\HFB$ & & $\A$ & $\A$ & \\
    \hline
    Rh-106 & 2.34 -- 1.63e+01 & 873 & & & $\A$ & $\A$ & & \scriptsize{\textit{FHT}} & $\A$ & $\A$ \\
     & 2.07e-01 -- 1.90e+01 & 1050 & $\A$ & $\A$ & $\A$ & $\A$ & $\A$ & $\A$ & $\A$ & $\A$ \\
     & 1.20e-01 -- 1.90e+01 & 1562 & $\A$ & $\A$ & $\A$ & $\A$ & $\A$ & $\A$ & $\A$ & $\HFB\T$ \\
     & 9.70e-02 -- 1.79e+01 & \textbf{1766}, 1797 & $\A$ & $\A$ & $\A$ & $\A$ & \scriptsize{\textit{F}}$\HFB\T$ & $\A$ & $\A$ & $\A$ \\
     & 1.61e-01 -- 1.72e+01 & 1927, \textbf{1988} & $\A$ & $\A$ & $\A$ & $\A$ & $\A$ & $\A$ & $\A$ & $\A$ \\
     & 1.61e-01 -- 1.72e+01 & 2113 & $\A$ & $\A$ & $\A$ & $\A$ & $\A$ & $\A$ & $\A$ & $\A$ \\
     & 1.09e-01 -- 1.78e+01 & 2366 & $\A$ & $\A$ & $\A$ & $\A$ & $\A$ & $\A$ & $\A$ & $\A$ \\
     & 2.80e-02 -- 1.73e+01 & 2406 & & & & & $\F$ & & & $\HFB$ \\
     & 3.40e-02 -- 1.46e+01 & 2705, \textbf{2710} & $\A$ & $\A$ & $\A$ & $\A$ & $\A$ & $\T$ & $\A$ & $\A$ \\
     & 2.40e-02 -- 1.24e+01 & 3037 & $\T$ & & $\A$ & & \scriptsize{\textit{F}}$\HFB\T$ & & & $\A$ \\
    %Sn-126 & 3.93e+01-2.54e+06 & th & $\A$ & $\A$ & $\A$ & $\A$ & $\A$ & $\A$ & $\A$ & $\A$ \\
    \hline
    Sb-125 & 9.90e-02 -- 4.25e+01 & 176 & $\A$ & $\A$ & $\A$ & $\A$ & $\A$ & $\A$ & $\A$ & $\A$ \\
     & 3.19e-01 -- 4.51e+01 & \textbf{428}, 463 & $\A$ & $\A$ & $\A$ & $\A$ & $\A$ & & $\T$ & $\A$ \\
     & 6.27e-01 -- 3.91e+01 & \textbf{601}, 636 & $\A$ & $\A$ & $\A$ & $\A$ & $\A$ & & $\F\T$ & $\A$ \\
    \hline
    Sb-126 & 4.36e+01 -- 2.54e+06 & 415 & $\A$ & $\A$ & $\A$ & $\A$ & $\A$ & $\A$ & $\A$ & $\A$ \\
     & 3.77e+01 -- 2.92e+06 & \textbf{666}, 695, 697 & $\A$ & $\A$ & $\A$ & $\A$ & $\A$ & $\A$ & $\A$ & $\A$ \\
     & 1.35e+01 -- 1.17e+06 & 1213 & $\A$ & $\A$ & $\F\T$ & & $\A$ & & & $\A$ \\
     & 1.57e+01 -- 3.36e+05 & 1477 & $\A$ & $\F\T$ & $\F$ & & $\A$ & & & $\A$ \\

\end{longtable*}

For the weak $r$ process, lines from Rb-88, Ru-103, Rh-106, Sn-125 and Sb-125 are found to be above threshold across all astrophysical variations and nuclear models for the post-merger timescale of days to years. Of these most robustly reported cases, only Rh-106 and Sb-125 are also reported to emit above threshold during remnant timescales beyond ten years, with Sb-125 signals lasting as long as several decades. In the case of Sb-126, while its half life is only 12.4 days, it is directly $\beta$-fed by Sn-126 which has a half-life of 230 kiloyears. Sn-126 does not currently have decay radiation reported in ENDF above 0.1 MeV, while Sb-126 has many lines above above 0.1 MeV, making Sb-126 the decay that is ultimately detectable. We note that Sn-126 was highlighted by the peak finder on remnant timescales, but only for theoretical lines which are not reported in the tables. Above $A\geq 130$, there are not many nuclei that produce signals in the weak $r$-process conditions considered here, notable exceptions being I-131 and I-132, which were only reported in HFB27 calculations for only one astrophysical condition (2), as well as La-140, which is still reported for two conditions despite having a lower relative abundance ($\sim$10$^{-7}$).

All of the lighter nuclei reported in the neutron star merger mass-weighted scenario (Fe-59, Co-60, Ga-72, Ge-77, Rb-88, Nb-95, Ru-103, Rh-106, Ag-112, Sn-125, Sb-125, Sb-126, Sb-128, I-131, I-132) are also found in Tables~\ref{tab:long_weak} and \ref{tab:rem_weak}. Several of these isotopes have emission lines commonly predicted across all model variations. Note however that the exact number of lines / length of time that they remain above threshold varies across models and conditions. Given astrophysical variations several isotopes appear that were not previously reported in the mass weighted calculations: Na-24, K-42, Cu-66, Cu-67, Zn-72, Ge-77, Kr-85, Kr-88, Y-91, Zr-95, Sb-127, and Sb-129, none of which are robustly produced across weak $r$-process variations for all nuclear models. We note that across the main and weak $r$-process results in Tables~\ref{tab:long_main}, \ref{tab:rem_main}, \ref{tab:long_weak}, and \ref{tab:rem_weak} Rb-92, Pd-112, Sn-126, Cd-127, Te-137, Bi-213, and Ra-225 were also highlighted by our peak finder but in these cases the ENDF decay spectra include theoretical emission which we do not report in our tables.

\begin{figure}
    \centering
    \includegraphics[width=0.9\linewidth]{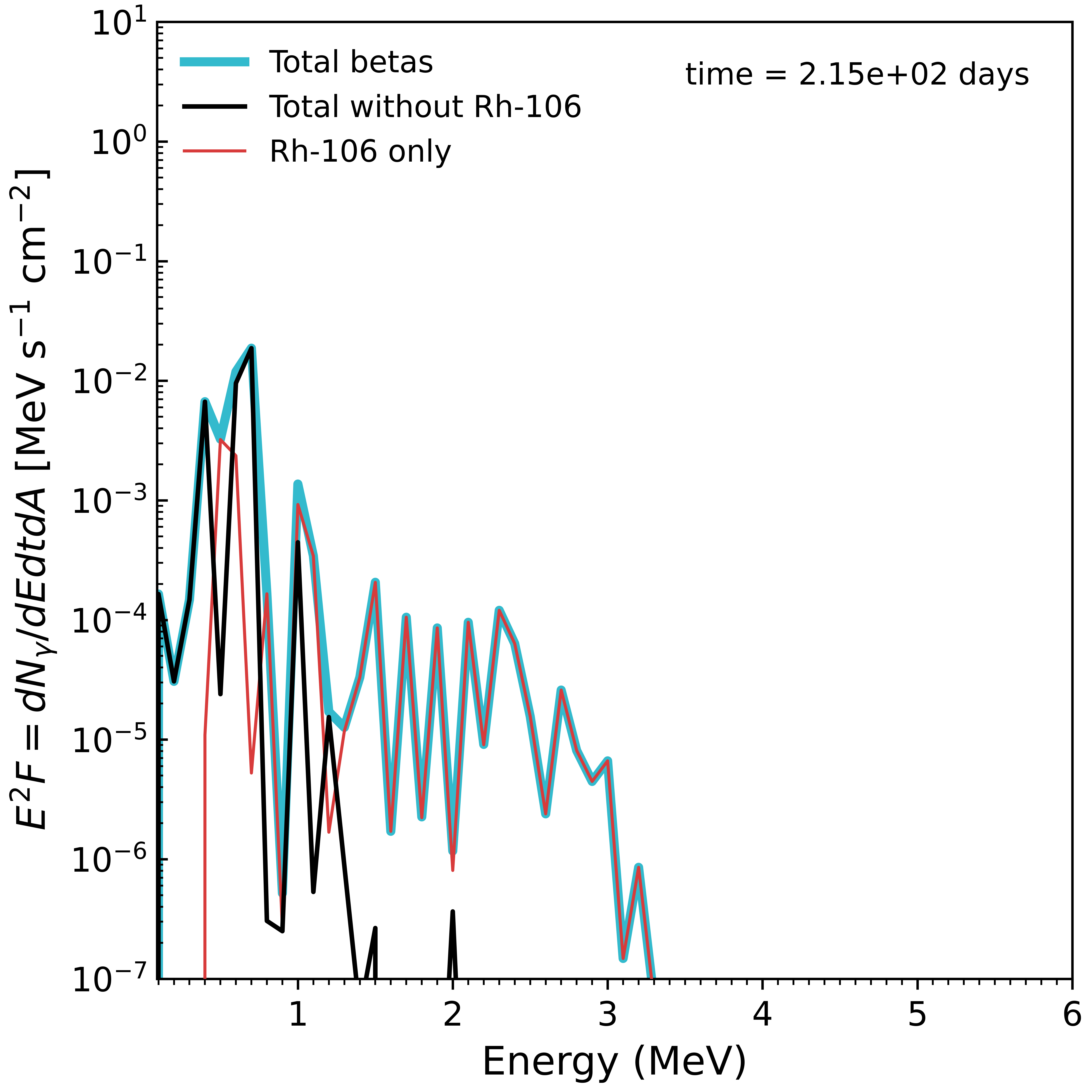}
    \hspace{0.5cm}
    \includegraphics[width=0.9\linewidth]{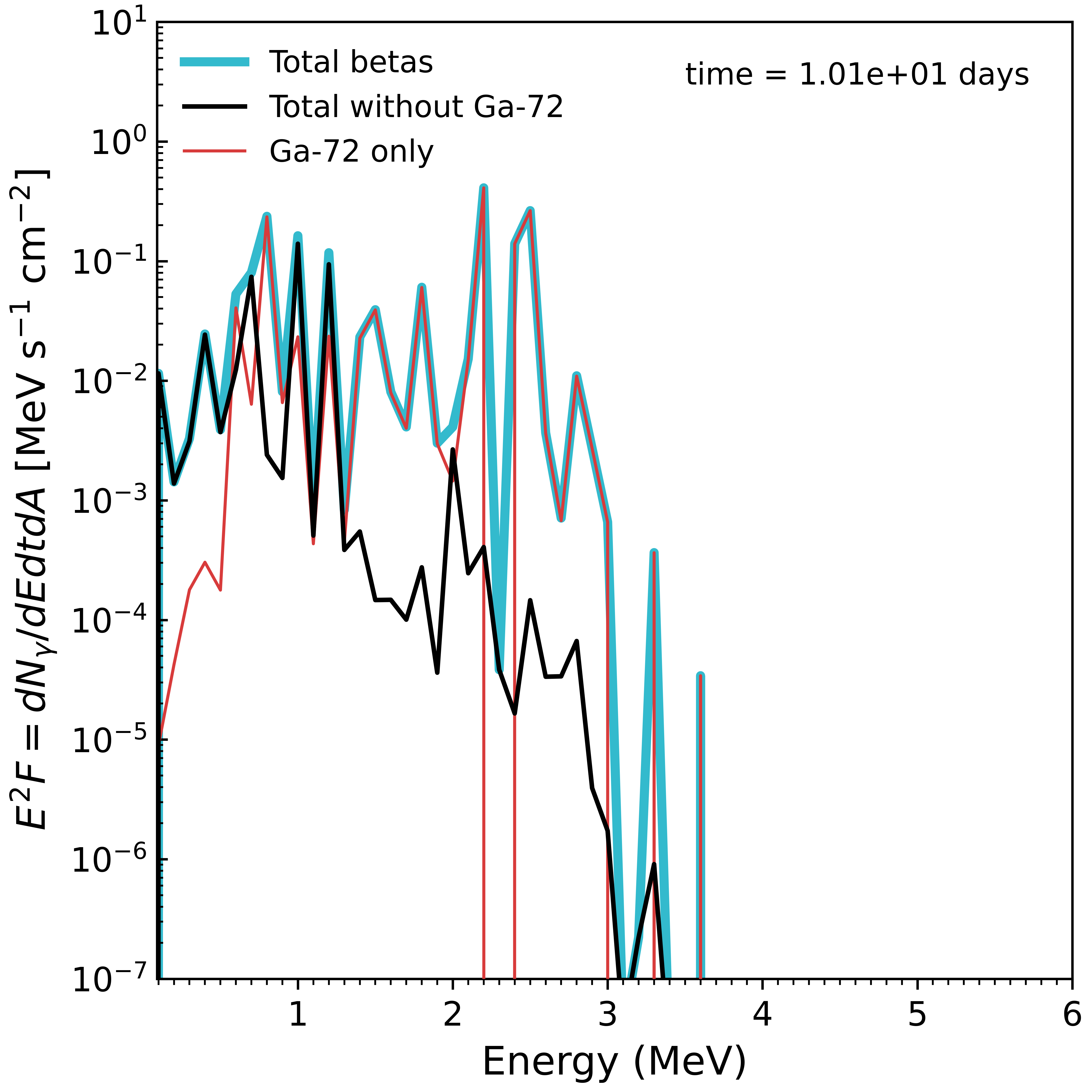}
    \caption{Total emission spectrum from 0.1 to 6 MeV compared with individual contributions from (top) Rh-106 lines (here at 215 days using FRDM2012 and condition 5) and (bottom) Ga-72 lines (here at 10 days using FRDM2012 and condition 7). In both calculations solely weak $r$-process nuclei are produced, and the spectrum from $\sim$1 MeV beyond becomes identical to that of (top) Rh-106 for tens of years or (bottom) Ga-72 for tens of days.}
    \label{fig:rh106}
\end{figure}

One of the most featured isotopes is Rh-106, with a total of 10 lines of interest between 0.5 MeV and 3.1 MeV in Table~\ref{tab:rem_weak}, 7 of which are commonly found across calculation variations. In comparison, there are only very few cases where this isotope appears when considering main $r$-process ejecta in Tables~\ref{tab:long_main}, \ref{tab:rem_main} or a mass weighted total in Table~\ref{tab:massweight_days}, \ref{tab:massweight_years}. When we consider a combination of weak plus main $r$-process ejecta as in Sec.~\ref{sec:nsmw3models} only 1 line (2.366 MeV) of Rh-106 appears which is solely reported for the TF model. This is due to Rh-106 emission being obscured by other lines arising from main $r$-process isotopes. Importantly, we find that in the case of a weak $r$-process only, numerous Rh-106 lines are robustly reported for all nuclear and astrophysical variations across both post-merger and remnant timescales. Additionally, numerous Rh-106 lines can last for longer than 10 years. The influence of Rh-106 emission at certain time windows for the weak $r$-process is exemplified in Fig~\ref{fig:rh106} where the entire spectrum above 1 MeV follows the Rh-106 spectrum itself. Therefore the observation of numerous Rh-106 lines could serve as an implication that a site was dominated by weak $r$-process production rather than being a significant astrophysical source of main $r$-process nuclei. Following the same logic, we find Ga-72 to report numerous lines in Table~\ref{tab:long_weak} for about half of the astrophysical conditions considered. As is shown in Fig.~\ref{fig:rh106}, a similar argument to Rh-106 spectral takeover on the order of tens of years can be made for Ga-72 at earlier emission times between a few to tens of days. Thus our procedure identified two possible chances to discern weak from main $r$-process production, via Ga-72 around tens of days and Rh-106 on longer timescales of tens of years.

\section{Discussion of Tl-208 signal robustness and competition}\label{sec:tl208}

Tl-208 has a particularly strong 2.6 MeV line in its emission spectrum first discussed as an astrophysical signal in \cite{VasshTl208}. Thus we next carefully consider the impact of the nuclear physics and astrophysical variations in this work on the visibility of a signal from Tl-208.  We investigate specifically the energy window of $2.5-2.8$ MeV in order to focus on decaying species whose emission may interfere with Tl-208 emission at some point in time. This energy window is selected to include the full broadened peak post-radiation transfer.

\begin{figure}
    \centering
    \includegraphics[width=0.95\linewidth]{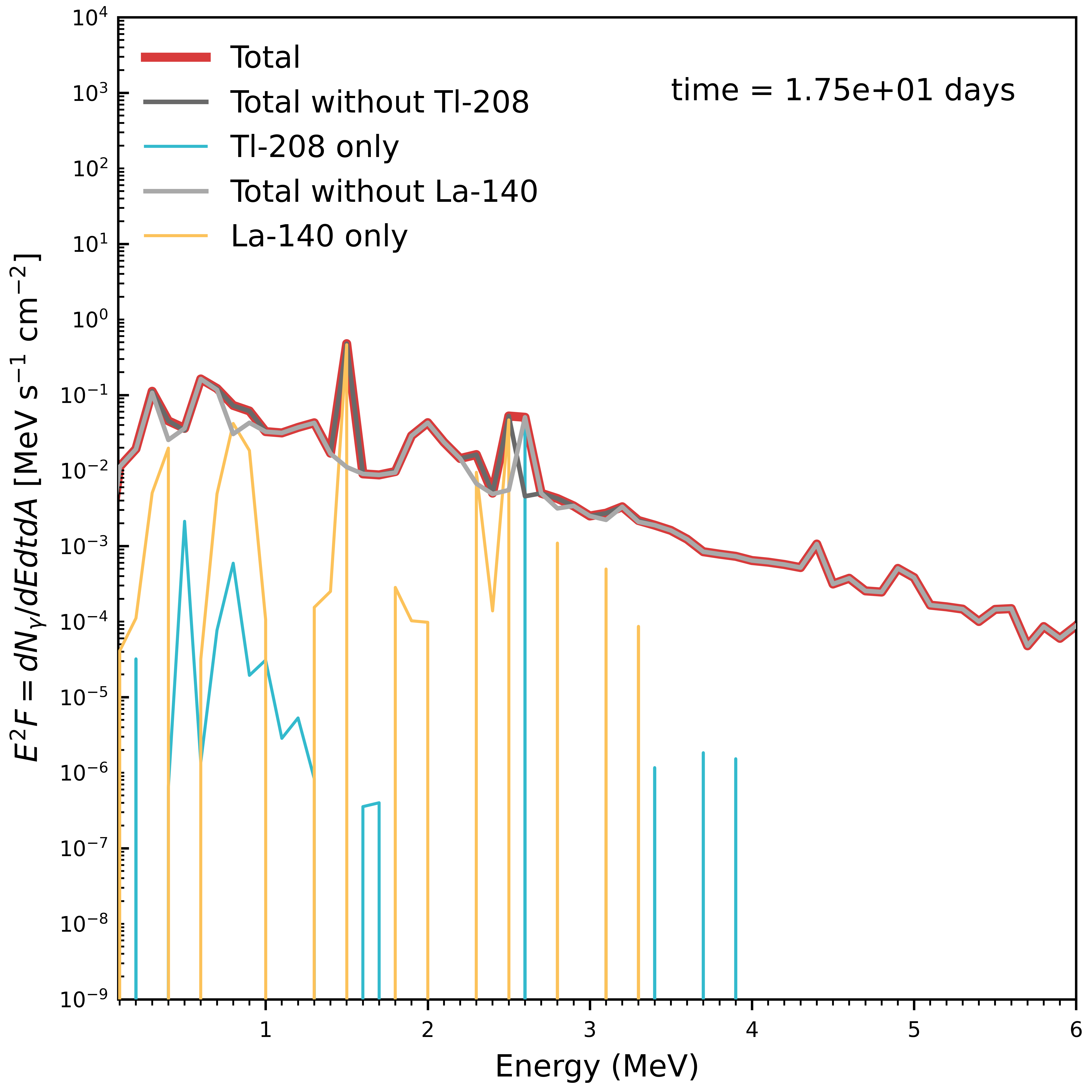}
    \includegraphics[width=0.95\linewidth]{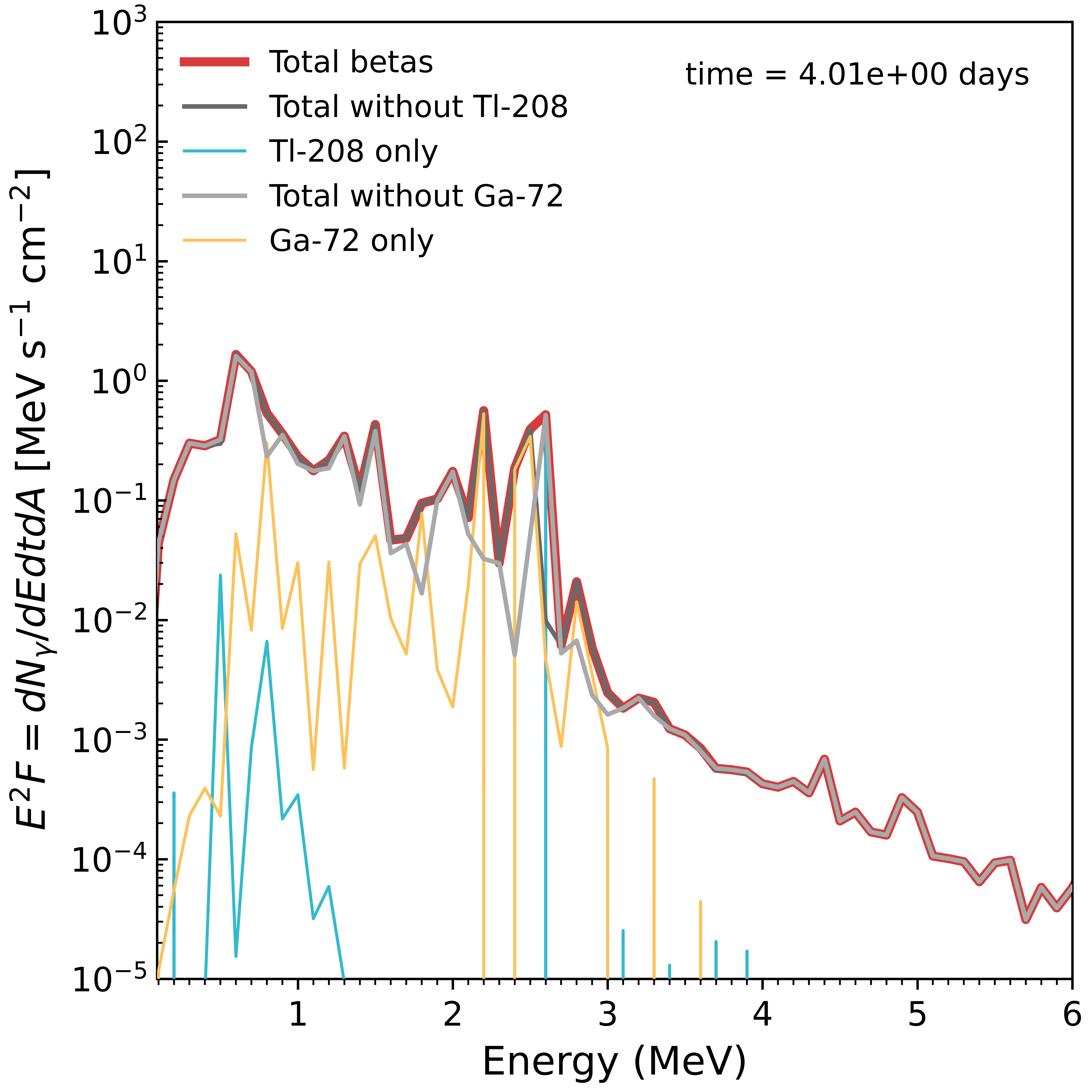}
    \caption{(Top) Total emission spectrum from 0.1 to 6 MeV compared to individual contributions from La-140 and Tl-208 at 17.5 days (assuming astrophysical condition $Y_e=0.01$, $s/k=10$; shown after radiation transfer with a velocity $0.1c$). This time shows when La-140 takes over emission in the 2.5-2.8 MeV range. (Bottom) Total emission spectrum along with individual contributions from Ga-72 and Tl-208. Here we show the neutron star merger total ejecta scenario from Sec.~\ref{sec:nsmw3models} to consider a case where the entire $r$-process abundance pattern is produced, from the first peak (including Ga-72) to beyond the third peak (including Tl-208). This snapshot shows the time when Ga-72 comes closest to taking over the 2.5 - 2.8 MeV emission region (4 days) in this scenario. The example considered here shows results before radiation transfer. Both top and bottom panels demonstrate results with the FRDM2012 mass model and M\"{o}ller et al. $\beta$-decay rates.}
    \label{fig:spectra_LavsTl}
\end{figure}

We first carefully tease out the decaying species whose emission may interfere with a Tl-208 signal at some point in any of our calculations. Two specific nuclei of interest are found: Ga-72  (T$_{1/2}$ = 14 h)  and La-140 (T$_{1/2}$ = 40.3 h). In particular the 2.508 MeV emission line of Ga-72 and the 2.521 MeV emission line of La-140 could possibly interfere with the 2.6 MeV signal from Tl-208. However, as can be seen in Fig.~\ref{fig:spectra_LavsTl}, there are additional strong lines for these species that would indicate their presence: the 2.202 MeV line of Ga-72 and the 1.596 MeV line of La-140. We only see Ga-72 competition in our mass weighted calculations featured in Sec.~\ref{sec:nsmw3models}, and so both a first and third $r$-process peak are required for Ga-72 and Tl-208 to compete. Note however that in the specific mass weighted scenario considered here, the 2.5 MeV line from Ga-72 never overtakes the Tl-208 signal at early times in the case of the FRDM2012 mass model and M\"{o}ller et al $\beta$-decay, as can be explicitly seen in Fig.~\ref{fig:spectra_LavsTl}. Note that the predicted relative amounts of weak versus main $r$-process ejecta can vary depending on the specifics of the merger (e.g. progenitor masses) and predicted ratios of these ejecta types in simulations can vary greatly depending on neutrino transport treatments. In contrast, every main $r$-process calculation we performed produced both La-140 and Tl-208 to some degree. Thus since La-140 more consistently appears with a competitive abundance across astrophysical conditions producing Tl-208, we next focus on the details of La-140 and Tl-208 competition. With La-140 lines being reported on the order of days, around these times emission combinations of strong $\sim$1.5 MeV / 2.5 MeV lines could be tied to La-140 alone. Interestingly, a $\sim$1.5 MeV / 2.6 MeV strong line combination can also be produced by Tl-208 alongside Tl-209, but only on longer timescales of tens to hundred of years. Note however when $\sim$1.5 MeV / 2.5 MeV lines are due to La-140 alone their relative ratio is fixed by their spectral intensities, whereas the relative height of a Tl-208 emission line to that from Tl-209 will vary over time due to their being populated by different $\alpha$-decay chains (see Appendix Table~\ref{tab:BemittersAFdec}).

\begin{figure*}[t!]
    \centering
    \includegraphics[width=0.9\linewidth]{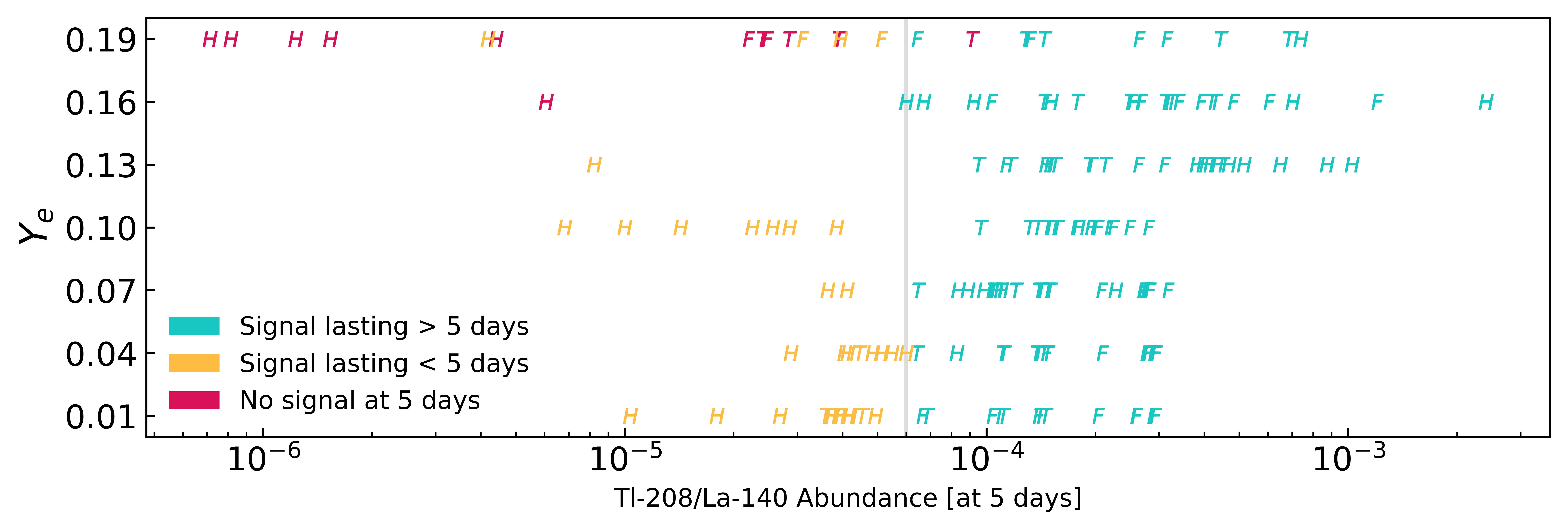}
    \hspace{0.5cm}
    \includegraphics[width=0.9\linewidth]{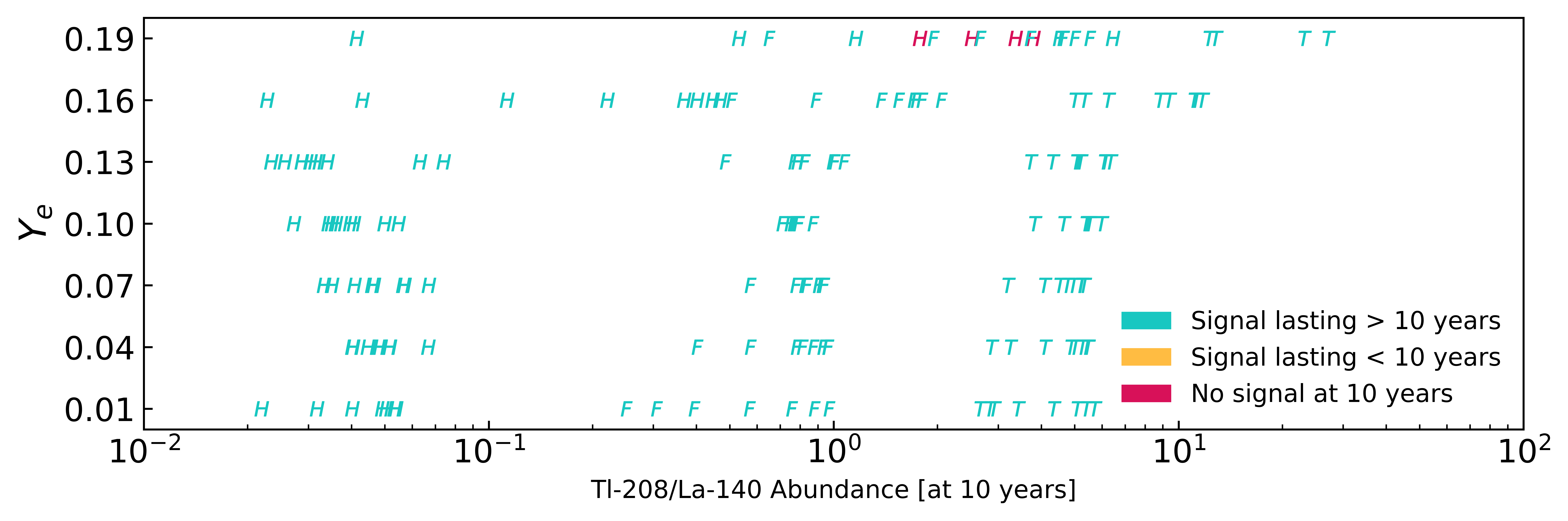}
    \caption{(Top) The Tl-208/La-140 abundance ratio at 5 days for all the parametrized trajectories in the extended set. Results with the three nuclear models considered are labeled with H, T, or F. The cases with a signal longer than 5 days are in teal, those in yellow have a signal shorter than 5 days and in red are the cases which do not produce a Tl-208 signal on the order of days. (Bottom) Same as above, but at 10 years. Teal shows cases with a Tl-208 2.6 MeV signal lasting beyond 10 years. Red shows cases which never report a signal around 10 years.}
    \label{fig:thresholdratio}
\end{figure*}

For a more extensive Tl-208 and La-140 comparison, we expand the grid of astrophysical calculations in the previous section to include 7 $Y_e$ values between 0.01 and 0.19 (0.01, 0.04, 0.07, 0.10, 0.13, 0.16, 0.19) and 8 entropies per baryon ($s/k =$ 10, 18, 24, 32, 42, 56, 75, 100). This corresponds to a total of 56 trajectories per mass model instead of 15 (168 total compared to 45). We consider this larger dataset here in order to better discern trends regarding Tl-208 production and emission in the specific $2.5-2.8$ MeV energy window of interest.  That is, the new grid provides a larger set of possible relative abundances of Tl-208 to other nuclei who may compete in its emission window. 

With the peak finder reporting local maxima at every time step, we identify and show when the La-140 line (2.5--2.6 MeV bin) starts rising above the Tl-208 line (2.6--2.7 MeV bin) explicitly in Fig.~\ref{fig:spectra_LavsTl}. When La-140 emission competes, its contribution typically rises to take over the Tl-208 signal rather than obscuring the entire possible emission window. For the trajectory shown here with $Y_e=0.01$, $s/k=10$, La-140 starts peaking above Tl-208 at roughly 17.5 days, however this take over time varies across the parametrized trajectory grid considered. In calculations using the FRDM2012 model, the rise in La-140 typically starts later, leaving Tl-208 visible to 15 days on average. TF tends to follow that trend as well with an average end time for the earlier Tl-208 signal being 13 days. However for HFB27 the rise in La-140 occurs earlier, making the Tl-208 signal end on average around 9.6 days. The spread in the time we find Tl-208 emission to be visible over La-140 is also larger for HFB27, with many conditions having a signal for Tl-208 shorter than 5 days, but some longer than 20 days. Overall, although the predicted times for which the signal may be detectable vary widely, the Tl-208 signal is distinguishable at some point across almost all nuclear models and astrophysical conditions. 

We use the results of our extended data set to quantify the impact of the relative Tl-208 to La-140 abundance. In Fig.~\ref{fig:thresholdratio} we show the abundance ratio of Tl-208/La-140 at both 5 days and 10 years, with each calculation labeled as to whether Tl-208 dominates at this time. Interestingly we observe there to be a threshold of $5.5\times10^{-5}$ for the Tl-208/La-140 ratio which separates cases that will have a visible Tl-208 signal around $\sim$5 days and those that will not. Importantly this implies that ten thousand more times La-140 than Tl-208 is required in order for it to overtake the $\sim$2.6 MeV emission window.

\begin{figure}
    \centering
    \includegraphics[width=1.0\linewidth]{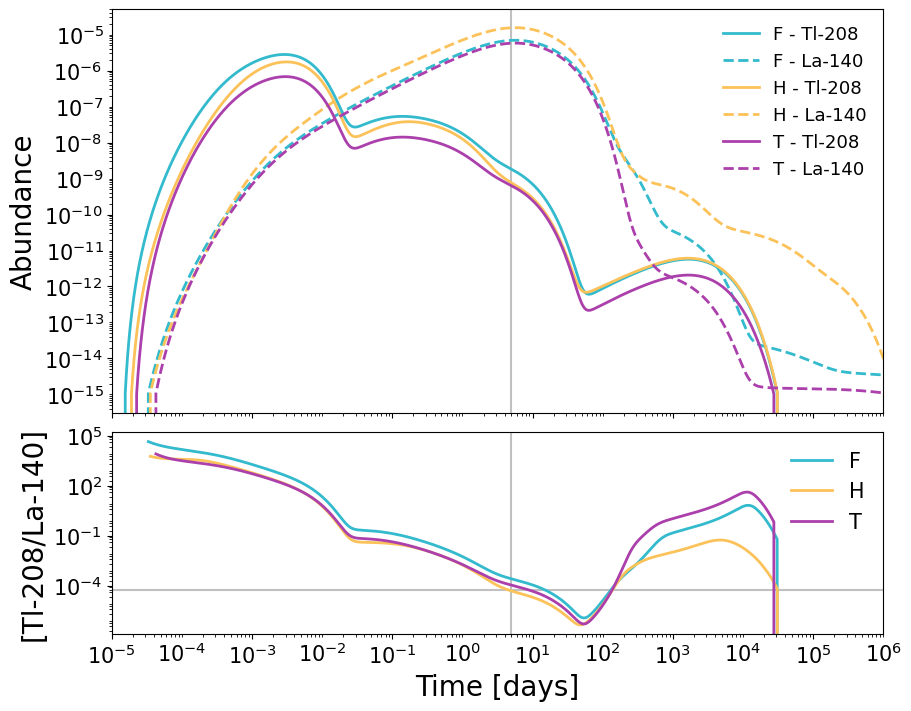}
    \caption{(Top) Individual abundance of La-140 and Tl-208 across time for the $Y_e=0.01$, $s/k=10$ case given all three nuclear models (teal for FRDM2012, yellow for HFB27 and purple for TF). (Bottom) Ratio of Tl-208 to La-140 abundances across time. The gray vertical line is a point of reference for 5 days while the horizontal line shows the abundance ratio of $5.5\times10^{-5}$ found to be a threshold of sorts in Fig.~\ref{fig:thresholdratio}.}
    \label{fig:lavstl}
\end{figure}

We next consider the dynamical evolution of Tl-208 and La-140 abundances in our network calculations. We show the abundance ratio of Tl-208 to La-140 as a function of time in Fig.~\ref{fig:lavstl} for all three models for the $Y_e=0.01$, $s/k=10$ case. Calculations using HFB27 produce Tl-208 on comparable levels to those with FRDM2012 and TF but La-140 abundances are the highest using HFB27, both at early and late times. Note that the higher abundance of La-140 at late times in this case is produced by fission fragments from the heavier actinide species that HFB fission barriers allow. Utilizing the $5.5\times10^{-5}$ ratio threshold found, we can see that for the astrophysical condition considered here Tl-208 is sufficiently populated in all models such that it produces a stronger signal than La-140 at some point during both timescales of days and years, even with the higher La-140 abundances of HFB27.

\begin{figure}
    \centering
    \includegraphics[width=1.0\linewidth]{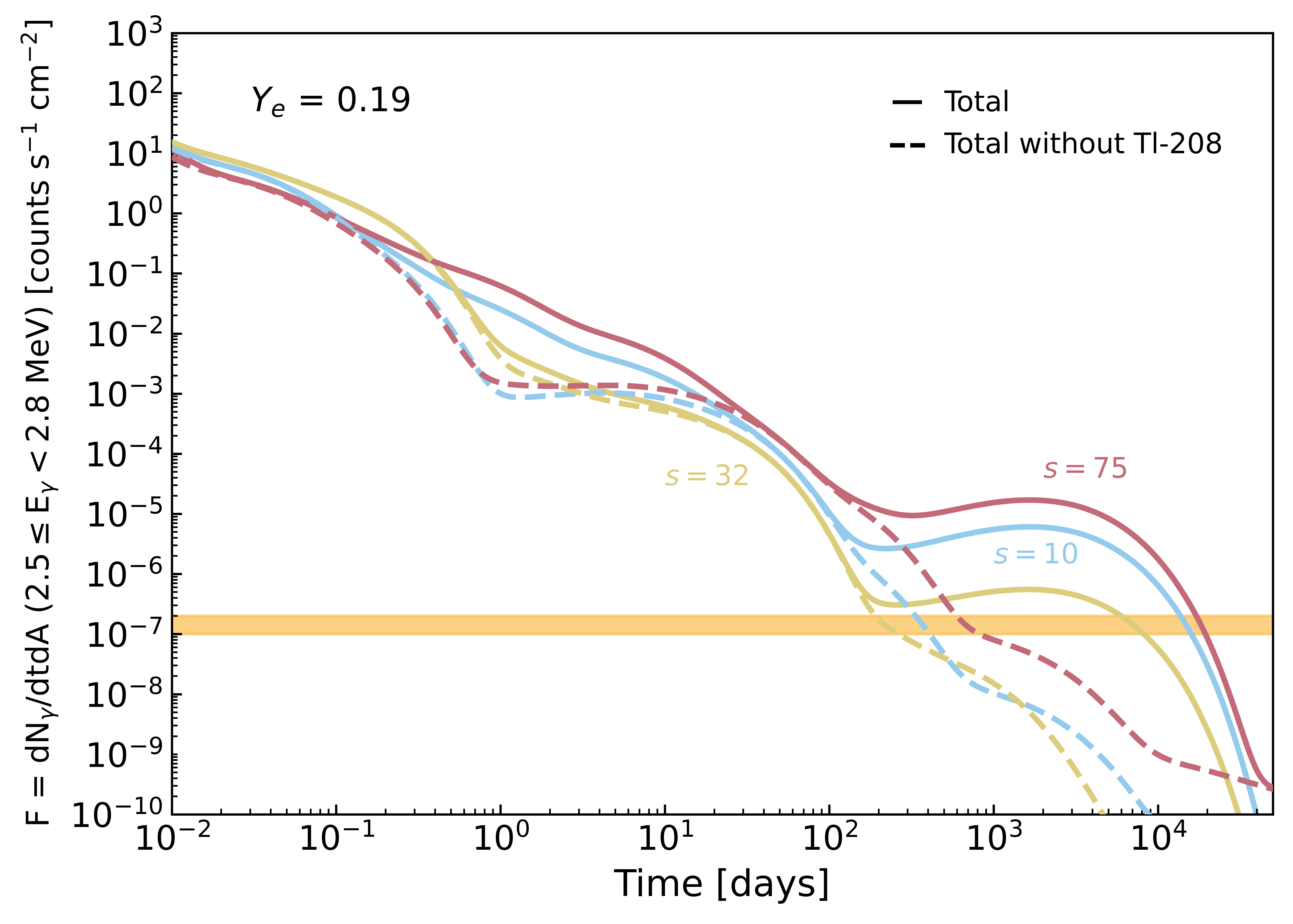}
    \caption{The evolution of the total light curve in the energy range of relevance for the 2.6 MeV line of Tl-208 for three distinct astrophysical conditions with $Y_e=0.19$: one for which actinides are strongly produced ($s/k=75$, red), another with a level of actinides typical of merger dynamical ejecta ($s/k=10$, blue), and a third with lower lead / actinide production ($s/k=32$, yellow). Lightcurve calculations are shown both with and without Tl-208 emission. The orange vertical band shows the detection threshold assumed ($10^{-7}$).}
    \label{fig:leadprod_tsignal}
\end{figure}

We lastly explicitly discuss how the strength of Tl-208 emission correlates with the abundance of isotopes near the third $r$-process peak at $A\sim195$ and beyond such as the lead spike ($A\sim$ 206, 207, 208) and the actinides. In Fig.~\ref{fig:leadprod_tsignal} we highlight results with three astrophysical conditions (all with $Y_e=0.19$): one case where actinides are strongly produced ($s/k=75$), another with a level of actinides typical of merger dynamical ejecta ($s/k=10$) and lastly a case with low lead / actinide production ($s/k=32$). Only the case with low lead sees little influence from Tl-208 emission on the order of days while the other two cases show that their spectrum is significantly impacted by Tl-208 from $\sim 0.5-30$ days. Importantly, the late time signal from Tl-208 on the order of years is found to be above the sensitivity limit and robust, with no competition from other species, even in the case considered here which produces low lead. Therefore whether the 2.6 MeV emission line from Tl-208 is predicted to be visible on the order of days depends on many factors, but we consistently find the late time signal on the order of years to robustly shine through, even in cases with low production of third peak species.

\section{Discussion of the high energy ($>$ 3.5 MeV) regime}\label{sec:fgam}

In previous work \cite{Wang_fgam} reported that emission in the $\geq 3.5$ MeV region is dominated by prompt fission gammas in cases where actinide species fission at late times. Additionally, this work demonstrated that emission from fission overtakes delayed gamma emission from $\beta$-decays of neutron-rich fission daughters. Since our peak finder calculations were applied to the total spectrum deduced by considering $\beta$ flow, we reported lines from three cases (Sb-134, I-136, and In-128) emitting above $\sim3.5$ MeV which are such fission daughter products, with half-lives on the order of only seconds or milliseconds. Two of these cases report emission lines above $6$ MeV (Sb-134, I-136), with In-128 lines at 3.520 and 4.298 MeV. None of these delayed gammas from the $\beta$-decay of fission daughters are visible over the prompt fission gammas background. Overall across all tables in this work, we find 8 isotopes undergoing $\beta$-decay which emit above 3.5 MeV: Na-24 (3.866 MeV), Ga-72 (3.679 MeV), Rb-88 (4.036, 4.742 MeV), In-128 (3.520, 4.298 MeV), Sb-134 (6.451, 6.687, 6.820 MeV), I-136 (5.800, 6.104 MeV), La-142 (3.612, 3.633 MeV) and Tl-208 (3.708, 3.960 MeV). 

\begin{figure}
    \centering
    \includegraphics[width=0.9\linewidth]{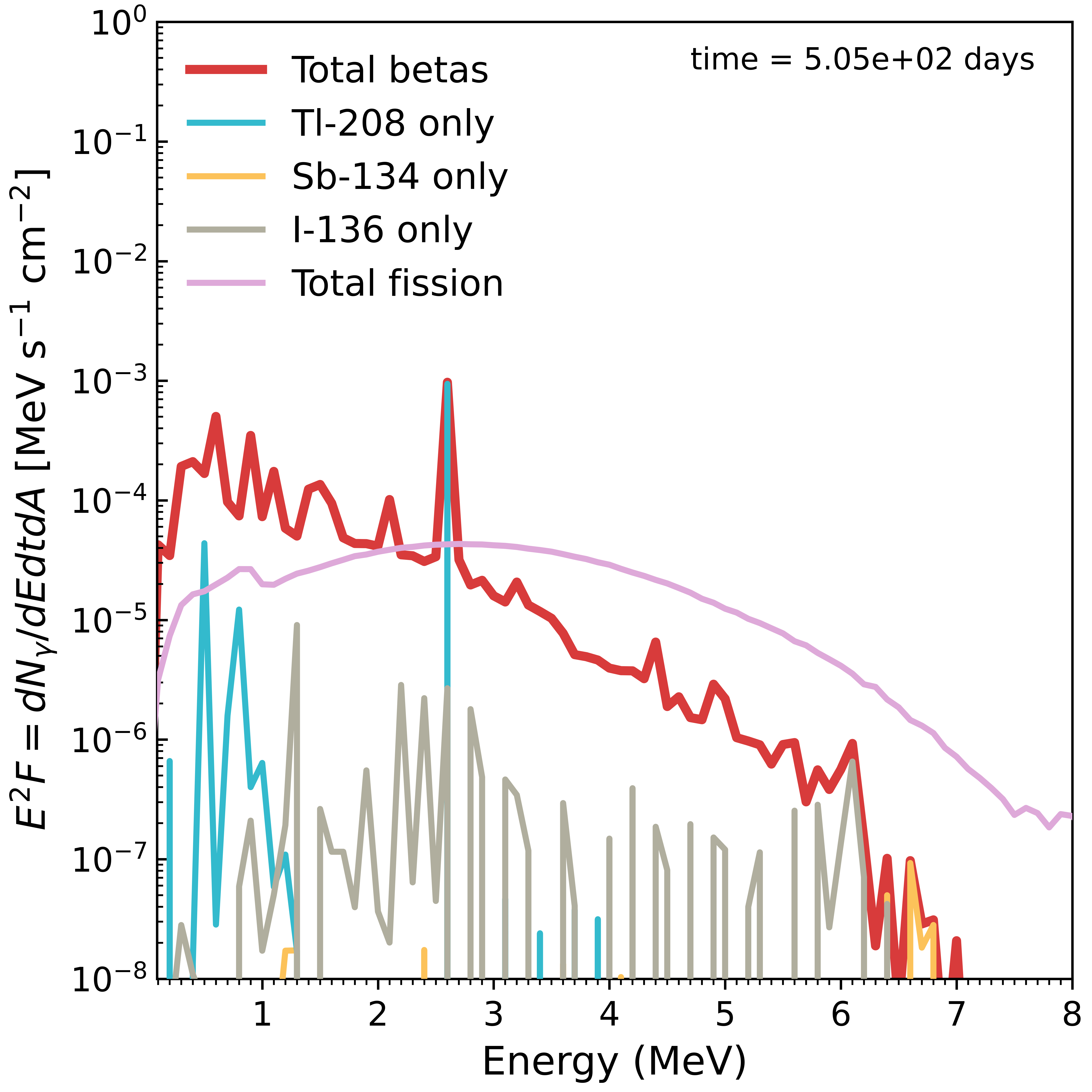}
    \caption{Total emission spectrum from 0.1 to 8 MeV compared to individual contributions from Tl-208 and prompt fission gamma emission at 500 days (assuming FRDM12 nuclear model and astrophysical condition $Y_e=0.01$, $s/k=10$). Note the dominance of fission gammas over the reported lines from Sb-134 and I-136 around 6 MeV from the $\beta$ decay of these isotopes (additionally the presence of these isotopes is ultimately due to late-time fission (eg. Cf-254)). }
    \label{fig:fgam_500days}
\end{figure}

\begin{figure*}
\centering
        \includegraphics[scale=0.5]{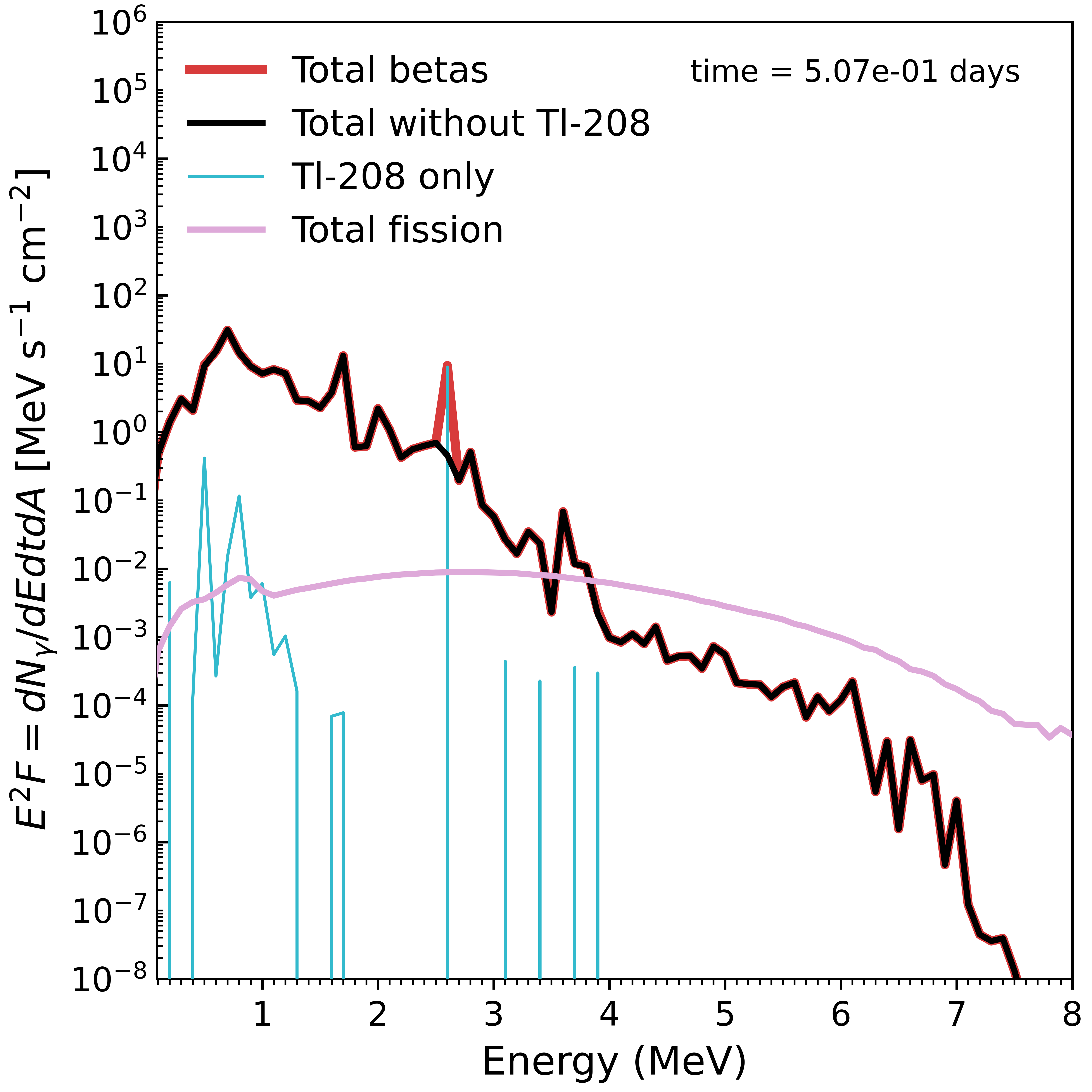}
        \hspace{0.25cm}
        \includegraphics[scale=0.5]{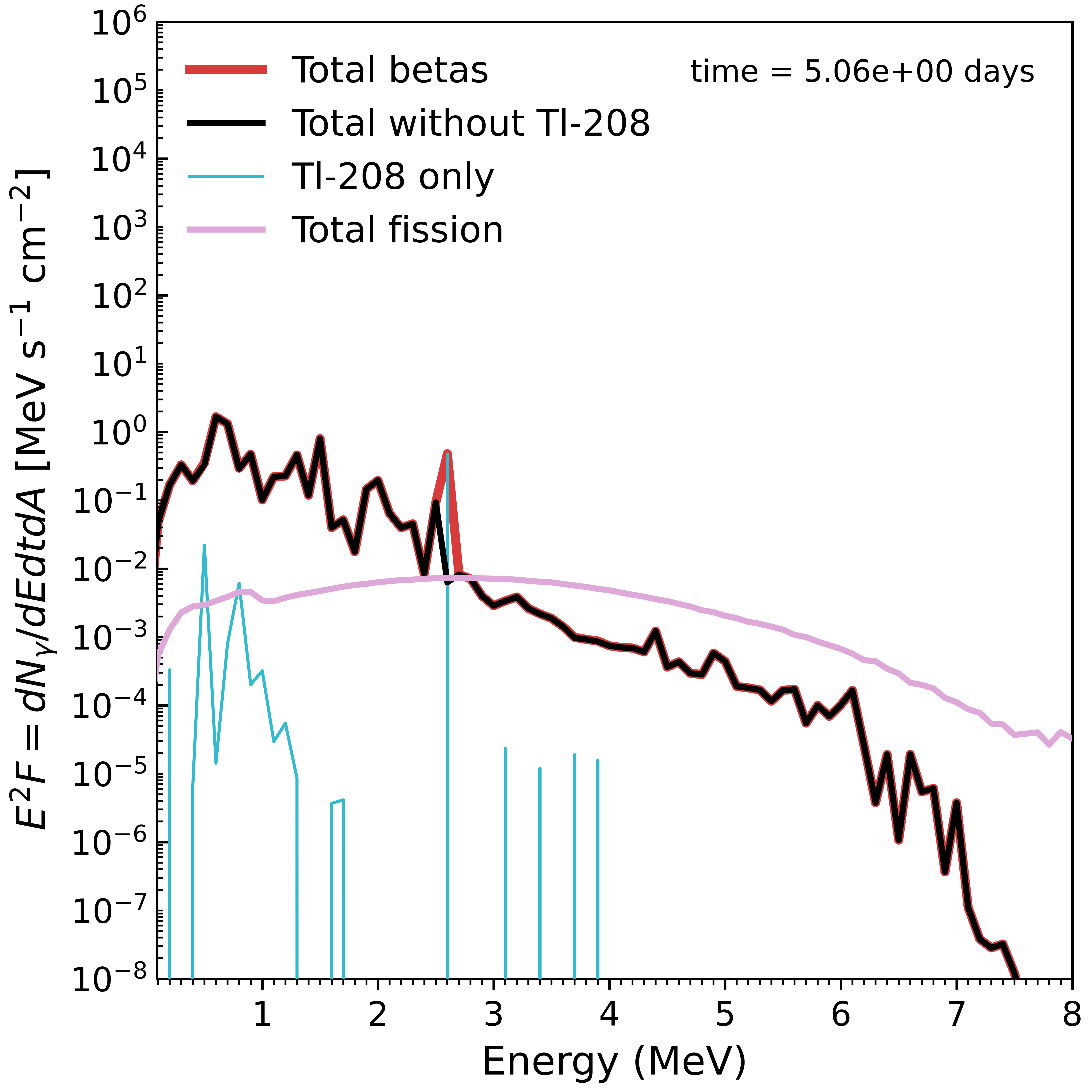}

        \includegraphics[scale=0.5]{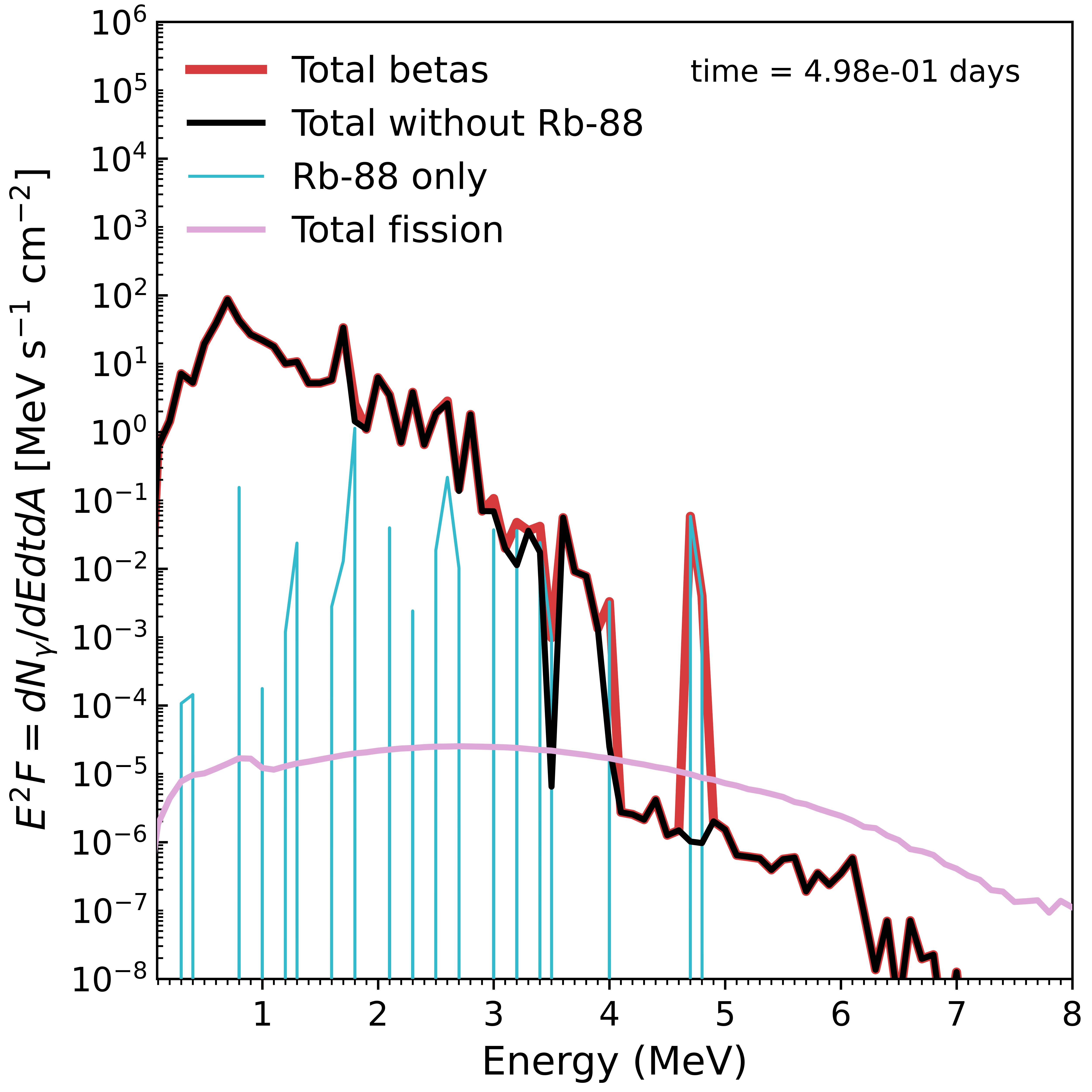}
        \hspace{0.25cm}
        \includegraphics[scale=0.5]{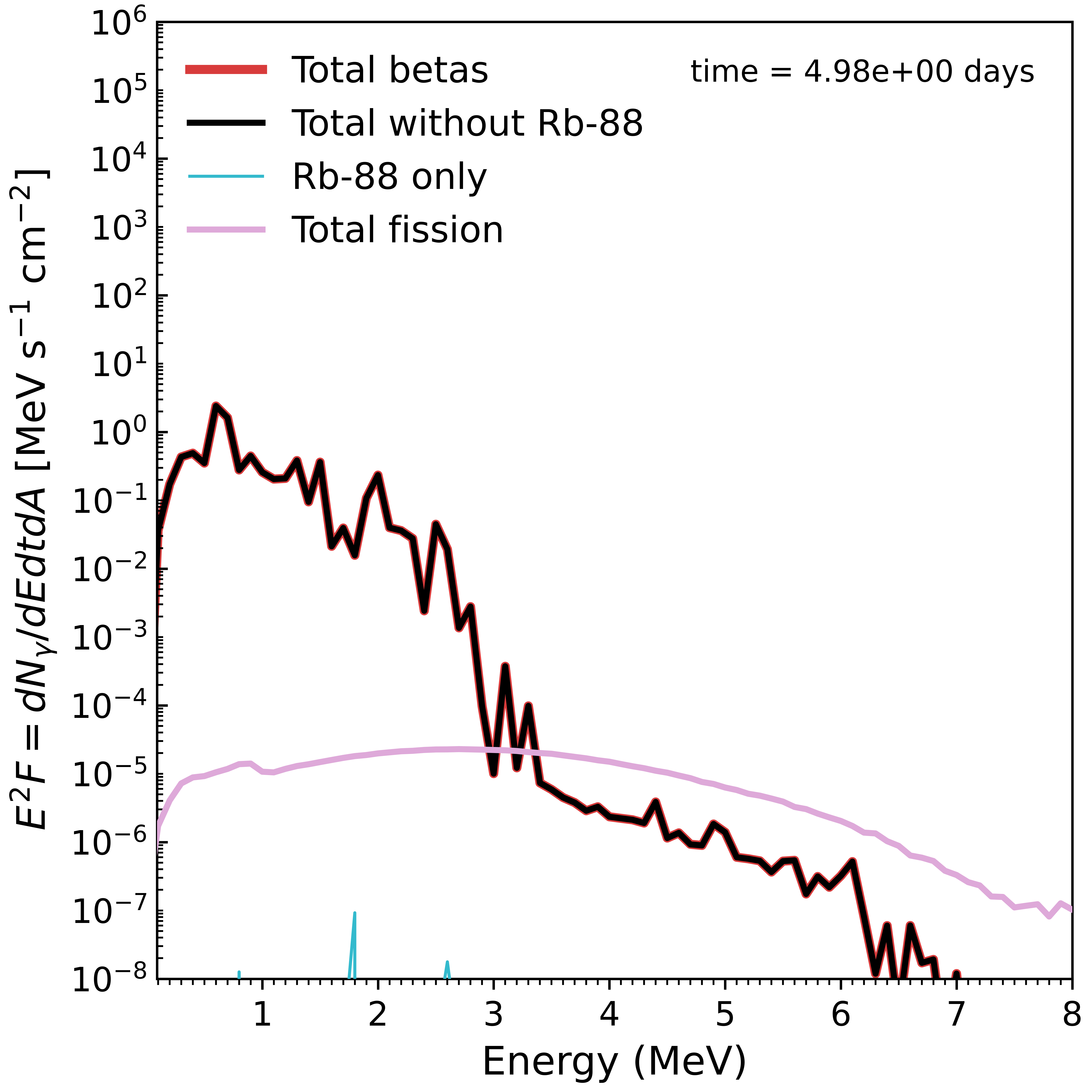}

        \includegraphics[scale=0.5]{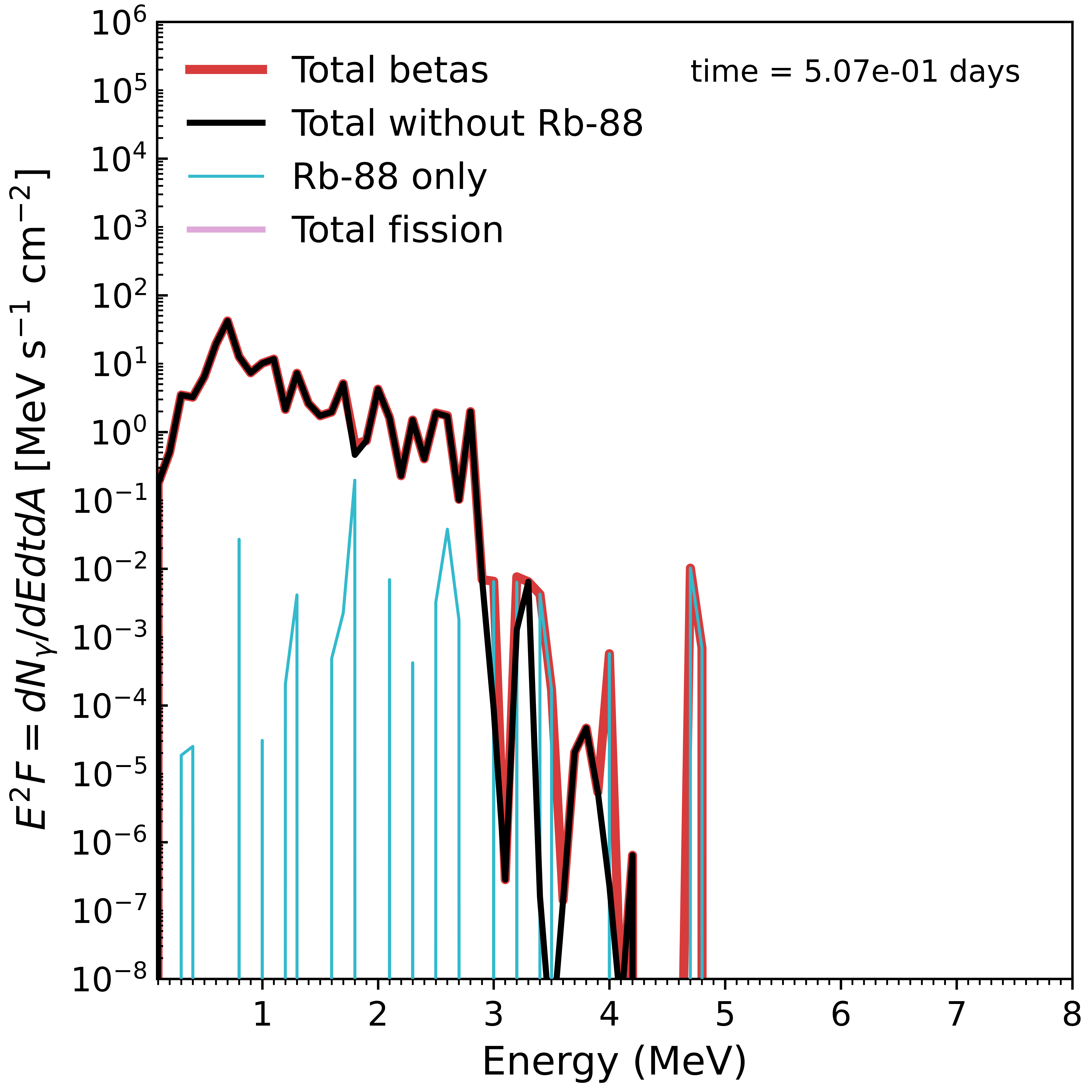}
        \hspace{0.25cm}
        \includegraphics[scale=0.5]{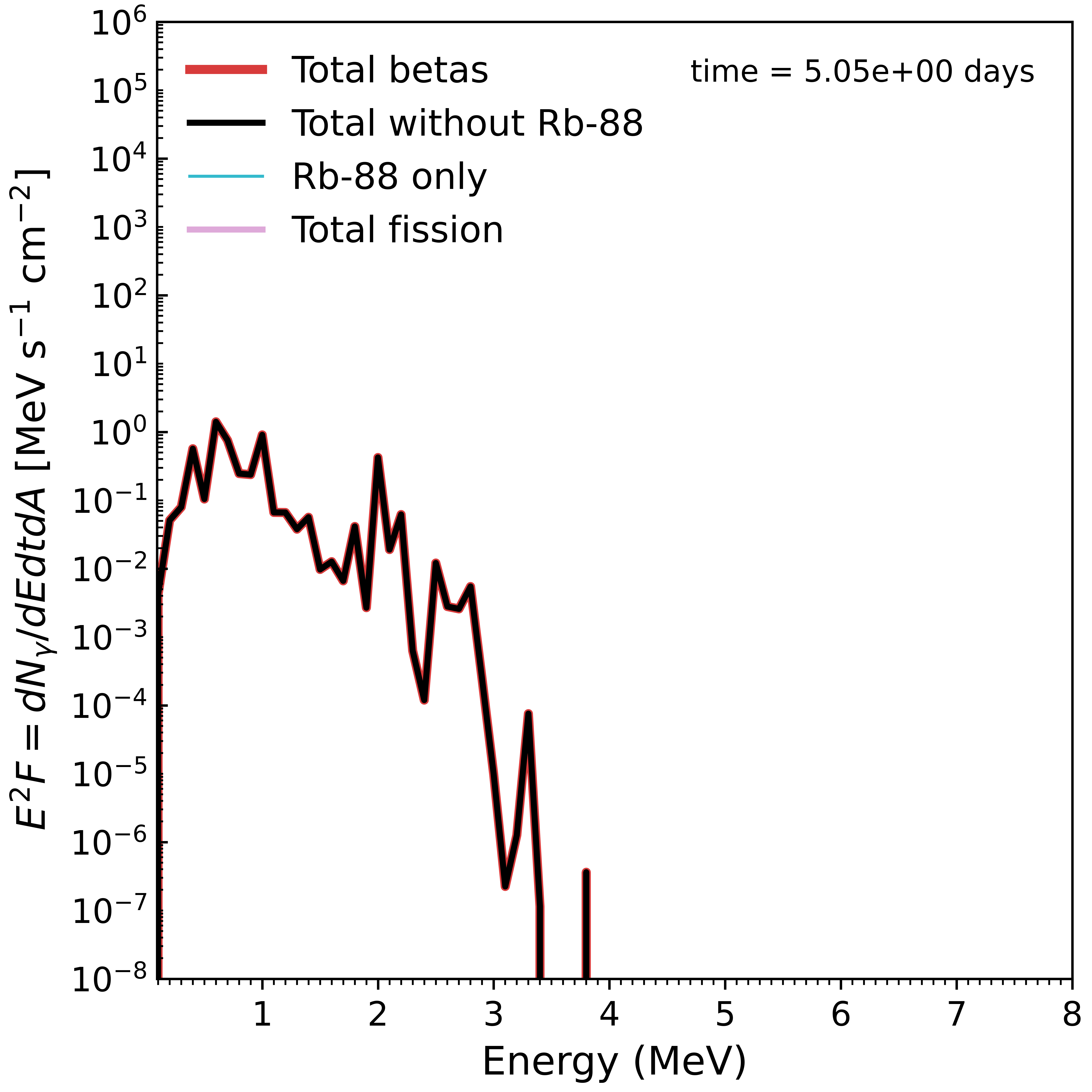}

    \caption{Total emission spectrum from 0.1 to 8 MeV at both 0.5 days (left) and 5 days (right) for three different astrophysical conditions with distinct reach in mass number: (top) a strongly fission cycling case with $Y_e=0.01$, $s/k=10$, (middle) the $Y_e=0.19$, $s/k=32$ case which produces both Rb-88 and fissioning species, and (bottom) a weak $r$-process case with $Y_e = 0.29$, $s/k = 75$ that never populates fissioning species. Comparisons with either individual contributions from Tl-208 (top) or Rb-88 (middle, bottom) are also shown. All examples here consider the FRDM2012 nuclear model.}
    \label{fig:fgamvstl3p5}
\end{figure*}

We start by discussing the three particularly neutron-rich species In-128, Sb-134, and I-136. All three have short half-lives (816 ms, 674 ms, and 83.4 s, respectively), and no long-lived $\beta$ or $\alpha$ feeders. They are, however, predicted to be fission yield products of long-lived fissioning species such as Cf-254. Fig.~\ref{fig:fgam_500days} shows the $\beta$ gamma spectrum at 500 days compared to the fission gammas spectrum. As demonstrated in the figure, any $\beta$ gamma emission above $\sim$3.5 MeV resulting from the In-128, Sb-134, and I-136 fission daughters will ultimately be overtaken by the prompt gammas from the fission itself. The dominance of Tl-208 and fission gammas at energies above 2.6 MeV around 500 days is evident. A comparison of the spectra at earlier times on the order of days is shown in the top panel of Fig.~\ref{fig:fgamvstl3p5}, where the same conclusion holds.

We next discuss the other 5 nuclei emitting above $3.5$ MeV (Na-24, Ga-72, Rb-88, La-142 and Tl-208), all of which have lines in the $\sim$3.5--4.7 MeV range. To first examine earlier emission at $\sim$ days, Fig.~\ref{fig:fgamvstl3p5} looks at the earliest time (0.5 days) for which we report a Tl-208 signal in the 2.6 MeV range as well as later at 5 days. The Tl-208 lines above $3.5$ MeV are not visible over the fission gammas spectrum at either time, nor are the lines above $3.5$ MeV from La-142. Emission above $3.5$ MeV is seen in weak $r$-process cases as well, but solely earlier on the order of days, mostly via Ga-72 and Rb-88. Note that Na-24 is only reported in very few weak $r$-process variations and disappears quickly. Regarding higher energy emission at later timescales on the order of hundreds of days to years, it is only the two Tl-208 lines at 3.708 and 3.960 MeV which are not due to fission products. Therefore, an astrophysical condition (e.g. 13 and 14) could potentially produce actinide species such as Ra-224 and Ra-228 which populate Tl-208 late, but then not significantly produce the long-lived fissioning species (such as Cf-254) responsible for prompt gamma emission emerging above 3.5 MeV. Note that these higher energy lines in the Tl-208 spectrum have a significantly lower intensity than its 2.6 MeV line however so are not likely to be detected above threshold at late times (see Fig.~\ref{fig:fgam_500days}). Importantly, we find that given conditions which produce a main $r$-process with actinides which fission at late times, the fission gamma emission will correspondingly dominate above $\sim$ 3.5 MeV. 

One particularly strong line above $3.5$ MeV is found in the Rb-88 (T$_{1/2}$ = 17.8 min) spectrum at 4.7 MeV. To compare this to the strength and signature of prompt fission gammas emission, we consider a case which produces both weak $r$-process nuclei alongside fissioning nuclei and a weak $r$-process condition that does not produce fissioning species. As can be seen in Fig.~\ref{fig:fgamvstl3p5}, at earlier times of $\sim$0.5 days, Rb-88 lines appear over the fission gammas background, observable as a clear emission line over the fission gammas continuum. However the 4.036 and 4.742 MeV Rb lines that shine through at early times fall below threshold at 2.4 days, leaving no trace above the fission gammas spectrum at 5 days (middle right panel of Fig.~\ref{fig:fgamvstl3p5}). Looking at the bottom of Fig.~\ref{fig:fgamvstl3p5} we can also see Rb-88 line emission above $3.5$ MeV in the weak $r$-process case but this time it is the only emitter in the energy range with no competing fission gammas spectrum. Even in the weak $r$-process case, however, Rb-88 emission does not survive past $\sim$2 days. Ga-72 is the other weak $r$-process species with a line $>$3.5 MeV predicted to be visible on the order of days in the mass weighted case (see Table~\ref{tab:massweight_days}) but like Rb-88 is not reported on longer timescales and thus does not compete at late times with prompt fission gammas.

\section{Conclusions}

In this work we report on nucleosynthesis network calculations that predict the total MeV emission spectrum from a neutron star merger event. We choose several distinct nuclear models to account for the impact of data uncertainties. We apply a peak finding algorithm across the spectrum in the 0.1 to 20 MeV range to identify when a particular emission line from $\beta$-decay dominates the light curve in the energy bins of relevance. We also require that emission be above threshold flux values inspired by future MeV missions such as COSI and MeVGRO. By applying this peak search algorithm across the spectrum and across time for all models, we identify features that can be uniquely tied back to dominant emitters and report the timescale on which emission is predicted to be visible. We additionally apply Monte-Carlo radiation transport calculations at two different ejecta velocities ($0.1c$ and $0.3c$) to assess whether the peaks we find remain distinguishable features in the spectrum. 

We first apply our algorithm to calculations for the mass weighted total ejecta from a neutron star merger. For remnant emission, we find potentially observable signals from only 9 species (Co-60, Rh-106, Sb-125, Sb-126, Ir-194, Tl-208, Tl-209, Pb-214, and Ac-228) with most cases visible on the order of tens to hundreds of years, with a few emitters lasting beyond 10,000 years (Co-60, Sb-126, and Tl-209). We then consider astrophysical variations in the form of parameterized trajectory variations for both weak and main $r$-process cases to account for both (1) natural site variations (e.g. progenitor mass) and (2) uncertainties in simulation inputs (e.g. equation of state, neutrino treatment) which make relying on a single mass weighted prediction insufficient to highlight all the isotopes that may ultimately be detectable from a nearby merger event. Exploring astrophysical variations reveals 7 additional isotopes of relevance to remnant emission (K-42, Kr-85, Eu-155, Ta-182, Pb-211, Pu-243, and Am-246) all of which show their signals to be highly dependent on the nuclear model and astrophysical conditions assumed.

In the investigation of weak $r$-process ejecta, we identify a particular isotope of interest---Rh-106---that emerges in differentiating weak $r$-process production from that of the heavier main $r$-process species. We find in weak $r$-process-only ejecta the total emission spectrum above 1 MeV from $\sim$ 0.2 to $\sim$17 years is identical to that of Rh-106 (identifiable by the numerous Rh-106 lines consistently reported in Table~\ref{tab:long_weak}). We note similar behavior from Ga-72, taking over the majority of the spectrum on the order of days when solely weak $r$-process species are produced. Such results exemplify the utility of our peak search approach since this behavior lasts over only a certain time window but is able to be explicitly identified here. Importantly, this behavior of Rh-106 is observed across nuclear model and astrophysical variations, indicating robust predictions for its impact on the spectrum. Therefore we find that Rh-106 could serve a unique role in illuminating the $r$-process nucleosynthetic reach in ejecta like that from mergers for future MeV telescope campaigns such as COSI.

Looking at signatures of ejecta producing the heaviest species (i.e. actinides such as Cf-254), \cite{Wang_fgam} found the higher energy ($>$3.5 MeV) portion of the spectrum to be dominated by prompt fission gamma emission. Here we explore this further by applying our peak finding algorithm across the MeV spectrum from $\beta$-decay, including the $>$3.5 MeV regime. We find a handful of isotopes emitting above 3.5 MeV, with most (In-128, Sb-134, and I-136) ultimately being fission products and not producing delayed emission visible over that from prompt fission gammas. The isotopes Na-24, Ga-72, Rb-88, La-142 and Tl-208 are exceptions, emitting above $3.5$ MeV while not being ultimately populated as a fission daughter. However lines from these nuclei are found to be potentially visible only at early times $\sim$ days, with the spectra above $\sim$3.5 MeV at later times either dominated by fission gammas or cleared out if fissioning species were not reached, supporting the findings in \cite{Wang_fgam}.

At lower energies, and lower mass numbers, we previously reported Tl-208 to be a real-time indicator of lead production due to its strong 2.6 MeV emission line \citep{VasshTl208}. Ultimately Tl-208 production is tied to Pb-212 ($\sim$12 hours), Ra-224 ($\sim$days), and Ra-228 ($\sim$years) so it too can be an indicator of not just lead but actinides. Here we report on the robustness of the Tl-208 2.6 MeV line given our nucleosynthesis calculation variations in nuclear model and astrophysical conditions. We use our peak finding algorithm to identify potential competing emitters and the timescale on which they might overtake Tl-208 emission in the 2.6 MeV bin. We find two species to be specifically of interest: Ga-72 and La-140. Still, in cases where the emission from these species overtakes the Tl-208 signal, we find that Tl-208 typically dominates early ($\sim$0.5 days) before being overtaken around $\sim$2 days (depending on nuclear model and astrophysical assumptions). We find that all cases for which the Tl-208 abundance was at least 1/10,000 that of La-140 saw Tl-208 dominate emission in the 2.6 MeV regime at some point on the order of days. Additionally, we find that emitters such as Ga-72 and La-140 have other strong lines at distinct energies which can be used to readily differentiate them from Tl-208 emission. Importantly we see no competing isotopes for Tl-208 2.6 MeV emission on the timescale of years, and find this to be present across all nuclear model and astrophysical variations. 

As several previous works by distinct research groups have considered MeV emission from nuclear decays in mergers, we note that outside of the emission from Tl-208 and prompt fission gammas, many of the emission lines reported here have also been mentioned in works by other groups (see Appendix~\ref{sec:litcomp}). However, it is important to note that we highlight species not seen in some studies, and likewise the literature highlights cases which we do not report in our tables (in some cases due to our implemented flux thresholds or our requirement that emission be dominant in the energy bin in question). Note that no study produces an identical set of isotopes to what was presented in works by other groups. This variation in isotopes reported across groups demonstrates the sensitivity of predictions to nuclear data assumptions and analysis approaches, and also helps to identify species which are robustly predicted across calculation assumptions (such as K-42, Fe-59, Cu-66, Cu-67, Zn-72, Ga-72, Ge-77, Kr-85, Kr-88, Zr-95, Nb-95, Ru-103, Rh-106, Ag-112, Sn-125, Sn-127, Sb-125, Sb-126, Sb-127, Sb-128, Sb-129, I-131, I-132, I-133, I-135, Xe-133, La-140, Ir-194, Pb-214, and Am-246). Such robustly predicted cases could be of particular interest to future MeV telescope COSI and next-generation observatories like MeVGRO, eAstrogam, GRAMS, etc.

Our study was built with MeV gamma-ray telescope campaigns in mind, with our peak search procedure allowing us to systematically explore how robustly all reported emitters are found to dominate the spectrum across several nuclear model and astrophysical variations. The nuclei that are most consistently predicted to participate across variations are Co-60, Ga-72, Rb-88, Nb-95, Ru-103, Rh-106, Sn-125, Sb-125, Sb-126, Sb-128, I-132, La-140, La-142, Eu-156, Ir-194, Tl-208, Tl-209, and Pb-214. Additionally, it is interesting to highlight cases that can be mostly tied to a given nuclear model only, such as emission from nuclei around A$\sim$180 with HFB27 (i.e., Hf-181, Ta-182, Ta-184, and Re-188) due to the enhancement in predicted abundances at this mass number in the HFB27 calculations relative to those using other models. Therefore the observation of signals from such isotopes would point to the need for nuclear structure features just prior to the $N=126$ shell closure to be more aligned with HFB27 predictions, or the lack of such signals could instead disfavor such models.

Regarding nuclear data considerations, we find that varying nuclear models changes which emitters we see shine through, how long a signal can last, as well as how isotopes compete in a given energy window. Thus pushing the boundaries of known data for $\beta$ decays, captures, and masses contributes directly to efforts aiming to tease out astrophysical signals from nuclear decay lines. Additionally we suggest that remeasurements of the emission spectra for species highlighted to be dominating emitters could also impact the predictions discussed in this work, given that sometimes relative intensities of lines can be refined upon reinvestigation. In Appendix~\ref{sec:appendpopmech} we note several species which are known to have half-lives of relevance to gamma-ray emission but the spectra of their $\beta$-decaying daughter is presently not reported to have a high intensity line which stands out in present calculations. Experiments at current facilities such as ARIEL at TRIUMF and FRIB are well poised to refine our present knowledge of line emission from such species.

Overall we demonstrate the sensitivity of predictions to details of both the nuclear model and astrophysical conditions. Our predictions would directly benefit from developments in hydrodynamic simulations and nuclear physics modeling. Furthermore, we highlight the important role that nuclear experiments can play by refining our knowledge of these important long-lived, emitting isotopes. We aspire to connect directly to MeV gamma-ray observations by seeking to identify meaningful features in the spectrum, implementing thresholds inspired by current detectors, and checking that the isotope emission highlighted is still above threshold as a peak in the spectrum after considering radiation transport. This highlights the concerted effort needed to drive predictions towards the precision needed to be of the best use to next-generation MeV telescope campaigns.

\section{acknowledgments}

M.L. and N.V. acknowledge the support of the Natural Sciences and Engineering Research Council of Canada (NSERC) via Early Career Grant SAPIN-2022-00022. The work of M.L. was also supported by the NSERC Canada Graduate Research Scholarship (CGRS) program. N.V. acknowledges support from the National Research Council (NRC) of Canada via their contribution agreement with TRIUMF. The work of Y.D. and X.W. is supported by the National Natural Science Foundation of China (Grant Nos. 12494570, 12494574, 12521005), the National Key R$\&$D Program of China (2021YFA0718500), and China’s Space Origins Exploration Program. R.S. is supported in part by the National Science Foundation Grant No. PHY-2020275 (Network for Neutrinos, Nuclear Astrophysics and Symmetries) as well as the U.S. Department of Energy under contract numbers DE-FG0295-ER40934 and LA22-ML-DEFOA-2440.

\bibliography{ref}
\bibliographystyle{aasjournalv7}

\appendix

\section{Populating long-lived $\beta$ emitters and comparison to long-lived $\alpha$-decay emission}\label{sec:appendpopmech}

Here we elaborate on how each of the emitting isotopes identified in the text is populated on observable timescales. Most cases are populated at later times due to their own long $\beta$-decay half-life or the long half-life of their $\beta$ feeder, as listed in Table~\ref{tab:BemittersBdec}.

\begin{figure}[!h]
    \centering
    \includegraphics[width=0.35\linewidth]{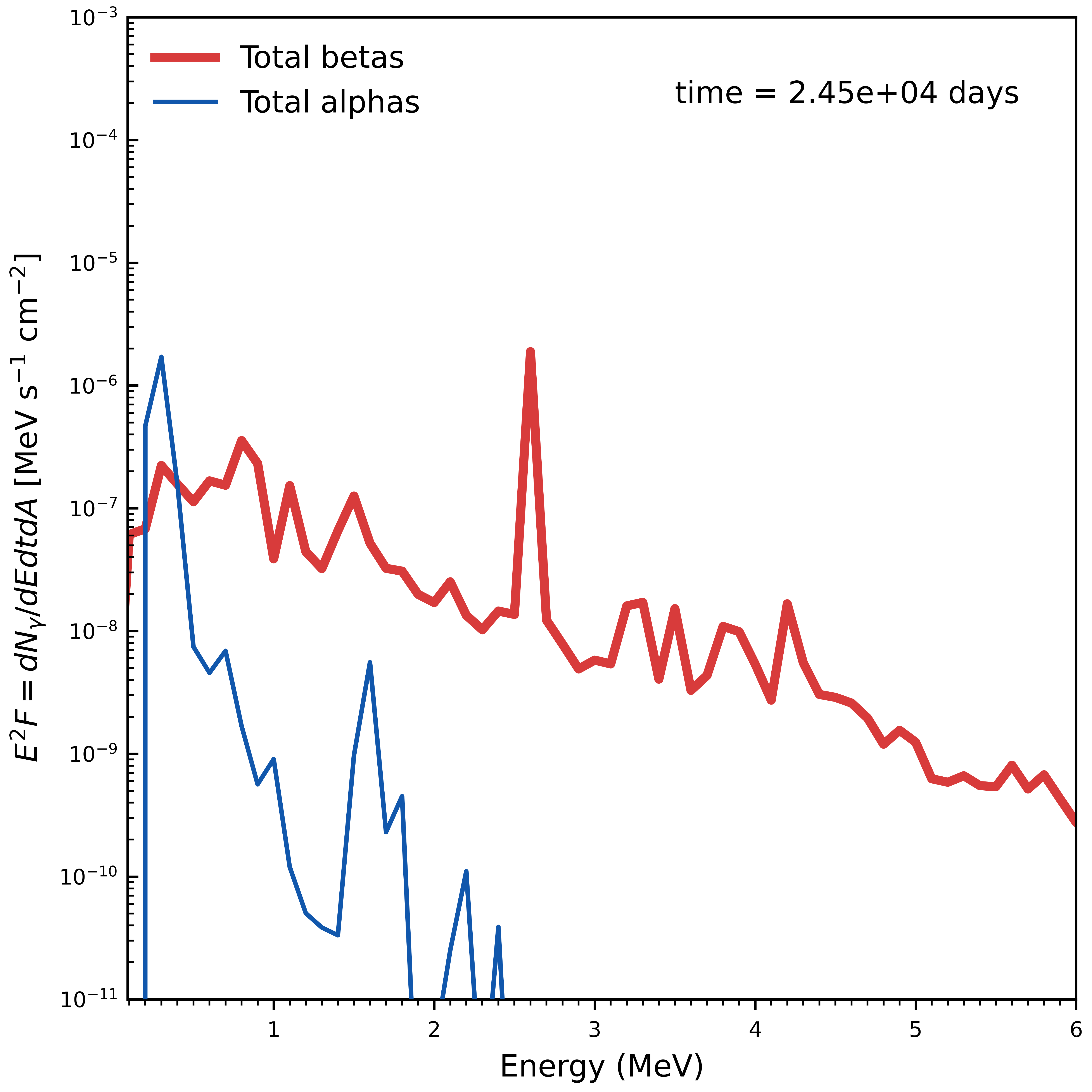}
    \includegraphics[width=0.35\linewidth]{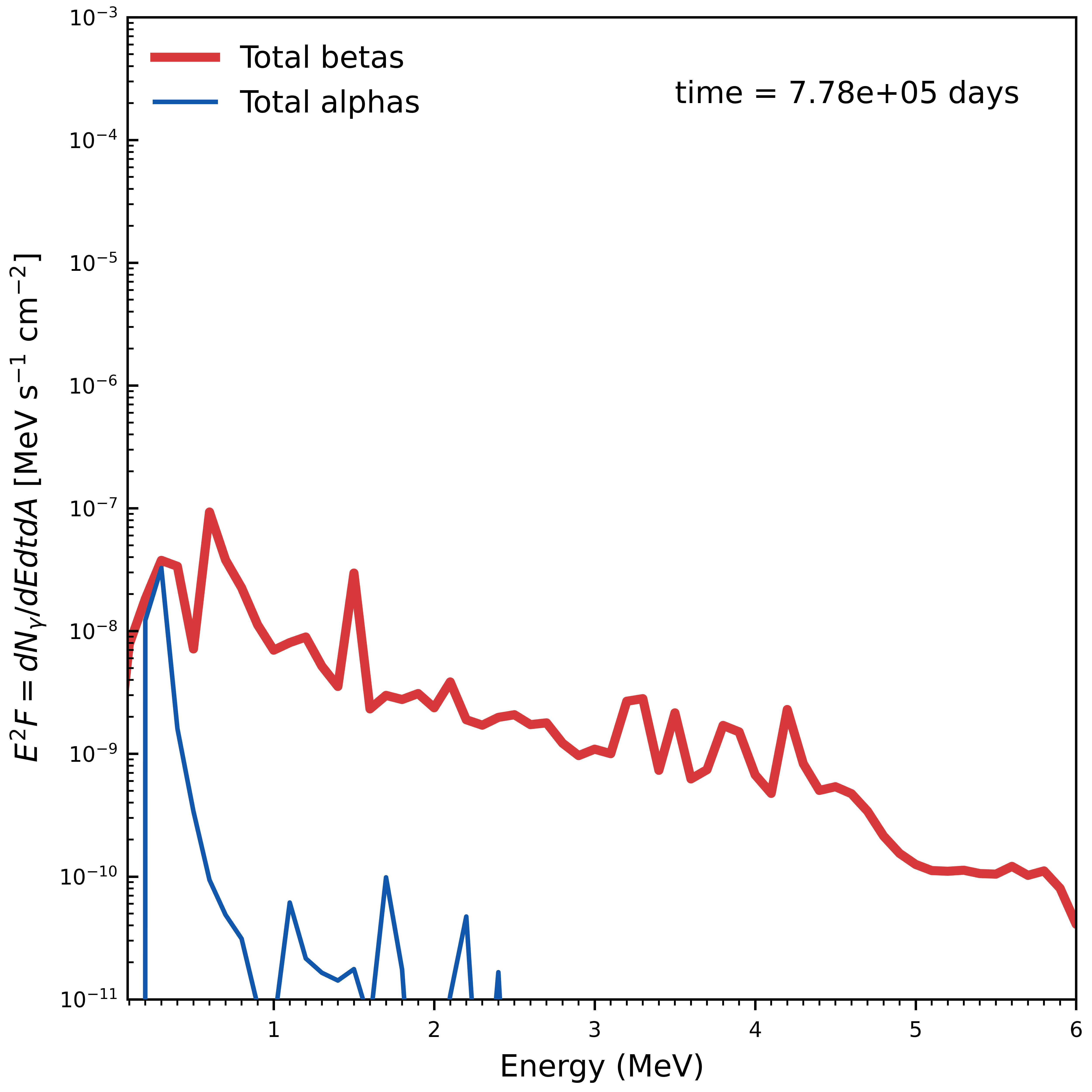}
    \caption{Comparison between emission from $\beta$-decay and $\alpha$-decay at two times ((left) 25000 days (68.5 years) and (right) 778,000 days (2131.5 years)) for the extremely neutron-rich, low entropy case ($Y_e=0.01$ and $s=10$) which heavily populates actinide nuclei. The left panel shows the time where $\alpha$-decay begins to dominate emission at low energies ($<0.5$ MeV). The right panel shows when $\alpha$-decay ceases to be the dominant channel and $\beta$-decay again overtakes $\alpha$-decay emission.}
    \label{fig:falpha}
\end{figure}

To consider nuclei whose late-time production relies on a long-lived $\alpha$-decay, we first compare $\alpha$-decay and $\beta$-decay emission in the $0.1-20$ MeV range considered in this work. We consider our most neutron-rich condition $Y_e=0.01$, $s/k=10$ which has high $\alpha$-decay flows. We show snapshots of the predicted spectrum comparing the emission from these two decay channels in Fig.~\ref{fig:falpha}. Even for this very neutron rich case which populates heavy nuclei strongly we find $\beta$-decay to dominate the emission spectrum until 68 years, where the $\alpha$-decay contribution begins to peak through at lower energies only ($<0.5$ MeV). After roughly 2131 years gamma emission from the $\beta$-decays of long-lived parents retake their dominance of the spectrum even at these lower energies.

We next highlight how long lived $\alpha$-decays still significantly impact our results by eventually producing a $\beta$-decaying species along their decay chain. It is via this mechanism that our calculations suggest $\alpha$-decays can contribute the most to strong, visible emission in the $0.1-20$ MeV range. In Table~\ref{tab:BemittersAFdec} we detail the decay chain which produces the dominant emitters reported in the tables of this work which are not explained by $\beta$-decay alone. 

\begin{longtable*}[]{|l|l|l|l|}
    \caption{Isotopes found to be dominant emitters of spectral lines identified by our peak finder that are populated by a long-lived $\beta$-decaying species.}\label{tab:BemittersBdec} \\
    \hline
    Z & Isotope & $T_{1/2}$ &  Decay chain responsible for observable emission \\
    \endfirsthead
    
    \multicolumn{4}{l}
    {{\bfseries \tablename\ \thetable{} -- continued from previous page}} \\
    \hline 
    Z & Isotope & $T_{1/2}$ &  Decay chain responsible for observable emission \\
    \hline
    \endhead

    \hline\endlastfoot
    \hline
    11 & Na-24  &   15 h     &   Self \\
    19 & K-42    &   12.4 h &    $^{42}$Ar ($\beta$, 32.9 y) \\
    26 & Fe-59  &    44.5 d  &  Self \\
    27 & Co-60  &    5.3 y  &    $^{60}$Fe ($\beta$, 2.6 My) \\
    29 & Cu-66  &   5.1 m  &    $^{66}$Ni ($\beta$, 54.6 h) \\
    29 & Cu-67  &   61.8 h  &   Self \\
    30 & Zn-72   &   46.5 h  &   Self \\
    31 & Ga-72   &  14 h    &    $^{72}$Zn ($\beta$, 46.5 h) \\
    32 & Ge-77   &  11.2 h   &   Self \\
    36 & Kr-85   &   10.7 y   &   Self \\
    36 & Kr-88   &    2.8 h   &    Self \\
    37 & Rb-88  &   17.8 m  &   $^{88}$Kr ($\beta$, 2.8 h) \\
    39 & Y-91    &    58.5 d  &   Self \\
    40 & Zr-95   &    64 d    &    Self \\
    41 & Nb-95   &   35 d    &    $^{95}$Zr ($\beta$, 64.0 d) \\
    44 & Ru-103  &  39.2 d  &   Self \\
    45 & Rh-106 &   30.1 s  &   $^{106}$Ru ($\beta$, 371.8 d) \\
    47 & Ag-112  &   3.1 h  &    $^{112}$Pd ($\beta$, 21 h) \\
    50 & Sn-125  &   9.6 d &     Self \\
    50 & Sn-127  &   2.1 h &     Self \\
    51 & Sb-125  &   2.8 y  &     Self \\
    51 & Sb-126  &   12.4 d  &   $^{126}$Sn ($\beta$, 230 ky) \\
    51 & Sb-127  &   3.9 d   &    Self \\
    51 & Sb-128  &   9.1 h    &   Self \\
    51 & Sb-129  &   4.4 h    &   Self \\
    53 & I-131     &    8 d    &     Self \\
    53 & I-132     &    2.3 h   &   $^{132}$Te ($\beta$, 3.2 d) \\
    53 & I-133     &    20.8 h  &  Self \\
    53 & I-135     &    6.6 h   &   Self \\
    54 & Xe-133  &    5.2 d    &   Self \\
    57 & La-140   &    40.3 h &   $^{140}$Ba ($\beta$, 12.8 d) \\
    57 & La-142   &    91.1 m &   Self \\
    59 & Pr-144    &    17.3 m &  $^{144}$Ce ($\beta$, 284.9 d) \\
    63 & Eu-155   &    4.7 y    &   Self \\
    63 & Eu-156   &    15.2 d   &  Self \\
    72 & Hf-181    &    42.4 d  &   Self \\
    73 & Ta-182   &     14.7 d  &  $^{182}$Hf ($\beta$, 8.9 My) \\
    73 & Ta-184   &     8.7 h    &  $^{184}$Hf ($\beta$, 4.1 h) \\
    75 & Re-188   &    17 h    &   $^{188}$W ($\beta$, 69.8 d) \\
    77 & Ir-194      &   19.4 h  &   $^{194}$Os ($\beta$, 6.0 y) \\
    89 & Ac-228    &   6.2 h &  $^{228}$Ra ($\beta$, 5.75 y) \\
    
\end{longtable*}

\begin{longtable*}[]{|l|l|l|l|}
    \caption{Isotopes found to be dominant emitters of spectral lines identified by our peak finder that are populated by long-lived $\alpha$-decays or fission.}\label{tab:BemittersAFdec} \\
    \hline
    Z & Isotope & $T_{1/2}$ &  Decay chain responsible for observable emission \\
    \endfirsthead
    
    \multicolumn{4}{l}
    {{\bfseries \tablename\ \thetable{} -- continued from previous page}} \\
    \hline 
    Z & Isotope & $T_{1/2}$ &  Decay chain responsible for observable emission \\
    \hline
    \endhead

    \hline\endlastfoot
    \hline
    49 & In-128  &   816 ms &  fission product \\
     51 & Sb-134  &   674 ms &  fission product \\
    53 & I-136     &    83.4 s   &  fission product \\
    81 & Tl-208     &   3.05 m  &  $^{224}$Ra ($\alpha$, 3.6 d) $\rightarrow$ $^{220}$Rn ($\alpha$, 55.6 s) $\rightarrow$ $^{216}$Po ($\alpha$, 144.0 ms) $\rightarrow$ $^{212}$Pb ($\beta$, 10.6 h) $\rightarrow$ $^{212}$Bi ($\alpha$, 60.6 m)  \\
          &            &           &          $^{228}$Ra ($\beta$, 5.8 y) $\rightarrow$ $^{228}$Ac ($\beta$, 6.2 h) $\rightarrow$ $^{228}$Th ($\alpha$, 1.9 y) $\rightarrow$ $^{224}$Ra ($\alpha$, 3.6 d) …(cont. as above) \\    
    81 & Tl-209     &   2.2 m   &    $^{233}$U ($\alpha$, 159.2 ky) $\rightarrow$ $^{229}$Th ($\alpha$, 7.9 ky) $\rightarrow$ $^{225}$Ra ($\beta$, 14.8 d) $\rightarrow$ $^{225}$Ac ($\alpha$, 9.9 d) $\rightarrow$ (cont. below) \\
               &       &       &                  $^{221}$Fr ($\alpha$, 4.8 m) $\rightarrow$ $^{217}$At ($\alpha$, 32.6 ms) $\rightarrow$ $^{213}$Bi ($\alpha$, 45.6 m) \\
    82 & Pb-211   &    36.2 m  &  $^{227}$Ac ($\beta$, 21.8 y) $\rightarrow$ $^{227}$Th ($\alpha$, 18.7 d) $\rightarrow$ $^{223}$Ra ($\alpha$, 11.4 d), $^{219}$Rn ($\alpha$,4 s) $\rightarrow$ $^{215}$Po ($\alpha$, 1.8 ms)     \\
    82 & Pb-214   &    27.1 m  &  $^{226}$Ra ($\alpha$, 1.6 ky) $\rightarrow$ $^{222}$Rn ($\alpha$, 3.8 d) $\rightarrow$ $^{218}$Po ($\alpha$, 3.1 min)    \\
    94 & Pu-243    &   5 h  &   $^{247}$Cm ($\alpha$, 15.6 My) \\
    95 & Am-246   &   39 m &  $^{246}$Pu ($\beta$, 10.8 d) + theoretical ($\beta$,$\alpha$) ($\sim$10-100 y)  \\
    
\end{longtable*}

Next we bring forward cases known to decay on timescales of relevance for astrophysical observables but were found here to not have special features in their currently reported emission spectrum which shine through and dominate over competing emitters in our calculations. We report these cases since remeasurements of decay spectra can refine the reported intensity of a given line, and in cases where spectra are not presently reported new lines of interest could be identified. In Table~\ref{tab:longBdec} we highlight experimentally established $\beta$-decays with half-lives on the order of days or longer, note whether these cases were reported in our tables, and state whether these species were populated with a high abundance ($>10^{-7}$) at some point during any of the calculations for the neutron star merger scenario presented in Sec. \ref{sec:nsmw3models} (that is we consider the abundance range from all 6 nuclear data variations). Thus both the lists of isotopes outlined in Table~\ref{tab:BemittersBdec} found by our peak searching algorithm and those given in Table~\ref{tab:longBdec} (specifically the cases marked as populated during our calculation) would be good candidates for gamma spectra remeasurements. We also examine all $\alpha$-decays with half-lives longer than 1 day in in Table~\ref{tab:longAdec}, report which have a daughter product which $\beta$-decays, note whether this daughter was already featured in the tables in the main text, and note whether our nucleosynthesis calculations in Sec. \ref{sec:nsmw3models} populated this $\alpha$-decaying parent (and if so during what time interval).

\begin{longtable*}[]{|l|l|l|l|l|}
    \caption{All isotopes with measured $\beta$-decay half-lives longer than 1 day (assuming NUBASE2020 decay data).}\label{tab:longBdec} \\
    \hline
    ($Z$, $A$) & $T_{1/2}$ [days] & Branchings ($\beta$, $\alpha$, SF) & ($Z$, $A$) in Tables? & Populated w/ abund. $>10^{-7}$? \\
    & & & & (When? [days]) \\
    \endfirsthead
    
    \multicolumn{5}{l}
    {{\bfseries \tablename\ \thetable{} -- continued from previous page}} \\
    \hline 
    ($Z$, $A$) & $T_{1/2}$ [days]  & Branchings ($\beta$, $\alpha$, SF) & ($Z$, $A$) in Tables? & Populated w/ abund. $>10^{-7}$? \\
    & & & & (When? [days]) \\
    \hline
    \endhead

    \hline\endlastfoot
    \hline
(1, 3) & 4.50e+03 & (100, 0, 0) & N &  N \\
(4, 10) & 5.06e+08 & (100, 0, 0) & N &  N \\
(6, 14) & 2.08e+06 & (100, 0, 0) & N &  N \\
(14, 32) & 5.73e+04 & (100, 0, 0) & N &  N \\
(15, 32) & 1.43e+01 & (100, 0, 0) & N &  N \\
(15, 33) & 2.54e+01 & (100, 0, 0) & N &  N \\
(16, 35) & 8.74e+01 & (100, 0, 0) & N &  N \\
(17, 36) & 1.10e+08 & (98.1, 0, 0) & N &  N \\
(18, 39) & 9.78e+04 & (100, 0, 0) & N &  N \\
(18, 42) & 1.20e+04 & (100, 0, 0) & N &  N \\
(19, 40) & 4.56e+11 & (89.28, 0, 0) & N &  N \\
(20, 45) & 1.63e+02 & (100, 0, 0) & N &  N \\
(20, 47) & 4.54 & (100, 0, 0) & N &  N \\
(21, 46) & 8.38e+01 & (100, 0, 0) & N &  N \\
(21, 47) & 3.35 & (100, 0, 0) & N &  N \\
(21, 48) & 1.82 & (100, 0, 0) & N &  N \\
(25, 54) & 3.12e+02 & (0.93e-4, 0, 0) & N &  N \\
(26, 59) & 4.45e+01 & (100, 0, 0) & Y &  Y (2.31e-04 $-$ 2.80e+02) \\
(26, 60) & 9.56e+08 & (100, 0, 0) & N &  Y (2.31e-04 $-$ 8.18e+09) \\
(27, 60) & 1.92e+03 & (100, 0, 0) & Y &  N \\
(28, 63) & 3.69e+04 & (100, 0, 0) & N &  Y (2.31e-04 $-$ 3.45e+05) \\
(28, 66) & 2.27 & (100, 0, 0) & N &  Y (2.31e-04 $-$ 2.14e+01) \\
(29, 67) & 2.58 & (100, 0, 0) & Y &  Y (2.31e-04 $-$ 2.43e+01) \\
(30, 72) & 1.94 & (100, 0, 0) & Y &  Y (2.31e-04 $-$ 1.87e+01) \\
(33, 74) & 1.78e+01 & (34, 0, 0) & N &  N \\
(33, 76) & 1.09 & (100, 0, 0) & N &  Y (1.50e-02 $-$ 2.10) \\ 
(33, 77) & 1.62 & (100, 0, 0) & N & Y (4.81e-04 $-$ 1.96e+01) \\
(34, 79) & 1.19e+08 & (100, 0, 0) & N &  Y (2.31e-04 $-$ 1.27e+09) \\
(35, 82) & 1.47 & (100, 0, 0) & N &  Y (1.32e-03 $-$ 6.62) \\
(36, 85) & 3.92e+03 & (100, 0, 0) & Y &  Y (2.31e-04 $-$ 3.70e+04) \\
(37, 84) & 3.28e+01 & (3.9, 0, 0) & N &  N \\
(37, 86) & 1.86e+01 & (100, 0, 0) & N &  N \\
(37, 87) & 1.81e+13 & (100, 0, 0) & N &  Y (2.31e-04 $-$ 3.65e+11) \\
(38, 89) & 5.06e+01 & (100, 0, 0) & N &  Y (5.66e-04 $-$ 4.20e+02) \\
(38, 90) & 1.06e+04 & (100, 0, 0) & N &  Y (2.31e-04 $-$ 9.58e+04) \\
(39, 90) & 2.67 & (100, 0, 0) & N &  N \\
(39, 91) & 5.85e+01 & (100, 0, 0) & Y &   Y (2.78e-03 $-$ 5.06e+02) \\
(40, 93) & 5.88e+08 & (100, 0, 0) & N &  Y (5.17e-03 $-$ 5.51e+09) \\
(40, 95) & 6.40e+01 & (100, 0, 0) & Y &  Y (2.31e-04 $-$ 5.80e+02) \\
(41, 94) & 7.45e+06 & (100, 0, 0) & N &  N \\
(41, 95) & 3.50e+01 & (100, 0, 0) & Y &  Y (2.42e-01 $-$ 5.96e+02) \\
(42, 99) & 2.75 & (100, 0, 0) & N &  Y (2.31e-04 $-$ 2.10e+01) \\
(43, 98) & 1.53e+09 & (100, 0, 0) & N &  N  \\
(43, 99) & 7.71e+07 & (100, 0, 0) & N &  Y (2.33e-02 $-$ 6.18e+08) \\
(44, 103) & 3.92e+01 & (100, 0, 0) & Y &  Y (2.31e-04 $-$ 3.48e+02) \\
(44, 106) & 3.72e+02 & (100, 0, 0) & N &  Y (2.31e-04 $-$ 3.31e+03) \\
(45, 102) & 2.07e+02 & (22, 0, 0) & N &  N \\
(45, 105) & 1.47 & (100, 0, 0) & N &  Y (4.15e-03 $-$ 1.33e+01) \\
(46, 107) & 2.37e+09 & (100, 0, 0) & N &  Y (1.09e-03 $-$ 2.06e+10) \\
(47, 111) & 7.43 & (100, 0, 0) & N &  Y (2.69e-04 $-$ 6.27e+01) \\
(48, 113) & 2.93e+18 & (100, 0, 0) & N &  Y (2.88e-03 $-$ 3.65e+11) \\
(48, 115) & 2.23 & (100, 0, 0) & N &  Y (3.44e-04 $-$ 1.74e+01) \\
(49, 115) & 1.61e+17 & (100, 0, 0) & N &  Y (3.39e-02 $-$ 3.65e+11) \\
(50, 121) & 1.13 & (100, 0, 0) & N &  Y (2.31e-04 $-$ 9.38) \\
(50, 123) & 1.29e+02 & (100, 0, 0) & N &  Y (2.31e-04 $-$ 1.60e+03) \\
(50, 125) & 9.63 & (100, 0, 0) & Y &  Y (2.31e-04 $-$ 1.02e+02) \\
(50, 126) & 8.40e+07 & (100, 0, 0) & N &   Y (2.31e-04 $-$ 9.99e+08) \\
(51, 122) & 2.72 & (97.59, 0, 0) & N &  N \\
(51, 124) & 6.02e+01 & (100, 0, 0) & N &  N \\
(51, 125) & 1.01e+03 & (100, 0, 0) & Y &  Y (1.12e-02 $-$ 1.06e+04) \\
(51, 126) & 1.23e+01 & (100, 0, 0) & Y &  N \\
(51, 127) & 3.85 & (100, 0, 0) & Y &  Y (2.31e-04 $-$ 4.70e+01) \\ 
(52, 132) & 3.20 & (100, 0, 0) & N &  Y (2.31e-04 $-$ 4.29e+01) \\
(53, 126) & 1.29e+01 & (47.3, 0, 0) & N &  N \\
(53, 129) & 5.89e+09 & (100, 0, 0) & N &  Y (4.56e-03 $-$ 7.41e+10) \\
(53, 131) & 8.02 & (100, 0, 0) & Y &  Y (9.00e-04 $-$ 1.29e+02) \\
(54, 133) & 5.25 & (100, 0, 0) & Y &  Y (4.59e-03 $-$ 7.31e+01) \\ 
(55, 132) & 6.48 & (1.87, 0, 0) & N &  N \\
(55, 134) & 7.54e+02 & (100, 0, 0) & N &  N \\
(55, 135) & 4.86e+08 & (100, 0, 0) & N &  Y (2.33e-02 $-$ 5.16e+09) \\
(55, 136) & 1.30e+01 & (100, 0, 0) & N &  N \\
(55, 137) & 1.10e+04 & (100, 0, 0) & N &  Y (2.31e-04 $-$ 1.18e+05) \\
(56, 140) & 1.28e+01 & (100, 0, 0) & N &  Y (2.31e-04 $-$ 1.36e+02) \\
(57, 138) & 3.76e+13 & (34.5, 0, 0) & N & N \\
(57, 140) & 1.68 & (100, 0, 0) & Y &  Y (1.83e-02 $-$ 9.68e+01) \\
(58, 141) & 3.25e+01 & (100, 0, 0) & N &  Y (5.60e-03 $-$ 3.07e+02) \\
(58, 143) & 1.38 & (100, 0, 0) & N &  Y (2.83e-04 $-$ 1.07e+01) \\
(58, 144) & 2.85e+02 & (100, 0, 0) & N &  Y (2.31e-04 $-$ 2.55e+03) \\
(59, 143) & 1.36e+01 & (100, 0, 0) & N &  Y (2.36e-02 $-$ 1.12e+02) \\
(60, 147) & 1.10e+01 & (100, 0, 0) & N &  Y (4.57e-04 $-$ 8.96e+01) \\
(61, 146) & 2.02e+03 & (34.0, 0, 0) & N &  N \\
(61, 147) & 9.58e+02 & (100, 0, 0) & N &  Y (9.01e-02 $-$ 7.68e+03) \\
(61, 148) & 5.37 & (100, 0, 0) & N &  N \\
(61, 149) & 2.21 & (100, 0, 0) & N &  Y (1.82e-03 $-$ 1.78e+01) \\
(61, 151) & 1.18 & (100, 0, 0) & N &  Y (2.46e-04 $-$ 9.66) \\
(62, 151) & 3.45e+04 & (100, 0, 0) & N &  Y (1.64e-02 $-$ 2.87e+05) \\
(62, 153) & 1.93 & (100, 0, 0) & N &  Y (2.83e-04 $-$ 1.48e+01) \\
(63, 152) & 4.93e+03 & (27.92, 0, 0) & N &  N \\
(63, 154) & 3.14e+03 & (99.982, 0, 0) & N &  N \\
(63, 155) & 1.73e+03 & (100, 0, 0) & Y &  Y (6.76e-04 $-$ 1.31e+04) \\
(63, 156) & 1.52e+01 & (100, 0, 0) & Y &  Y (2.02e-03 $-$ 1.31e+02) \\
(65, 158) & 6.57e+04 & (16.6, 0, 0) & N &  N \\
(65, 160) & 7.23e+01 & (100, 0, 0) & N &  N \\
(65, 161) & 6.95 & (100, 0, 0) & N &  Y (2.31e-04 $-$ 5.47e+01) \\
(66, 166) & 3.40 & (100, 0, 0) & N &  Y (2.31e-04 $-$ 2.99e+01) \\
(67, 166) & 1.12 & (100, 0, 0) & N &  Y (7.85e-03 $-$ 2.64e+01) \\
(68, 169) & 9.39 & (100, 0, 0) & N &  Y (2.46e-04 $-$ 8.00e+01) \\
(68, 172) & 2.05 & (100, 0, 0) & N &  Y (2.31e-04 $-$ 1.78e+01) \\
(69, 168) & 9.31e+01 & (0.010, 0, 0) & N &  N \\
(69, 170) & 1.29e+02 & (99.869, 0, 0) & N &  N \\
(69, 171) & 7.01e+02 & (100, 0, 0) & N &  Y (3.20e-03 $-$ 5.61e+03) \\
(69, 172) & 2.65 & (100, 0, 0) & N &  Y (8.68e-03 $-$ 2.87e+01) \\
(70, 175) & 4.18 & (100, 0, 0) & N &  Y (4.00e-04 $-$ 3.72e+01) \\
(71, 176) & 1.35e+13 & (100, 0, 0) & N &  N \\
(71, 177) & 6.64 & (100, 0, 0) & N &  Y (1.40e-03 $-$ 5.99e+01) \\
(72, 181) & 4.24e+01 & (100, 0, 0) & Y &  Y (3.20e-04 $-$ 3.38e+02) \\
(72, 182) & 3.25e+09 & (100, 0, 0) & N &  Y (2.31e-04 $-$ 2.61e+10) \\
(73, 182) & 1.15e+02 & (100, 0, 0) & Y &  N \\
(73, 183) & 5.10 & (100, 0, 0) & N &  Y (1.21e-03 $-$ 3.99e+01) \\
(74, 185) & 7.51e+01 & (100, 0, 0) & N &  Y (1.75e-03 $-$ 6.12e+02) \\
(74, 188) & 6.98e+01 & (100, 0, 0) & N &  Y (2.31e-04 $-$ 6.77e+02) \\
(75, 186) & 3.72 & (92.53, 0, 0) & N &  N \\
(75, 187) & 1.52e+13 & (100, 0, 0) & N &  Y (5.48e-03 $-$ 3.65e+11) \\
(75, 189) & 1.01 & (100, 0, 0) & N &  Y (3.45e-04 $-$ 1.03e+01) \\
(76, 191) & 1.50e+01 & (100, 0, 0) & N &  Y (2.31e-04 $-$ 1.58e+02) \\
(76, 193) & 1.24 & (100, 0, 0) & N &  Y (2.31e-04 $-$ 1.38e+01) \\
(76, 194) & 2.19e+03 & (100, 0, 0) & N &  Y (2.31e-04 $-$ 2.54e+04) \\
(77, 192) & 7.38e+01 & (95.24, 0, 0) & N &  N \\
(78, 202) & 1.83 & (100, 0, 0) & N &  Y (2.31e-04 $-$ 1.01e+01) \\
(79, 196) & 6.17 & (7.0, 0, 0) & N &  N \\
(79, 198) & 2.69 & (100, 0, 0) & N &  N \\
(79, 199) & 3.14 & (100, 0, 0) & N &  Y (5.28e-04 $-$ 2.33e+01) \\
(80, 203) & 4.66e+01 & (100, 0, 0) & N &  Y (2.83e-04 $-$ 2.28e+02) \\
(81, 204) & 1.38e+03 & (97.08, 0, 0) & N &  N \\
(82, 210) & 8.10e+03 & (100, 1.9e-6, 0) & N &  Y (2.96e-04 $-$ 4.21e+05) \\
(83, 210) & 5.01 & (100, 13.2e-5, 0) & N &  N \\
(88, 225) & 1.48e+01 & (100, 0, 0) & N &  Y (9.48e-04 $-$ 9.25e+01) \\
(88, 228) & 2.10e+03 & (100, 0, 0) & N &  Y (2.31e-04 $-$ 1.54e+04) \\
(89, 226) & 1.22 & (83, 0.006, 0) & N &  N \\
(89, 227) & 7.95e+03 & (98.62, 1.38, 0) & N &  Y (2.60e-03 $-$ 4.97e+04) \\
(90, 231) & 1.06 & (100, 0, 0) & N &  Y (9.48e-04 $-$ 7.13) \\
(90, 234) & 2.41e+01 & (100, 0, 0) & N &  Y (2.31e-04 $-$ 1.85e+02) \\
(91, 230) & 1.74e+01 & (7.8, 0.0032, 0) & N &  N \\
(91, 232) & 1.32 & (100, 0, 0) & N &  N \\
(91, 233) & 2.70e+01 & (100, 0, 0) & N &  Y (1.57e-03 $-$ 1.85e+02) \\
(92, 237) & 6.75 & (100, 0, 0) & N &  Y (1.10e-03 $-$ 4.79e+01) \\
(93, 236) & 5.59e+07 & (13.5, 0.16, 0) & N &  N \\
(93, 238) & 2.10 & (100, 0, 0) & N &  N \\
(93, 239) & 2.36 & (100, 0, 0) & N &  Y (5.38e-03 $-$ 1.91e+01) \\
(94, 241) & 5.23e+03 & (100, 0.00245, 0) & N &  Y (1.40e-03 $-$ 5.06e+04) \\
(94, 246) & 1.08e+01 & (100, 0, 0) & N &  Y (2.31e-04 $-$ 8.25e+01) \\
(94, 247) & 2.27 & (100, 0, 0) & N &  Y (2.31e-04 $-$ 1.60e+01) \\
(97, 249) & 3.27e+02 & (100, 0.00145, 47e-9) & N &  Y (4.74e-03 $-$ 3.94e+04) \\
(98, 253) & 1.78e+01 & (99.69, 0.31, 0) & N &  Y (2.93e-03 $-$ 3.33e+02) \\
(99, 254) & 2.76e+02 & (1.74e-4, 100, 0) & N &  N \\
(99, 255) & 3.98e+01 & (92.0, 8.0, 0.0041) & N &  Y (6.37e-03 $-$ 2.39e+02) \\
(99, 257) & 7.70 & (100, 0, 0) & N &  Y (5.99e-03 $-$ 4.19e+01) \\
\end{longtable*}

\begin{longtable*}[]{|l|l|l|l|l|l|l|l|}
    \caption{All isotopes with measured $\alpha$-decay half-lives longer than 1 day (assuming NUBASE2020 decay data).}\label{tab:longAdec} \\
    \hline
    ($Z$, $A$) & $T_{1/2}$ [days] & Branchings & (Z$_d$, A$_d$)  & (Z$_d$, A$_d$) $\beta$ & (Z$_d$, A$_d$) $T_{1/2}$ & In Tab.? & ($Z$, $A$) pop.? (When?) \\
    \endfirsthead
    
    \multicolumn{8}{l}
    {{\bfseries \tablename\ \thetable{} -- continued from previous page}} \\
    \hline 
    ($Z$, $A$) & $T_{1/2}$ [days] & Branchings & (Z$_d$, A$_d$)  & (Z$_d$, A$_d$) $\beta$ & (Z$_d$, A$_d$) $T_{1/2}$ & In Tab.? & ($Z$, $A$) pop.? (When?) \\
    \hline
    \endhead

    \hline\endlastfoot
    \hline
(60, 144) & 8.36e+17 & (0, 100, 0) & (58, 140) & stable & - & - & - \\
(61, 145) & 6.46e+03 & (0, 2.8e-7, 0) & (59, 141) & stable & - & - & - \\
(62, 146) & 2.48e+10 & (0, 100, 0) & (60, 142) & stable & - & - & - \\
(62, 147) & 3.89e+13 & (0, 100, 0) & (60, 143) & stable & - & - & - \\
(62, 148) & 2.30e+18 & (0, 100, 0) & (60, 144) & 0 & - & - & - \\
(63, 147) & 2.41e+01 & (0, 0.0022, 0) & (61, 143) & 0 & - & - & - \\
(63, 148) & 5.45e+01 & (0, 9.4e-7, 0) & (61, 144) & 0 & - & - & - \\
(63, 151) & 1.68e+21 & (0, 100, 0) & (61, 147) & 100 & 9.58e+02 & N & Y (2.01e+02 $-$ 3.65e+11) \\
(64, 148) & 2.60e+04 & (0, 100, 0) & (62, 144) & stable & - & - & - \\
(64, 149) & 9.28 & (0, 4.3e-4, 0) & (62, 145) & 0 & - & - & - \\
(64, 150) & 6.53e+08 & (0, 100, 0) & (62, 146) & 0 & - & - & - \\
(64, 151) & 1.24e+02 & (0, 1.1e-6, 0) & (62, 147) & 0 & - & - & - \\
(64, 152) & 3.94e+16 & (0, 100, 0) & (62, 148) & 0 & - & - & - \\
(66, 154) & 1.10e+09 & (0, 100, 0) & (64, 150) & 0 & - & - & - \\
(72, 174) & 7.30e+17 & (0, 100, 0) & (70, 170) & stable & - & - & - \\
(74, 180) & 5.80e+20 & (0, 100, 0) & (72, 176) & stable & - & - & - \\
(76, 186) & 7.30e+17 & (0, 100, 0) & (74, 182) & stable & - & - & - \\
(78, 188) & 1.02e+01 & (0, 2.6e-5, 0) & (76, 184) & 0 & - & - & - \\
(78, 190) & 1.76e+14 & (0, 100, 0) & (76, 186) & 0 & - & - & - \\
(82, 210) & 8.10e+03 & (100, 1.9e-6, 0) & (80, 206) & 100 & 5.78e-03 & N & Y (2.96e-04 $-$ 4.21e+05) \\
(83, 209) & 7.34e+21 & (0, 100, 0) & (81, 205) & stable & - & - & - \\
(83, 210) & 5.01 & (100, 13.2e-5, 0) & (81, 206) & 100 & 2.92e-03 & N & N \\
(84, 206) & 8.80 & (0, 5.45, 0) & (82, 202) & 0 & - & - & - \\
(84, 208) & 1.06e+03 & (0, 100, 0) & (82, 204) & stable & - & - & - \\
(84, 209) & 4.53e+04 & (0, 99.546, 0) & (82, 205) & 0 & - & - & - \\
(84, 210) & 1.38e+02 & (0, 100, 0) & (82, 206) & stable & - & - & - \\
(86, 222) & 3.82 & (0, 100, 0) & (84, 218) & 0.020 & 2.15e-03 & N & Y (7.04e-04 $-$ 2.25e+01) \\
(88, 223) & 1.14e+01 & (0, 100, 0) & (86, 219) & 0 & - & - & - \\
(88, 224) & 3.63 & (0, 100, 0) & (86, 220) & 0 & - & - & - \\
(88, 226) & 5.84e+05 & (0, 100, 0) & (86, 222) & 0 & - & - & - \\
(89, 225) & 9.92 & (0, 100, 0) & (87, 221) & 0.0048 & 3.33e-03 & N & Y (3.22e-01 $-$ 1.06e+02) \\
(89, 226) & 1.22 & (83, 0.006, 0) & (87, 222) & 100 & 9.86e-03 & N & N \\
(89, 227) & 7.95e+03 & (98.62, 1.38, 0) & (87, 223) & 100 & 1.53e-02 & N & Y (2.60e-03 $-$ 4.97e+04) \\
(90, 227) & 1.87e+01 & (0, 100, 0) & (88, 223) & 0 & - & - & - \\
(90, 228) & 6.98e+02 & (0, 100, 0) & (88, 224) & 0 & - & - & - \\
(90, 229) & 2.89e+06 & (0, 100, 0) & (88, 225) & 100 & 1.48e+01 & N & Y (3.86e-03 $-$ 1.30e+09) \\
(90, 230) & 2.75e+07 & (0, 100, 0) & (88, 226) & 0 & - & - & - \\
(90, 232) & 5.11e+12 & (0, 100, 1.1e-9) & (88, 228) & 100 & 2.10e+03 & N & Y (5.18e-04 $-$ 3.65e+11) \\
(91, 229) & 1.55 & (0, 0.49, 0) & (89, 225) & 0 & - & - & - \\
(91, 230) & 1.74e+01 & (7.8, 0.0032, 0) & (89, 226) & 83 & 1.22 & N & N \\
(91, 231) & 1.19e+07 & (0, 100, 0) & (89, 227) & 98.62 & 7.95e+03 & N & Y (2.61e-02 $-$ 8.03e+07) \\
(92, 230) & 2.02e+01 & (0, 100, 0) & (90, 226) & 0 & - & - & - \\
(92, 231) & 4.20 & (0, 0.004, 0) & (90, 227) & 0 & - & - & - \\
(92, 232) & 2.52e+04 & (0, 100, 2.7e-12) & (90, 228) & 0 & - & - & - \\
(92, 233) & 5.81e+07 & (0, 100, 0) & (90, 229) & 0 & - & - & - \\
(92, 234) & 8.96e+07 & (0, 100, 1.64e-9) & (90, 230) & 0 & - & - & - \\
(92, 235) & 2.57e+11 & (0, 100, 7e-9) & (90, 231) & 100 & 1.06 & N & Y (2.99e-03 $-$ 3.65e+11) \\
(92, 236) & 8.55e+09 & (0, 100, 9.4e-8) & (90, 232) & 0 & - & - & - \\
(92, 238) & 1.63e+12 & (0, 100, 5.44e-5) & (90, 234) & 100 & 2.41e+01 & N & Y (5.65e-04 $-$ 3.65e+11) \\
(93, 235) & 3.96e+02 & (0, 0.00260, 0) & (91, 231) & 0 & - & - & - \\
(93, 236) & 5.59e+07 & (13.5, 0.16, 0) & (91, 232) & 100 & 1.32 & N & N \\
(93, 237) & 7.83e+08 & (0, 100, 0) & (91, 233) & 100 & 2.70e+01 & N & Y (9.14e-02 $-$ 7.93e+09) \\
(94, 236) & 1.04e+03 & (0, 100, 1.9e-7) & (92, 232) & 0 & - & - & - \\
(94, 237) & 4.56e+01 & (0, 0.0042, 0) & (92, 233) & 0 & - & - & - \\
(94, 238) & 3.20e+04 & (0, 100, 1.9e-7) & (92, 234) & 0 & - & - & - \\
(94, 239) & 8.80e+06 & (0, 100, 3.1e-10) & (92, 235) & 0 & - & - & - \\
(94, 240) & 2.40e+06 & (0, 100, 5.796e-6) & (92, 236) & 0 & - & - & - \\
(94, 241) & 5.23e+03 & (100, 0.00245, 0) & (92, 237) & 100 & 6.75 & N & Y (1.40e-03 $-$ 5.06e+04) \\
(94, 242) & 1.37e+08 & (0, 100, 5.510e-4) & (92, 238) & 0 & - & - & - \\
(94, 244) & 2.97e+10 & (0, 99.877, 0.123) & (92, 240) & 100 & 5.88e-01 & N & Y (2.31e-04 $-$ 2.70e+11) \\
(95, 240) & 2.12 & (0, 1.9e-4, 0) & (93, 236) & 13.5 & 5.59e+07 & N & N \\
(95, 241) & 1.58e+05 & (0, 100, 3.6e-10) & (93, 237) & 0 & - & - & - \\
(95, 243) & 2.68e+06 & (0, 100, 3.7e-9) & (93, 239) & 100 & 2.36 & N & Y (5.60e-03 $-$ 2.29e+07) \\
(96, 240) & 3.04e+01 & (0, 100, 3.9e-6) & (94, 236) & 0 & - & - & - \\
(96, 241) & 3.28e+01 & (0, 1.0, 0) & (94, 237) & 0 & - & - & - \\
(96, 242) & 1.63e+02 & (0, 100, 6.2e-6) & (94, 238) & 0 & - & - & - \\
(96, 243) & 1.06e+04 & (0, 100, 5.3e-9) & (94, 239) & 0 & - & - & - \\
(96, 244) & 6.61e+03 & (0, 100, 1.37e-4) & (94, 240) & 0 & - & - & - \\
(96, 245) & 3.01e+06 & (0, 100, 6.1e-7) & (94, 241) & 100 & 5.23e+03 & N & Y (3.44e-02 $-$ 2.86e+07) \\
(96, 246) & 1.72e+06 & (0, 99.97, 0.026) & (94, 242) & 0 & - & - & - \\
(96, 247) & 5.69e+09 & (0, 100, 0) & (94, 243) & 100 & 2.06e-01 & Y & Y (4.94e-02 $-$ 4.57e+10) \\
(96, 248) & 1.27e+08 & (0, 91.61, 8.39) & (94, 244) & 0 & - & - & - \\
(97, 245) & 4.95 & (0, 0.12, 0) & (95, 241) & 0 & - & - & - \\
(97, 247) & 5.04e+05 & (0, 100, 0) & (95, 243) & 0 & - & - & - \\
(97, 249) & 3.27e+02 & (100, 0.00145, 47e-9) & (95, 245) & 100 & 8.54e-02 & N & Y (4.74e-03 $-$ 3.94e+04) \\
(98, 246) & 1.49 & (0, 100, 2.4e-4) & (96, 242) & 0 & - & - & - \\
(98, 248) & 3.34e+02 & (0, 100, 0.0029) & (96, 244) & 0 & - & - & - \\
(98, 249) & 1.28e+05 & (0, 100, 5.0e-7) & (96, 245) & 0 & - & - & - \\
(98, 250) & 4.77e+03 & (0, 99.923, 0.077) & (96, 246) & 0 & - & - & - \\
(98, 251) & 3.28e+05 & (0, 100, 0) & (96, 247) & 0 & - & - & - \\
(98, 252) & 9.66e+02 & (0, 96.8972, 3.1028) & (96, 248) & 0 & - & - & - \\
(98, 253) & 1.78e+01 & (99.69, 0.31, 0) & (96, 249) & 100 & 4.45e-02 & N & Y (2.93e-03 $-$ 3.33e+02) \\
(98, 254) & 6.05e+01 & (0, 0.31, 99.69) & (96, 250) & ? & - & - & - \\
(99, 251) & 1.38 & (0, 0.5, 0) & (97, 247) & 0 & - & - & - \\
(99, 252) & 4.72e+02 & (0, 78, 0) & (97, 248) & ? & - & - & - \\
(99, 253) & 2.05e+01 & (0, 100, 8.7e-6) & (97, 249) & 100 & 3.27e+02 & N & Y (3.63e-01 $-$ 3.91e+02) \\
(99, 254) & 2.76e+02 & (1.74e-4, 100, 0) & (97, 250) & 100 & 1.34e-01 & N & N \\
(99, 255) & 3.98e+01 & (92.0, 8.0, 0.0041) & (97, 251) & 100 & 3.86e-02 & N & Y (6.37e-03 $-$ 2.39e+02) \\
(100, 252) & 1.06 & (0, 100, 0.0023) & (98, 248) & 0 & - & - & - \\
(100, 253) & 3.00 & (0, 12, 0) & (98, 249) & 0 & - & - & - \\
(100, 257) & 1.00e+02 & (0, 99.790, 0.210) & (98, 253) & 99.69 & 1.78e+01 & N & Y (3.69e-01 $-$ 6.26e+02) \\
(101, 258) & 5.16e+01 & (0, 100, 0) & (99, 254) & 1.74e-4 & 2.76e+02 & N & N \\
\end{longtable*}

\section{Comparison to isotopes highlighted in other literature}\label{sec:litcomp}
We lastly report on whether the literature on MeV gamma emission from mergers have previously highlighted the species we reported in our tables. We first consider five works which use nuclear data and nucleosynthesis tools that are distinct from those employed here: [1] - \cite{Li2019_MeV}, [2] - \cite{Wu_remnants}, [3] - \cite{Korobkin2020_MeV}, [4] - \cite{Chen2021_MeV}, [5] - \cite{Terada2022_MeV}. We then compare to a more recent work ([6] - \cite{Gross2025}) which uses similar nuclear data and the same nucleosynthesis network as that employed by our calculations, as well as our recent work which first highlighted Tl-208 \cite{VasshTl208} (which focused on the 2.5-2.8 MeV energy window only).

In considering references [1]-[5], 31 were isotopes commonly reported across our work and this literature (K-42, Fe-59, Cu-66, Cu-67, Zn-72, Ga-72, Ge-77, Kr-85, Kr-88, Zr-95, Nb-95, Ru-103, Rh-106, Ag-112, Sn-125, Sn-127, Sb-125, Sb-126, Sb-127, Sb-128, Sb-129, I-131, I-132, I-133, I-135, Xe-133, La-140, Ir-194, Pb-214, and Am-246). This highlights that some emission lines are robust across input data, astrophysical variations, network set-ups and analysis techniques. Our tables reported on 21 species that were not reported in references [1]-[5] (Na-24, Co-60, Rb-88, Y-91, In-128, Sb-134 I-136, La-142, Pr-144, Eu-155, Eu-156, Hf-181, Ta-182, Ta-184, Re-188, Tl-208, Tl-209, Pb-211, Ac-228, and Pu-243). However it is important to note that these references also report emission from species that we do not see to be a dominant emitter with our peak finding method: (Sc-48, Br-82, Nb-96, Cs-136, Pm-148) [1], (I-129, Hf-182, Pb-206, Th-230, Pa-231, U-233, U-234, U-235, Np-237, Pu-239, Pu-240, Pu-242, Am-243, Pu-244, Cm-245, Cm-247, Cm-250) [2], (Y-92, Nb-97, Xe-135, Ce-141, Pt-197, Bi-214, Pa-233, Np-239, Am-243, Am-241, Cm-245, Bk-250) [3], (Mn-56, Ga-73, Se-83, Kr-87, Sr-91, Ru-105, Sb-130, Te-132, Xe-135, Bi-211) [4], and (Sc-44, Zn-65, Kr-81, Nb-92, Nb-94, I-129, Os-194, Bi-214, Th-227, Pa-233, Np-239, Am-241, Am-243) [5]. Most of the species reported in [2] that are not featured in this work undergo $\alpha$-decay (e.g. Cm-245) and do not get highlighted in our tables likely due to their having subdominant, lower energy emission relative to $\beta$-decay until later times when signals become fainter (e.g. Fig.~\ref{fig:falpha}) (thus these may not be reported given our threshold criterion). Regarding the differences between our work and refs. [1], [3], [4], and [5], most of these species listed undergo $\beta$-decay but were not reported in our tables due to not being highlighted by our peak finding procedure, either because their spectral features were overtaken by other emitters or they were only significantly populated at a time earlier than the smallest time considered in our tables of 0.001 day (1.44 min). The differences across the literature well represent the impact of different nuclear data treatments as well as analysis assumptions.

When we compare to \cite{VasshTl208}, here Rh-106, Ag-112, Sn-127, La-140, La-142, and Tl-208 were all noted as nuclei of interest for emission in the 2.5-2.8 MeV range of Tl-208 emission, and all 6 of these species are reported in the tables of this work. Lastly comparing to the recent work of \cite{Gross2025} [6] which uses similar data and the same network as our calculation, we find that nearly all species reported here were also reported to be emitters in [6], with only Sb-134, I-136, Pu-243 being uniquely reported by this current work. However, similarly to references [1]-[5], the more recent work of ref. [6] also reports numerous nuclei that we do not find meet our visibility over threshold / dominance criterion and so are not listed in our tables.

\end{document}